\documentclass[11pt,letterpaper]{article}

\pdfoutput=1

\PassOptionsToPackage{nosort}{cite}

\usepackage{textcomp}
\usepackage[british]{chet}
\usepackage{mathtools}
\usepackage{amssymb,amsmath,amsfonts,mathrsfs,braket}
\usepackage[T1]{fontenc}
\usepackage{tikz}
\usepackage{diagbox}
\usepackage{tcolorbox}
\usepackage{bm}
\usepackage[colorinlistoftodos,textsize=tiny]{todonotes}
\usepackage{hyperref}

\usetikzlibrary{external,arrows.meta,decorations.markings}
\tikzset{external/figure name={ancb2-figure}}

\makeatletter
\renewcommand{\todo}[2][]{\tikzexternaldisable\@todo[#1]{#2}\tikzexternalenable}
\makeatother

\tcbset{shield externalize}
\tcbuselibrary{theorems,skins,breakable}

\usepackage{float}
\usepackage{subcaption}
\usepackage[font=small,format=hang,labelfont={sf,bf}]{caption}
\usepackage{pgfplots}
\usepackage{helvet}
\usepackage[eulergreek]{sansmath}
\pgfplotsset{compat=1.18,
    xlabel shift=-2pt,
    ylabel shift=-4pt,
    every picture/.append style={trim axis left, trim axis right},
    every axis label/.append style={font=\sansmath\sffamily\footnotesize},
    every axis title/.append style={font=\sansmath\sffamily\footnotesize, yshift=-0.75ex},
    every tick label/.append style={font=\sansmath\sffamily\footnotesize},
    every axis legend/.append style={font=\sansmath\sffamily\footnotesize}
}
\usepgfplotslibrary{fillbetween}
\usepgfplotslibrary{statistics}
\definecolor{forestgreen}{rgb}{0.0, 0.27, 0.13}
\definecolor{darkblue}{rgb}{0.0, 0.0, 0.55}
\definecolor{darkred}{rgb}{0.55, 0.0, 0.0}
\hypersetup{colorlinks,
           linkcolor={forestgreen},
           citecolor={darkblue},
           urlcolor={darkred},
           pdftitle={Neural Spectral Bias and Conformal Correlators II: Modular and Annulus Bootstrap},
           pdfdisplaydoctitle
}

\newtcolorbox{cluster}[1][]{
  colback=blue!5!white,
  colframe=blue!75!black,
  title={$\triangleright$ Cluster runs},
  #1
}

\makeatletter
\g@addto@macro\bfseries{\boldmath}
\makeatother

\newcommand{\lsp}{\hspace{0.5pt}}

\renewcommand{\ge}{\geqslant}
\renewcommand{\le}{\leqslant}

\def\one{{\,\hbox{1\kern-.8mm l}}}

\allowdisplaybreaks

\DeclareMathOperator{\Ima}{Im}

\def\ZZ{{\cal Z}}

\def\ZZL{{\rm ZZ}}
\def\FZZ{{\rm FZZ}}

\def\beq{\begin{equation}}
\def\eeq{\end{equation}}
\newcommand{\bea}{\begin{eqnarray}}
\newcommand{\eea}{\end{eqnarray}}
\def\bal{\begin{align}}
\def\eal{\end{align}}

\definecolor{emerald}{rgb}{0.31, 0.78, 0.47}

\preprint{CCTP-2026-16  \\ ITCP-IPP 2026/16}

\title{Neural Spectral Bias and Conformal Correlators II\\[5pt]
Modular and Annulus Bootstrap} 

\author{Kausik Ghosh,$^{\tau,}$\email{kau.rock91@gmail.com}
Sidhaarth Kumar,$^{\tau,}$\email{sidhaarth.kumar@kcl.ac.uk}
Vasilis Niarchos,$^{\bar{\tau},}$\email{niarchos@physics.uoc.gr} 
Andreas Stergiou$^{\tau,}$\email{andreas.stergiou@kcl.ac.uk}}

\affiliation{$^\tau$Department of Mathematics, King's College London, Strand, London WC2R 2LS, United Kingdom\\
$^{\bar{\tau}}$Department of Physics, ITCP \& CCTP, University of Crete, 71003 Heraklion, Greece}

\abstract{We develop a neural network bootstrap framework for reconstructing partition functions of two-dimensional conformal field theories (CFTs) based on modular invariance and the Cardy condition, which are recast as crossing equations for four-point correlators. For torus partition functions, we use the twist-field representation in the symmetric-orbifold description to map modular $S$-invariance to four-point crossing and focus on the diagonal kinematics of four insertions on a line. For annulus partition functions, we formulate open/closed channel duality as crossing symmetry for mixed four-point functions of defect-changing operators in interface CFT. In both cases, the reconstruction problem is formulated in the anchored-bootstrap form, where the crossing constraints are supplemented by minimal spectral input (a gap) and anchor data. We solve this under-determined problem by using lightweight feed-forward neural networks to parametrise the correlators and their corresponding partition functions. A key ingredient of this approach is the spectral bias of the neural networks in the lazy training regime, which selects specific crossing-symmetric configurations. This reformulation unifies standard modular and annulus constraints in two dimensions with the anchored neural approach for CFT correlators, providing a new way to reconstruct full partition functions from sparse data with remarkable accuracy.}

\date{July 2026}

\begin{document}

\maketitle

\toc

\section{Introduction}
\label{intro}
The non-perturbative exploration of the space of consistent conformal field theories (CFTs) has become a central theme in modern theoretical physics, driven in large part by the success of bootstrap methods \cite{Poland:2018epd, RevModPhys.96.045004, Rychkov:2025zks}. Conventionally, in the conformal bootstrap programme, the search space is taken to be the space of CFT data, namely the spectrum of local operators together with their operator product expansion (OPE) coefficients. In unitary theories, the reality of OPE coefficients allows the crossing equations arising from four-point functions to be recast as a convex optimisation problem \cite{Rattazzi:2008pe} and, more specifically, as a positive semidefinite program \cite{Kos:2014bka, Simmons-Duffin:2016wlq}. This has led to a remarkably powerful way of carving out the space of consistent CFT data \cite{El-Showk:2012cjh,Chang:2024whx}.

A different approach was proposed in our previous work \cite{GKNS:0, GKNS:1}. Instead of directly searching over CFT data, we proposed to search directly in the space of crossing-symmetric correlators. The key idea was to use the spectral bias of neural networks, together with a small amount of physical input, to select the desired solution of the crossing equation. The strength of this approach lies in its generality. Since it does not rely on unitarity or positivity, it can be applied to both unitary and non-unitary CFTs, as well as to thermal correlators. Although at present the method does not provide rigorous error bounds on the predicted correlators, all test examples studied so far show remarkable accuracy in reconstructing the crossing-symmetric solutions of interest. As a notable example, we point out that the method gives specific predictions for the thermal two-point function in the three-dimensional Ising CFT, as well as for the four-point function of the energy operator in the same theory \cite{GKNS:0, GKNS:1}.

Given this generality, it is natural to ask whether the same philosophy can be applied to other consistency conditions. In two dimensions, the consistency conditions of CFT are especially stringent. Crossing symmetry of local correlators and modular invariance of the torus partition function are low-genus manifestations of the more general sewing constraints on Riemann surfaces. In rational conformal field theories, the Moore--Seiberg analysis shows that these constraints can be organised into algebraic consistency conditions ensuring compatibility of CFT data on Riemann surfaces of arbitrary genus \cite{Moore:1988qv}. Motivated by this structure, in the present paper we turn to the modular bootstrap and explore whether neural networks can be used to learn functions constrained by modular invariance.

One of the targets of this work is the modular invariance of the torus partition function,
\begin{equation}
    \mathcal{Z}_{\mathcal T}(\tau,\bar\tau)
    = \mathcal{Z}_{\mathcal T}\left(-\frac{1}{\tau},-\frac{1}{\bar\tau}\right).
    \label{eq:modular-inv-intro}
\end{equation}
We use the fact that this problem can be reformulated as an equivalent problem of a crossing-symmetric four-point function on the sphere. More precisely, the torus partition function of the CFT can be related to the four-point function $G(z,\bar z)$ of twist operators in the $\mathbb Z_2$ symmetric product orbifold of the theory. The kinematic restriction to an imaginary modular parameter $\tau=-\bar \tau$, which we choose to study in this paper, maps in this context to the diagonal limit of the corresponding four-point function. In this limit, $G(z,\bar z)$ becomes a single-variable function, which we denote by $G(z)$, and the modular crossing equation reads
\begin{equation}
    G(z)=\left(\frac{z}{1-z}\right)^{c/4}G(1-z)\,,\qquad z\in(0,1)\,.
    \label{eq:crossing-intro}
\end{equation}
Following the practice of our previous work \cite{GKNS:0,GKNS:1}, we factor out a gap-dependent prefactor from $G(z)$ and the neural network is then trained to learn the smoother left-over function. In addition, we provide as extra physical input the value of the correlator at a single anchor point $z_0\in(0,1)$. The precise value of $z_0$ is unimportant as long as $z_0$ is not too close to 0 or 1. In the reported examples we set $z_0=0.3$. With the corresponding anchor value given, the goal is to reconstruct the correlator over the full interval $0<z<1$. 

This reconstruction problem does not have a unique solution. There are many crossing-symmetric, or equivalently modular-invariant, functions that satisfy the same constraints, and the neural network could, in principle, have found many different solutions, the vast majority of which would not be the correlator or the partition function of a fully-fledged consistent CFT. Nevertheless, our previous study \cite{GKNS:0,GKNS:1} revealed a remarkable fact. In all examples studied there, the network always selected a solution close to a physical correlator. The main goal of the present paper is to present experimental evidence that this bias towards physical correlators extends to modular-invariant partition functions as well.

Besides the torus partition function, another target of our study is partition functions in the presence of a boundary. We focus on the annulus partition function, where modular consistency relates the open- and closed-channel descriptions. This relation can again be mapped to a crossing equation---in this case, the crossing equation of a mixed-correlator system. Similar systems of mixed four-point correlators were studied in our previous work \cite{GKNS:0,GKNS:1}. Therefore, we apply the same neural-network strategy to the annulus setup and find that the method continues to reconstruct the expected partition functions to remarkable accuracy in this case too.

Overall, these results are very encouraging. In the context of 2d CFTs, they suggest that neural networks provide a useful new way to navigate the space of modular-invariant functions. More broadly, they point toward a way of searching directly in the space of functions constrained by consistency conditions, rather than searching only in the space of CFT data. A complementary numerical approach was used in \cite{Benjamin:2026lbj}, where modular invariance was imposed as a loss function and optimised directly over candidate primary spectra.

The rest of the paper is organised as follows.  Section~\ref{sec:review} reviews the modular and Cardy crossing reformulations.  Section~\ref{sec:nn} introduces the neural network setup.  Sections~\ref{sec:minimal} and~\ref{sec:annulus-results} collect the evidence in the context of compact CFTs. In the main text, we present characteristic examples of torus and annulus reconstructions for minimal models and Wess--Zumino--Witten (WZW) models, relegating further cases to the Appendix \ref{app:extra} and the companion \href{https://github.com}{\texttt{GitHub}} repository \href{https://github.com/andstergiou/nn-cft}{\tt andstergiou/nn-cft}, archived on \href{https://zenodo.org/}{\texttt{Zenodo}}~\cite{nn_cft_code}. Section \ref{noncompact} is devoted to the study of torus and annulus partition functions in non-compact CFTs. As an illustration, we single out the free non-compact boson and the Liouville CFT in the case of the torus partition function and the ZZ and FZZ boundaries in Liouville theory in the case of the annulus partition functions. Section~\ref{sec:outlook} discusses the scope, limitations and future prospects of the approach.

\section{Review of Modular and Annulus Bootstrap}
\label{sec:review}

The full torus partition function depends on the complex modulus $\tau$, while the annulus partition function depends on a real modulus together with a choice of boundary conditions. Both obey basic consistency conditions. For the torus, it is modular invariance. For the annulus, it is Cardy consistency, or equivalently open/closed-channel duality. There is a rich literature on using standard bootstrap treatments of these constraints to bound admissible spectral data, e.g.~\cite{Hellerman:2009bu,Friedan:2013cba,Collier:2016cls, Hartman:2019pcd, Collier:2021ngi, Afkhami-Jeddi:2020hde, Erramilli:2026esf}. We review these conditions below, emphasising the geometric reformulation that turns the corresponding partition-function identities into crossing equations on $\mathbb{CP}^1$. The torus partition function is realised as a four-point function of identical twist fields on a branched cover. The annulus partition function is realised as a mixed four-point function of defect-changing operators (DCOs). In this formulation, Cardy consistency becomes the crossing equation of this mixed correlator.

\subsection{Modular Bootstrap}\label{subsec:modular}
We start with the torus partition function of a two-dimensional CFT with central charge $c$,
\begin{equation}
\mathcal{Z}(\tau,\bar{\tau})
=
\mathrm{Tr}_{\mathcal H}
\left(
q^{L_0-\frac{c}{24}}
\bar q^{\bar L_0-\frac{c}{24}}
\right),
\qquad
q=e^{2\pi i\tau},
\qquad
\bar q=e^{-2\pi i\bar\tau}.
\label{eq:torus-partition-function}
\end{equation}
Here $\tau$ is the complex modulus of the torus, the trace is over the Hilbert space $\mathcal{H}$ of the CFT quantised on a circle, and $L_0, \bar{L}_0$ are the Virasoro zero-mode generators. With the above convention for $q$ and $\bar q$, the trace expansion is naturally convergent when $\tau$ lies in the upper half-plane, $\mathbb{H}_+$, and $\bar\tau$ lies in the lower half-plane, $\mathbb{H}_-$. On the physical Euclidean slice these variables are related by complex conjugation, $\bar\tau=\tau^*$. It is nevertheless useful to complexify the discussion and treat $\tau$ and $\bar\tau$ as independent variables. In this sense, the torus partition function defines a holomorphic function on the complex plane, whose restriction to $\bar\tau=\tau^*$ gives the ordinary Euclidean partition function. Notice that when $\tau=-\bar \tau = i t$ is purely imaginary (a case we will study extensively below), $\ZZ(it,-it)=\mathrm{Tr}_{\mathcal H}\left( e^{-2\pi t (L_0+\bar L_0-\frac{c}{12})}\right)$ is the thermal partition function.

The torus partition function is modular invariant. The basic transformations
\begin{equation}
T:\tau\mapsto \tau+1\,,
\qquad
S:\tau\mapsto -\frac{1}{\tau}
\end{equation}
generate the modular group
\begin{equation}
\tau\mapsto \frac{a\tau+b}{c\tau+d}\,,
\qquad
\bar\tau\mapsto \frac{a\bar\tau+b}{c\bar\tau+d}\,,
\end{equation}
where
\begin{equation}
\begin{pmatrix}
a & b\\
c & d
\end{pmatrix}
\in SL(2,\mathbb Z)\,,
\end{equation}
and modular invariance is the statement that
\begin{equation}
\mathcal Z(\tau,\bar\tau)
=
\mathcal Z\left(
\frac{a\tau+b}{c\tau+d},
\frac{a\bar\tau+b}{c\bar\tau+d}
\right).
\end{equation}
Invariance under the $S$-transformation,
\begin{equation}
\mathcal Z(\tau,\bar\tau)
=
\mathcal Z\left(
-\frac{1}{\tau},
-\frac{1}{\bar\tau}
\right),
\end{equation}
will be the most relevant one for us in what follows.

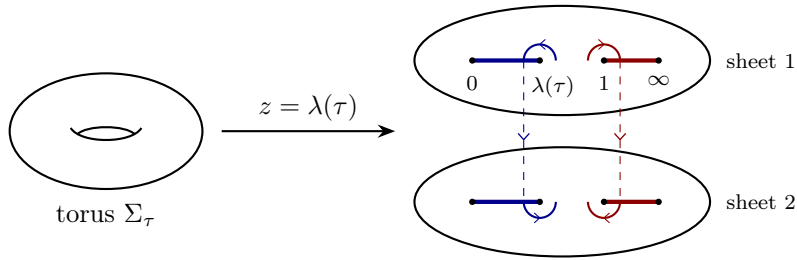
\begin{figure}[H]
\centering
\begin{tikzpicture}[>=Stealth, scale=0.85]
\begin{scope}[shift={(-3.5,0)}]
\draw[thick] (0,0) ellipse [x radius=1.5, y radius=0.9];
\draw[thick] (-0.55,0.05) .. controls (-0.4,-0.20) and (0.4,-0.20) .. (0.55,0.05);
\draw[thick] (-0.45,-0.05) .. controls (-0.30,0.10) and (0.30,0.10) .. (0.45,-0.05);
\node[font=\small] at (0,-1.30) {torus $\Sigma_{\tau}$};
\end{scope}
\draw[->, thick] (-1.7,0) -- (1.0,0) node[midway, above, font=\small] {$z=\lambda(\tau)$};
\begin{scope}[shift={(3.6, 1.10)}]
\draw[thick] (0,0) ellipse [x radius=2.3, y radius=0.85];
\draw[ultra thick, blue!55!black] (-1.40,0) -- (-0.35,0);
\draw[ultra thick, red!55!black] (0.65,0) -- (1.50,0);
\fill (-1.40,0) circle (1.4pt);
\fill (-0.35,0) circle (1.4pt);
\fill (0.65,0) circle (1.4pt);
\fill (1.50,0) circle (1.4pt);
\node[below=2pt, font=\scriptsize] at (-1.40,0) {$0$};
\node[below=2pt, font=\scriptsize, xshift=5pt] at (-0.35,0) {$\lambda(\tau)$};
\node[below=2pt, font=\scriptsize] at (0.65,0) {$1$};
\node[below=2pt, font=\scriptsize] at (1.50,0) {$\infty$};
\node[font=\scriptsize, anchor=west] at (2.40,0) {sheet 1};
\end{scope}
\begin{scope}[shift={(3.6, -1.10)}]
\draw[thick] (0,0) ellipse [x radius=2.3, y radius=0.85];
\draw[ultra thick, blue!55!black] (-1.40,0) -- (-0.35,0);
\draw[ultra thick, red!55!black] (0.65,0) -- (1.50,0);
\fill (-1.40,0) circle (1.4pt);
\fill (-0.35,0) circle (1.4pt);
\fill (0.65,0) circle (1.4pt);
\fill (1.50,0) circle (1.4pt);
\node[font=\scriptsize, anchor=west] at (2.40,0) {sheet 2};
\end{scope}
\tikzset{
  midarrow/.style={postaction={decorate, decoration={markings, mark=at position 0.5 with {
    \draw[line width=0.4pt, line cap=round] (0,0) -- (-2pt, 1.8pt);
    \draw[line width=0.4pt, line cap=round] (0,0) -- (-2pt,-1.8pt);
  }}}},
  middashed/.style={postaction={decorate, decoration={markings, mark=at position 0.56 with {
    \draw[line width=0.5pt, line cap=round] (0,0) -- (-3pt, 2.7pt);
    \draw[line width=0.5pt, line cap=round] (0,0) -- (-3pt,-2.7pt);
  }}}}
}
\draw[thick, blue!55!black, midarrow]
  ([shift={(3.6, 1.10)}]-0.10, 0)
  arc[start angle=0, end angle=180, radius=0.25];
\draw[dashed, thin, blue!55!black, middashed] (3.0, 1.10) -- (3.0, -1.10);
\draw[thick, blue!55!black, midarrow]
  ([shift={(3.6, -1.10)}]-0.60, 0)
  arc[start angle=180, end angle=360, radius=0.25];
\draw[thick, red!55!black, midarrow]
  ([shift={(3.6, 1.10)}]0.40, 0)
  arc[start angle=180, end angle=0, radius=0.25];
\draw[dashed, thin, red!55!black, middashed] (4.50, 1.10) -- (4.50, -1.10);
\draw[thick, red!55!black, midarrow]
  ([shift={(3.6, -1.10)}]0.90, 0)
  arc[start angle=0, end angle=-180, radius=0.25];
\end{tikzpicture}
\caption{The torus $\Sigma_{\tau}$ presented as a two-sheeted branched cover of $\mathbb{CP}^{1}$ with branch points $\{0,\lambda(\tau),1,\infty\}$.  The two sheets (top and bottom) are pairwise identified along the cuts $[0,\lambda(\tau)]$ (blue) and $[1,\infty)$ (red).  The blue and red loops illustrate the $\mathbb Z_2$ monodromy at the two cuts; encircling a branch point once on sheet 1 (upper arc) crosses the corresponding cut and continues on sheet 2 (lower arc), so a single revolution in the base swaps sheets.  Inserting four $\mathbb Z_2$ twist fields $\sigma_2$ at the branch points in $\mathrm{Sym}^{2}(\mathcal T)$ realises the path integral on $\Sigma_{\tau}$ as a four-point function on $\mathbb{CP}^{1}$.}
\label{fig:torus-cover-2}
\end{figure}

There is another useful way to think about the torus partition function. A torus can be realised as a double cover of the sphere, branched over four points.\footnote{See \cite{Lunin:2000yv,Cardy:2007mb,Maldacena:2015iua,Hartman:2019pcd} for detailed discussions of this map.} By a conformal transformation on the sphere, three of these branch points can be fixed to $0, 1,$ and $\infty$. The position of the fourth point is then the cross-ratio, which we denote by $z$. Thus, the complex structure of the torus can be encoded either by the modulus $\tau$, or equivalently by the cross-ratio $z$ of four branch points on $\mathbb{CP}^1$ (see Fig.\ \ref{fig:torus-cover-2}). In CFT, this branched-cover construction has a natural interpretation in the $\mathbb Z_2$ symmetric product orbifold of the original theory. The branch points are represented by twist fields. Going around a twist field exchanges the two copies of the CFT, the same way that going around a branch point exchanges the two sheets of the cover. Therefore, the torus path integral of the original CFT can be rewritten as a sphere four-point function of twist fields in the orbifold theory. More explicitly, the map between the two descriptions is controlled by the modular lambda function,
\begin{equation}
   z=\lambda(\tau)=\frac{\theta_2(\tau)^4}{\theta_3(\tau)^4}\,,
\end{equation}
where $\theta_2,\theta_3$ are Jacobi theta functions. Conversely, on the standard branch, the modulus is determined by the cross-ratio through
\begin{equation}
   \tau(z)
   =
   i\lsp\frac{K(1-z)}{K(z)}\,,
\end{equation}
where $K(z)$ is the complete elliptic integral of the first kind. Thus, the cross-ratio $z$ determines the complex structure of the double cover. In particular, for $z\in(0,1)$, the modular parameter is purely imaginary, $\tau\in i\mathbb{R}_{>0}$.

Under this map, the torus partition function is related to a four-point function
\begin{equation}
G(z,\bar z)
=
\langle \sigma_2(0)\sigma_2(z,\bar z)\sigma_2(1)\sigma_2(\infty) \rangle
\end{equation}
of identical $\mathbb Z_2$ twist fields with conformal dimension $\Delta_{\sigma_2}=c/8$. Here $G$ denotes the reduced correlator, in which the standard two-point kinematic factor $(x_{12}^2 x_{34}^2)^{\Delta_{\sigma_2}}$ has been stripped off. It is this stripped-off function that obeys the standard crossing relation below. The precise relation contains a universal prefactor fixed by the conformal anomaly and by the Weyl transformation between the two metrics. The important point for us is that modular transformations of the torus become crossing transformations of the four branch-point insertions on the sphere. In particular, the modular $S$-transformation corresponds to exchanging the relevant pair of branch points, which acts as $z\mapsto 1-z$ on the cross-ratio,
\begin{equation}
\mathcal Z(\tau,\bar\tau)
=
\mathcal Z\!\left(
-\frac{1}{\tau},
-\frac{1}{\bar\tau}
\right)
\qquad
\Longleftrightarrow
\qquad
G(z,\bar z)
=
\left|
\frac{z}{1-z}
\right|^{\frac{c}{4}}
G(1-z,1-\bar z) \,.
\label{eq:modular-crossing-map}
\end{equation}

 Our ultimate goal is to reconstruct $G(z,\bar z)$, or equivalently the torus partition function for any complex modular parameter, from a minimal set of inputs. As emphasised in \cite{GKNS:0,GKNS:1}, instead of directly targeting the full function $G(z,\bar z)$ on the complex plane, it is advantageous to restrict first to the diagonal kinematics,
\begin{equation}
\label{diag_kin}
z=\bar z\,,
\qquad
0<z<1 \,.
\end{equation}
On this slice, the crossing equation takes the form
\begin{equation}
G(z)
=
\left(\frac{z}{1-z}\right)^{c/4}
G(1-z)\,,
\qquad
0<z<1 \,.
\label{eq:diagonal-modular-crossing}
\end{equation}
This one-dimensional equation will be one of the key ingredients in the neural-network training that will be considered below. After learning the solution on the diagonal line, it is possible to move away from the line by learning the correlator along concentric circles centred at $z=\bar z=1/2$ at fixed radius $R<\frac{1}{2}$ \cite{GKNS:0,GKNS:1}. However, in this paper, we will not consider this extension to the plane. Our primary goal is to provide evidence of good reconstruction in the diagonal kinematics \eqref{diag_kin} corresponding to pure imaginary modular parameter $\tau=-\bar \tau$.

\subsection{Annulus Bootstrap}\label{subsec:annulus}
\begin{figure}[ht]
\centering
\begin{tikzpicture}[>=Stealth, scale=0.9]

\begin{scope}[shift={(-3.0,0)}]
  \fill[gray!12, even odd rule] (0,0) circle (1.5) (0,0) circle (0.55);
  \draw[thick] (0,0) circle (1.5);
  \draw[thick] (0,0) circle (0.55);
  \node[font=\small] at (0, 1.75) {$\beta$};
  \node[font=\small] at (0, 0.32) {$\alpha$};
  \draw[very thick, red!65!black] (0.55,0) -- (1.5,0);
  \draw[<->, thin, red!65!black] (0.57,-0.20) -- (1.48,-0.20)
    node[midway, below=1pt, font=\scriptsize, red!65!black] {$L$};
  \draw[->, thick, blue!60!black]
    (1.025,0) arc[start angle=0, end angle=320, radius=1.025];
  \node[below=12pt, font=\small\bfseries] at (0,-1.5) {open channel};
\end{scope}

\node[font=\Large] at (0,0) {$=$};

\begin{scope}[shift={(2.2,0)}]
  \fill[gray!12, even odd rule] (0,0) circle (1.5) (0,0) circle (0.55);
  \draw[thick] (0,0) circle (1.5);
  \draw[thick] (0,0) circle (0.55);
  \node[font=\small] at (0, 1.78) {$|\beta\rangle$};
  \node[font=\small] at (0, 0.15) {$|\alpha\rangle$};
  \draw[very thick, red!65!black] (0,0) circle (0.8);
  \node[font=\scriptsize, red!65!black] at (0.98,-0.42) {$S^1$};
  \draw[->, thick, blue!60!black] (0,0.57) -- (0,1.48)
    node[midway, right, font=\small, blue!60!black] {$L$};
  \node[below=12pt, font=\small\bfseries] at (0,-1.5) {closed channel};
\end{scope}

\end{tikzpicture}
\caption{The annulus partition function $\mathcal{Z}_{\alpha\beta}$ in the open and closed channels.  The open channel traces over $\mathcal H_{\alpha\beta}$, while the closed channel propagates between boundary states $|\alpha\rangle$ and $|\beta\rangle$.}
\label{fig:open-closed-channels}
\end{figure}
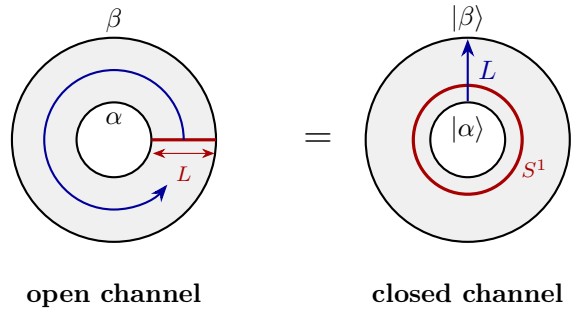

In the same spirit, one can also study partition functions in the presence of boundaries. The simplest example is the annulus partition function. We denote the two boundary conditions by $\alpha$ and $\beta$. The annulus has two natural channel decompositions. In the open channel, the spatial slice is an interval ending on the two boundaries as shown in Fig.\ \ref{fig:open-closed-channels}. The corresponding Hilbert space is the boundary Hilbert space $\mathcal H_{\alpha\beta}$, and the partition function is
\begin{equation}
\mathcal{Z}_{\alpha\beta}^{\rm open}(t)
=
\mathrm{Tr}_{\mathcal H_{\alpha\beta}}\!
\left[
\exp\left(
-2\pi t\left(L_0-\frac{c}{24}\right)
\right)
\right].
\label{eq:annulus-open-channel}
\end{equation}
Here $t$ is the annulus modulus. The same annulus can also be viewed in the closed channel. In this channel, the spatial slice is a circle, and bulk states propagate between two boundary states $|\alpha\rangle$ and $|\beta\rangle$. This gives
\begin{equation}
\mathcal{Z}_{\alpha\beta}^{\rm closed}(1/t)
=
\langle \beta |
\exp\!\left[
-\frac{\pi}{t}
\left(L_0+\bar L_0-\frac{c}{12}\right)
\right]
| \alpha\rangle \,.
\label{eq:annulus-closed-channel}
\end{equation}

\noindent The appearance of $1/t$ reflects the exchange of the two directions of the annulus. In the open channel, Euclidean time runs around the annulus. In the closed channel, Euclidean time runs across the annulus. Thus, the aspect ratio is inverted. The Cardy condition is the statement that the two decompositions describe the same path integral,
\begin{equation}
\mathcal{Z}_{\alpha\beta}^{\rm open}(t)
=
\mathcal{Z}_{\alpha\beta}^{\rm closed}(1/t)\,.
\label{eq:cardy-condition}
\end{equation}
This is the boundary analogue of modular invariance.

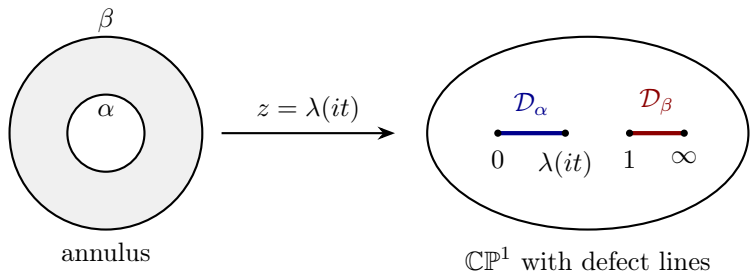
\begin{figure}[H]
\centering
\begin{tikzpicture}[>=Stealth, scale=0.85]
\begin{scope}[shift={(-3.5,0)}]
\fill[gray!12, even odd rule] (0,0) circle (1.5) (0,0) circle (0.6);
\draw[thick] (0,0) circle (1.5);
\draw[thick] (0,0) circle (0.6);
\node[font=\small] at (0, 1.75) {$\beta$};
\node[font=\small] at (0, 0.35) {$\alpha$};
\node[font=\small] at (0,-1.85) {annulus};
\end{scope}
\draw[->, thick] (-1.7,0) -- (1.0,0) node[midway, above, font=\small] {$z=\lambda(it)$};
\begin{scope}[shift={(4.0,0)}]
\draw[thick] (0,0) ellipse [x radius=2.5, y radius=1.5];
\draw[ultra thick, blue!55!black] (-1.40,0) -- (-0.35,0);
\draw[ultra thick, red!55!black] (0.65,0) -- (1.50,0);
\node[above=3pt, font=\small, blue!55!black] at (-0.875,0) {$\mathcal D_{\alpha}$};
\node[above=3pt, font=\small, red!55!black] at (1.075,0) {$\mathcal D_{\beta}$};
\fill (-1.40,0) circle (1.6pt);
\fill (-0.35,0) circle (1.6pt);
\fill (0.65,0) circle (1.6pt);
\fill (1.50,0) circle (1.6pt);
\node[below=3pt, font=\small] at (-1.40,0) {$0$};
\node[below=3pt, font=\small] at (-0.35,0) {$\lambda(it)$};
\node[below=3pt, font=\small] at (0.65,0) {$1$};
\node[below=3pt, font=\small] at (1.50,0) {$\infty$};
\node[font=\small] at (0,-1.95) {$\mathbb{CP}^{1}$ with defect lines};
\end{scope}
\end{tikzpicture}
\caption{The annulus with boundary conditions $\alpha,\beta$ mapped by $z=\lambda(it)$ to $\mathbb{CP}^{1}$ with defect segments on $[0,\lambda(it)]$ and $[1,\infty)$.  The endpoints carry the junction operators used in the mixed correlator.}
\label{fig:annulus-cover}
\end{figure}

As in the torus case, this condition has a useful geometric reformulation. Sewing two copies of the annulus gives a torus with modulus $\tau=it$. Equivalently, the annulus can be viewed as one sheet of the two-sheeted cover of the sphere \cite{Collier:2021ngi}. Under the covering map, the annulus is mapped to the complex plane with two intervals removed, which we may take to be $[0,z]$ and $[1,\infty)$ as shown in Fig.\ \ref{fig:annulus-cover}, with
\begin{equation}
z=\lambda(it)\,,\qquad t(z)=\frac{K(1-z)}{K(z)}\,.
\label{eq:annulus-lambda-map}
\end{equation}

\noindent The two boundary conditions are then represented by conformal defect lines ending on these intervals. The endpoint operators will be denoted by $\phi_\alpha$ and $\phi_\beta$. In this language, the annulus partition function is related to a mixed four-point function on the sphere,
\begin{equation}
\left\langle
\phi_\alpha(x_1)\phi_\alpha(x_2)
\phi_\beta(x_3)\phi_\beta(x_4)
\right\rangle
=
\left\langle \phi_\alpha(x_1)\phi_\alpha(x_2)\right\rangle
\left\langle \phi_\beta(x_3)\phi_\beta(x_4)\right\rangle
G_{\alpha\beta}(z)\,.
\label{eq:annulus-four-point-definition}
\end{equation}
With the standard choice of insertion points $0,z,1,\infty$, in the open channel the reduced correlator is related to the annulus partition function by
\begin{equation}
G^{(\mathrm{open})}_{\alpha\beta}(z)
=
B(z)
\mathcal{Z}_{\alpha\beta}^{\rm open}(t)\,,
\qquad B(z)=\left(
\frac{z^2}{2^8(1-z)}
\right)^{c/24},\qquad
z=\lambda(it)\,,
\label{eq:annulus-partition-four-point-map}
\end{equation}
while the closed channel gives an analogous correlator $G^{(\mathrm{closed})}_{\alpha\beta}(z)$ from the boundary-state decomposition of the annulus partition function at the same modulus $\tau=it$.
The important point for us is that the open-closed channel duality of the annulus becomes the crossing equation of the mixed four-point function. The endpoint operators have dimension
\begin{equation}
\Delta_{\phi_\alpha}=\Delta_{\phi_\beta}=\frac{c}{16}\,,
\label{eq:annulus-endpoint-dimension}
\end{equation}
for the identity endpoints associated with the defect lines. Open--closed channel duality is then the statement that the two representations agree,
\begin{equation}
G^{(\mathrm{open})}_{\alpha\beta}(z)
=
\left(
\frac{z}{1-z}
\right)^{c/8}
G^{(\mathrm{closed})}_{\alpha\beta}(1-z)\,,
\qquad
0<z<1\,,
\label{eq:annulus-crossing}
\end{equation}
the boundary analogue of the modular crossing equation.
Thus, the annulus bootstrap can be viewed as a mixed-correlator bootstrap problem. For the neural-network setup, this is the one-variable equation that we impose. The goal is to reconstruct $G_{\alpha\beta}(z)$, or equivalently the annulus partition function, from a minimal amount of input. In this way, the annulus problem is put in the same language as the torus problem: a consistency condition on a partition function is converted into a crossing equation for a four-point function.

\section{Neural Network Setup}
\label{sec:nn}
Our reconstruction of the line correlator \(G(z)\) proceeds by factoring out known endpoint structure and fitting only the remaining smooth function.  For torus amplitudes we first define
\begin{equation}
    B_{\rm t}(z)=\sqrt{t(z)}\,\bigl|\eta\bigl(\tau(z)\bigr)\bigr|^{2}\,,
    \label{eq:B-def}
\end{equation}
and define the transformed reduced correlator
\begin{equation}
    \widetilde{G}(z) = B_{\rm t}(z)\,G(z)\,.
    \label{eq:Gred-def}
\end{equation}
The prefactor \(B_{\rm t}(z)\) removes the universal Virasoro character denominator \(1/|\eta(\tau)|^2\) associated with the count of descendant states, effectively isolating the primary state contribution to the partition function.\footnote{For a Virasoro primary, the character takes the form \(\chi_h(\tau) = q^{h-(c-1)/24} P_h(q)/\eta(\tau)\), where \(P_h(q)\) accounts for null-state subtractions (\(P_h(q)=1\) for generic non-degenerate representations). The factor \(|\eta(\tau)|^2\) cancels the universal \(\eta\)-denominator of the descendants, and, together with the factor \(\sqrt{t}\), yields the reduced primary partition function \(\widetilde{\mathcal{Z}}(\tau,\bar\tau) = \sqrt{t}\,|\eta(\tau)|^2 \mathcal{Z}(\tau,\bar\tau)\).} Crucially, the combination \(\sqrt{t}\,|\eta(\tau)|^2\) is invariant under the modular \(S\)-transformation \(\tau \to -1/\tau\). Because the \(S\)-transformation maps to crossing on the sphere, \(B_{\rm t}(z)\) is crossing-symmetric, \(B_{\rm t}(z) = B_{\rm t}(1-z)\).
As a result, the transformed correlator still satisfies
\begin{equation}
    \widetilde{G}(z)=\left(\frac{z}{1-z}\right)^{c/4}\widetilde{G}(1-z)\,.
    \label{eq:Gred-crossing}
\end{equation}
Near \(z=0\),
\begin{equation}
    B_{\rm t}(z)\sim c_0\,z^{1/6}\sqrt{\log(16/z)}\,,\qquad G(z)\sim 1+\text{O}(z^{2\Delta_{\rm gap}})\,,\qquad
    c_0=\frac{16^{-1/6}}{\sqrt{\pi}}\,,
    \label{eq:B-asymp}
\end{equation}
so we split $\widetilde G(z)$ into a leading small-$z$ piece $L(z)$ and a gap-weighted neural correction $H(z)$, and set
\begin{equation}
    \widetilde{G}(z)=L(z)+H(z)\,,
    \label{eq:line-split}
\end{equation}
with
\begin{equation}
    L(z) = c_0\,z^{1/6}\sqrt{\log(16/z)}\,,\qquad
    H(z)=z^{2\Delta_{\rm gap}+\frac16}(1-z)^{\frac{2-3c}{12}}\,\mathrm{NN}_{\boldsymbol\theta}(z)\,,
    \label{eq:L-H-def}
\end{equation}
where \(\Delta_{\rm gap}\) is the lightest non-vacuum primary appearing in the chosen torus modular invariant.\footnote{An overall factor of $\sqrt{\log(16/z)}$ could also be included in $H(z)$ of \eqref{eq:L-H-def}, outside $\text{NN}_{\boldsymbol{\theta}}$. We have observed that this choice does not alter our reconstruction results in any essential way.}

The network, which is a fully connected single-input MLP with two hidden layers of width \(64\) and GELU activation functions,
is trained by minimising the loss
\begin{equation}
    \mathcal L(\boldsymbol\theta) = \mathcal L_{\mathrm{cross}}(\boldsymbol\theta) + \lambda_{\mathrm{anc}}\,\mathcal L_{\mathrm{anc}}(\boldsymbol\theta)\,,\qquad \lambda_{\mathrm{anc}}=100\,.
\end{equation}
This architecture and hyperparameters, including $\lambda_{\text{anc}}$, were also used in \cite{GKNS:1}.  The torus crossing term is evaluated on a uniform grid \(z_i\in[0.01,0.9]\) (for $i=1,\ldots, N$) with \(N=90\),
\begin{equation}
    \label{setupcc}
    \mathcal L_{\mathrm{cross}} = \frac{1}{N}\sum_{i=1}^N \Big\{ D(z_i)^{-1} \Big[ \widetilde{G}(z_i) - \left(\tfrac{z_i}{1-z_i}\right)^{c/4} \widetilde{G}(1-z_i) \Big] \Big\}^2,
\end{equation}
where
\begin{equation}
    D(z) = 1 + \bigl|\widetilde{G}(z)\bigr| + \Big|\left(\tfrac{z}{1-z}\right)^{c/4} \widetilde{G}(1-z)\Big|
\end{equation}
and
\begin{equation}
    \mathcal L_{\mathrm{anc}} = \bigl(\widetilde{G}(z_0) - B_{\rm t}(z_0)\,G_{\rm exact}(z_0)\bigr)^2,
\end{equation}
with one anchor at \(z_0=0.3\).

For annuli we train the two channel-transformed reduced correlators
\begin{equation}
\widetilde G^{(\mathrm o)}_{\alpha\beta}(z) = B_{\rm a}(z)\,G^{(\mathrm{open})}_{\alpha\beta}(z)\,,\qquad
\widetilde G^{(\mathrm c)}_{\alpha\beta}(z) = B_{\rm a}(z)\,G^{(\mathrm{closed})}_{\alpha\beta}(z)\,,
\label{eq:annulus-Gred-def}
\end{equation}
against the open/closed crossing equation of Section~\ref{subsec:annulus}.  Motivated by the torus factor $B_{\rm t}(z)=\sqrt{t(z)}\,|\eta(\tau(z))|^2$, which is the inverse of the $c=1$ non-compact free-boson partition function on the torus, we take \(B_{\rm a}(z)=\eta(\tau(z))\), the inverse of the $c=1$ non-compact free-boson partition function on the annulus. This choice of \(B_{\rm a}(z)\) is not \(S\)-invariant, since 
\begin{equation}
B_{\rm a}(1-z) = \eta\bigl(-1/\tau(z)\bigr) = \sqrt{t(z)}\,B_{\rm a}(z)\,,
\label{eq:B-annulus-crossing}
\end{equation}
so the reduced DCO crossing relation carries both the DCO weight and the extra factor from \(B_{\rm a}\),
\begin{equation}
\widetilde G^{(\mathrm o)}(z) = \left(\frac{z}{1-z}\right)^{c/8}\frac{\widetilde G^{(\mathrm c)}(1-z)}{\sqrt{t(z)}}\,.
\label{eq:C-annulus-crossing}
\end{equation}
Each channel is written as a leading-primary prefactor plus a gap-weighted neural correction, read off from its own character spectrum. In this paper, as a proof of concept we restrict to diagonal rational CFTs on the annulus, for which the Cardy construction supplies boundary states directly from the modular $S$-matrix,  $S_{ij}$. The Cardy boundary states read~\cite{Ishibashi:1988kg,Cardy:1989ir}
\begin{equation}
|\alpha\rangle = \sum_i \frac{S_{\alpha i}}{\sqrt{S_{\mathbf{1}i}}}\,|i\rangle\!\rangle\,.
\label{eq:cardy-state}
\end{equation}
The sum runs over the primaries $i$ of the chiral algebra of the CFT, and $|i\rangle\!\rangle$ is the Ishibashi state, i.e.\ the unique (up to normalisation) coherent state in the tensor product of the holomorphic and anti-holomorphic Verma modules of $i$ that satisfies the gluing condition 
\begin{equation}
    (L_n-\bar L_{-n})|i\rangle\!\rangle=0\,,\qquad \forall n\in\mathbb{Z}\,.
\end{equation}
The open-channel multiplicities are given by the Verlinde formula~\cite{Verlinde:1988sn},
\begin{equation}
N_{ij}{}^{k}=\sum_l \frac{S_{il}S_{jl}S^{*}_{kl}}{S_{\mathbf{1}l}}\,,
\label{eq:verlinde}
\end{equation}
which for the Cardy boundary states~\eqref{eq:cardy-state} gives $N_{\alpha i}{}^\beta$ for the orientation of~\eqref{eq:annulus-closed-channel}. The open-channel Verlinde decomposition and its closed-channel Cardy decomposition are then
\begin{equation}
Z^{(\mathrm{open})}_{\alpha\beta}(\tau)=\sum_i N_{\alpha i}{}^{\beta}\,\chi_i(\tau)\,,\qquad Z^{(\mathrm{closed})}_{\alpha\beta}(\tilde\tau)=\sum_i \frac{S_{\alpha i}S_{\beta i}^{*}}{S_{0i}}\,\chi_i(\tilde\tau)\,,\qquad \tilde\tau=-1/\tau\,,
\label{eq:open-closed-decomp}
\end{equation}
where the open decomposition contains the identity only for $\alpha=\beta$, whereas the closed one always contains the identity since $S_{\alpha 0}\,S_{\beta 0}^{*}/S_{00}>0$ in every diagonal unitary rational CFT.  Each channel $X\in\{\mathrm o,\mathrm c\}$ is written as an exact leading piece plus a neural correction,
\begin{equation}
\widetilde G^{X}(z) = n_0^{X}\,c_0^{X}\,z^{p_0^{X}}\;+\;z^{p_0^{X}+d^{X}}\,(1-z)^{p_1^{X,{\rm eff}}}\,\sqrt{\log\bigl(\tfrac{16}{1-z}\bigr)}\;\mathrm{NN}^{X}(z)\,,
\label{eq:annulus-ansatz}
\end{equation}
with $c_0^{X}=16^{-p_0^{X}}$ and the exponents fixed from CFT data,
\begin{equation}
p_0^{X} = 2 h_{\rm lead}^{X}+\tfrac1{12}\,,\qquad d^{X} = \min\bigl(1,\,2(h_{\rm next}^{X}-h_{\rm lead}^{X})\bigr)\,,
\label{eq:annulus-ansatz-exps}
\end{equation}
so $p_0^{\mathrm c}=\tfrac1{12}$ (fixed by $h_{\rm lead}^{\mathrm c}=0$).  The coefficient $n_0^{X}=m_{\rm lead}^{X}N_{\rm lead}^{X}$ combines the leading-character multiplicity with the degeneracy $N_{\rm lead}^{X}$ of the leading primary. Here $m_{\rm lead}^{X}$ is $N_{\alpha i}{}^{\beta}$ in the open channel and $S_{\alpha i}S_{\beta i}^{*}/S_{0i}$ in the closed channel. The quantity $h_{\rm lead}^{X}$ is the lightest primary with $n_0^{X}\ne 0$, and $h_{\rm next}^{X}$ is either the next-lightest primary or the kinematic correction from $B_{\rm a}(z)$.

The endpoint exponents follow from crossing at $z\to 1$,
\begin{equation}
p_1^{\mathrm o,{\rm raw}} = \tfrac{2-3c}{24}\,,\qquad p_1^{\mathrm c,{\rm raw}} = p_0^{\mathrm o}-\tfrac{c}{8} = 2 h_{\rm lead}^{\mathrm o}+\tfrac1{12}-\tfrac{c}{8}\,.
\end{equation}
We include the $(1-z)^{p_1^{X,{\rm raw}}}$ factor outside the NN only when it is divergent at $z=1$, and absorb it into the NN otherwise,
\begin{equation}
p_1^{X,{\rm eff}} = \min\bigl(p_1^{X,{\rm raw}},\,0\bigr)\,.
\label{eq:annulus-clip}
\end{equation}
In practice the open side switches at $c>2/3$, and the closed side whenever $2h_{\rm lead}^{\mathrm o}+\tfrac1{12}<c/8$.

Training minimises
\begin{equation}
\mathcal L^{(\rm ann)}(\boldsymbol\theta) = \mathcal L^{(\rm ann)}_{\rm cross}(\boldsymbol\theta) + \lambda^{(\rm ann)}_{\rm anc}\,\mathcal L^{(\rm ann)}_{\rm anc}(\boldsymbol\theta)\,,\qquad \lambda^{(\rm ann)}_{\rm anc} = 100\,,
\end{equation}
with the annulus crossing term evaluated on a uniform grid \(z_i\in[0.01,0.9]\) with \(N=90\) points,
\begin{equation}
\mathcal L^{(\rm ann)}_{\rm cross} = \frac{1}{N}\sum_{i=1}^N \Big\{ D^{(\rm ann)}(z_i)^{-1} \Big[ \widetilde G^{(\mathrm o)}(z_i) - \left(\tfrac{z_i}{1-z_i}\right)^{c/8}\widetilde G^{(\mathrm c)}(1-z_i)/\sqrt{t(z_i)} \Big] \Big\}^2,
\label{eq:annulus-loss}
\end{equation}
where
\begin{equation}
D^{(\rm ann)}(z) = 1 + \bigl|\widetilde G^{(\mathrm o)}(z)\bigr| + \Big|\left(\tfrac{z}{1-z}\right)^{c/8}\widetilde G^{(\mathrm c)}(1-z)/\sqrt{t(z)}\Big|
\label{eq:annulus-loss-D}
\end{equation}
and
\begin{equation}
\mathcal L^{(\rm ann)}_{\rm anc} = \bigl(\widetilde G^{(\mathrm o)}(z_0) - B_{\rm a}(z_0)\,G^{(\rm open)}_{\rm exact}(z_0)\bigr)^2 + \bigl(\widetilde G^{(\mathrm c)}(z_0) - B_{\rm a}(z_0)\,G^{(\rm closed)}_{\rm exact}(z_0)\bigr)^2,
\label{eq:annulus-loss-anc}
\end{equation}
with a single anchor at \(z_0=0.3\) for each of the two channels.

\paragraph{Training schedule.} For every torus and annulus example presented below, the network is trained with \texttt{Adam}~\cite{Kingma:2014} at learning rate \(5\times10^{-4}\), weight decay \(10^{-6}\), and a \texttt{StepLR} schedule multiplying the learning rate by \(\gamma=0.98\) every \(500\) epochs.  Runs last at most \(2\times10^5\) epochs and stop early after \(5\times10^3\) stagnant epochs. These choices are the same across all examples and coincide with those used in the companion work~\cite{GKNS:1}.

\paragraph{Reporting scheme.}
The rest of the paper analyses a variety of torus and annulus reconstructions.  Unless stated otherwise, results are based on an ensemble of $100$ independent runs with the single anchor $z_0=0.3$. Each figure summarising a torus reconstruction is laid out in three panels:
\begin{itemize}
\item[$(a)$] top-left: the ensemble mean of the reduced predicted correlator $\widetilde G^{\rm pred}(z)$ over the $100$ seeds as a solid blue curve, with a shaded blue $\pm 1$ standard-deviation band and the exact analytic correlator as a black dashed curve,
\item[$(b)$] top-right: the per-seed relative prediction error,
\begin{equation}
\text{Prediction relative error} = \frac{\widetilde G^{\rm pred}(z) - \widetilde G^{\rm exact}(z)}{1+|\widetilde G^{\rm exact}(z)|}\,,
\label{eq:reporting-relerr}
\end{equation}
shown as an ensemble mean (solid blue) with a shaded $\pm 1\sigma$ band. The $1$ in the denominator regularises the divergence when the exact value becomes small,
\item[$(c)$] bottom: a histogram of the $\widetilde G^{\rm pred}(z=0.5)$ values across seeds, with the histogram mean as a solid blue line and the exact reference value $\widetilde G^{\rm exact}(0.5)$ as a black dashed line.
\end{itemize}
For annulus reconstructions the same three-panel scheme is duplicated in a six-panel figure, with the open channel drawn in blue and the closed channel in red.  Additional details and the explicit \href{https://www.python.org}{\texttt{Python}} code that generated all the runs and figures are collected in the companion \href{https://github.com}{\texttt{GitHub}} repository \href{https://github.com/andstergiou/nn-cft}{\tt andstergiou/nn-cft}, archived on \href{https://zenodo.org/}{\texttt{Zenodo}}~\cite{nn_cft_code}.

\section{Torus Reconstructions in Compact CFTs}
\label{sec:minimal}

\subsection{Warm-Up: Two-Dimensional Ising Model \texorpdfstring{$\mathcal{M}(3,4)$}{M(3,4)}}
The 2D Ising model is the most basic A-type CFT with torus partition function
\begin{equation}
\mathcal{Z}_{\mathrm{Ising}}(\tau,\bar\tau)
=
\big|\chi_{\mathbf{1}}(\tau)\big|^{2}
+
\big|\chi_{\epsilon}(\tau)\big|^{2}
+
\big|\chi_{\sigma}(\tau)\big|^{2}\,,
\end{equation}
where \(\chi_{\mathbf{1}},\chi_{\epsilon},\chi_{\sigma}\) are the Virasoro characters associated with primaries \(\mathbf{1}\), \(\epsilon\), and \(\sigma\), respectively. In this particular case, characters admit compact expressions in terms of Jacobi theta functions, \(\theta_{2,3,4}(\tau)\), and the Dedekind eta function, $\eta(\tau)$. We also have closed form expressions under the map \eqref{eq:annulus-lambda-map},
\begin{align}
\chi_{\mathbf{1}}(\tau(z))
&=\frac12\left(\sqrt{\frac{\theta_{3}(\tau(z))}{\eta(\tau(z))}}+\sqrt{\frac{\theta_{4}(\tau(z))}{\eta(\tau(z))}}\right)=
2^{-\frac{5}{6}}\big(z(1-z)\big)^{-\frac{1}{24}}
\left(1+(1-z)^{\frac{1}{8}}\right),\\[4pt]
\chi_{\epsilon}(\tau(z))
&=\frac12\left(\sqrt{\frac{\theta_{3}(\tau(z))}{\eta(\tau(z))}}-\sqrt{\frac{\theta_{4}(\tau(z))}{\eta(\tau(z))}}\right)=
2^{-\frac{5}{6}}\big(z(1-z)\big)^{-\frac{1}{24}}
\left(1-(1-z)^{\frac{1}{8}}\right),\\[4pt]
\chi_{\sigma}(\tau(z))
&=\frac{1}{\sqrt{2}}\sqrt{\frac{\theta_{2}(\tau(z))}{\eta(\tau(z))}}=2^{-\frac{1}{3}}z^{\frac{1}{8}}\big(z(1-z)\big)^{-\frac{1}{24}}.
\end{align}
The associated twist field correlator is
\begin{equation}
    G(z) = \tfrac{1}{2} (1 - z)^{-1/8} (1 + z^{1/4} + (1 - z)^{1/4})\,.
\end{equation}

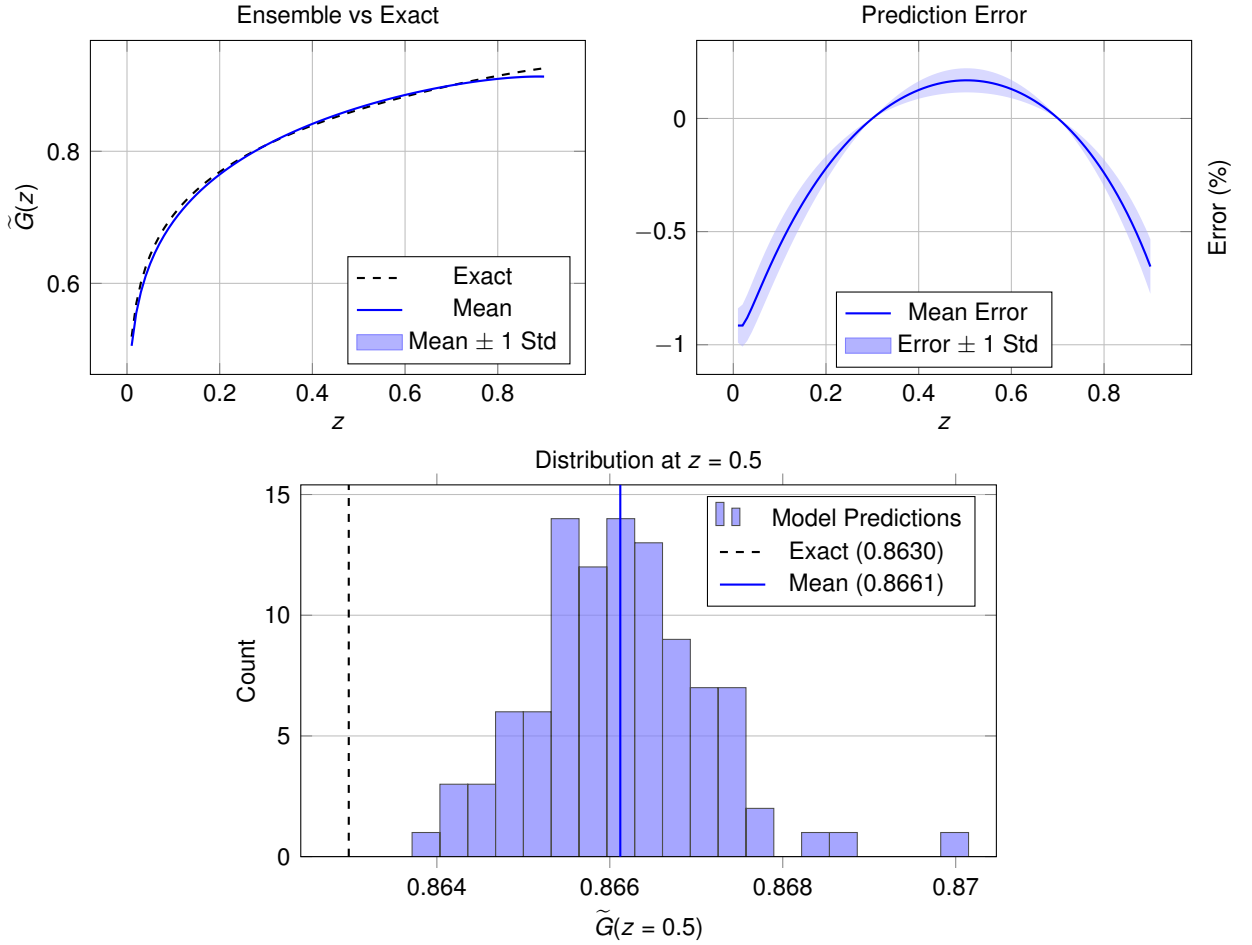
\begin{figure}[H]
    \centering
    \begin{subfigure}[b]{0.49\textwidth}
        \centering
        \begin{tikzpicture}
            \begin{axis}[
                width=\linewidth, height=6cm,
                xlabel={$z$},
                ylabel={$\widetilde{G}(z)$},
                title={Ensemble vs Exact},
                grid=major,
                legend pos=south east,
            ]
                \addplot [black, dashed, thick] table [x=z, y=Exact, col sep=space] {plots/m3_ensemble_comparison.dat};
                \addlegendentry{Exact}
                \addplot [blue, thick] table [x=z, y=Mean, col sep=space] {plots/m3_ensemble_comparison.dat};
                \addlegendentry{Mean}
                \addplot [forget plot, name path=upper, draw=none] table [x=z, y=Mean_plus_Std, col sep=space] {plots/m3_ensemble_comparison.dat};
                \addplot [forget plot, name path=lower, draw=none] table [x=z, y=Mean_minus_Std, col sep=space] {plots/m3_ensemble_comparison.dat};
                \addplot [forget plot, fill=blue!30, fill opacity=0.5, draw=none] fill between [of=upper and lower];
                \addlegendimage{legend image code/.code={\fill[blue!30, draw=blue!50] (0cm,-0.1cm) rectangle (0.6cm,0.1cm);}}
                \addlegendentry{Mean $\pm$ 1 Std}
            \end{axis}
        \end{tikzpicture}
    \end{subfigure}
    \hfill
    \begin{subfigure}[b]{0.49\textwidth}
        \centering
        \begin{tikzpicture}
            \begin{axis}[
                ylabel style={at={(axis description cs:1.10,0.5)}, anchor=south},
                width=\linewidth, height=6cm,
                xlabel={$z$},
                ylabel={Error (\%)},
                title={Prediction Error},
                grid=major,
                legend style={at={(0.5,0.02)}, anchor=south},
            ]
                \addplot [blue, thick] table [x=z, y=PctError, col sep=space] {plots/m3_percentage_error.dat};
                \addlegendentry{Mean Error}
                \addplot [forget plot, name path=upper, draw=none] table [x=z, y=PctError_plus_PctStd, col sep=space] {plots/m3_percentage_error.dat};
                \addplot [forget plot, name path=lower, draw=none] table [x=z, y=PctError_minus_PctStd, col sep=space] {plots/m3_percentage_error.dat};
                \addplot [forget plot, fill=blue!30, fill opacity=0.5, draw=none] fill between [of=upper and lower];
                \addlegendimage{legend image code/.code={\fill[blue!30, draw=blue!50] (0cm,-0.1cm) rectangle (0.6cm,0.1cm);}}
                \addlegendentry{Error $\pm$ 1 Std}
            \end{axis}
        \end{tikzpicture}
    \end{subfigure}

    \begin{subfigure}[b]{0.65\textwidth}
        \centering
        \begin{tikzpicture}
            \begin{axis}[
                width=\linewidth, height=6.5cm,
                xlabel={$\widetilde{G}(z=0.5)$},
                ylabel={Count},
                title={Distribution at $z=0.5$},
                ybar,
                bar width=0.000322,
                xmin=0.862426, xmax=0.870470,
                ymin=0,
                ymajorgrids=true,
                xmajorgrids=false,
                legend pos=north east,
                scaled ticks=false,
                xtick={0.864,0.866,0.868,0.87},
                xticklabel style={
                    /pgf/number format/fixed,
                    /pgf/number format/precision=4,
                    font=\sansmath\sffamily\footnotesize,
                }
            ]
                \addplot [fill=blue!50, draw=black, opacity=0.7] table [x=BinCenter, y=Count, col sep=space] {plots/m3_histogram_z0.500.dat};
                \addlegendentry{Model Predictions}
                \draw [black, dashed, thick] (axis cs:0.862981,\pgfkeysvalueof{/pgfplots/ymin}) -- (axis cs:0.862981,\pgfkeysvalueof{/pgfplots/ymax});
                \addlegendimage{legend image code/.code={\draw[black, dashed, thick] (0cm,0cm) -- (0.6cm,0cm);}}
                \addlegendentry{Exact ($0.8630$)}
                \draw [blue, thick] (axis cs:0.866121,\pgfkeysvalueof{/pgfplots/ymin}) -- (axis cs:0.866121,\pgfkeysvalueof{/pgfplots/ymax});
                \addlegendimage{legend image code/.code={\draw[blue, thick] (0cm,0cm) -- (0.6cm,0cm);}}
                \addlegendentry{Mean ($0.8661$)}
            \end{axis}
        \end{tikzpicture}
    \end{subfigure}

    \caption{NN-predicted reduced correlator $\widetilde G(z)$ for the $\mathcal M(3,4)$ Ising model ($c=1/2$, $\Delta_{\rm gap}=1/8$).  The NN prediction at $z=0.5$ is $\widetilde G^{\rm pred}(0.5)=0.86612\pm 0.00099$ against $\widetilde G^{\rm exact}(0.5)=0.86298$.}
    \label{fig:m3_summary}
\end{figure}

In the reduced-correlator parametrisation of Section~\ref{sec:nn}, our neural-network ansatz, specialised to Ising with $c=1/2$ and $\Delta_{\rm gap}=1/8$, reads
\begin{equation}
    \widetilde{G}(z) = c_0\,z^{1/6}\,\sqrt{\log\bigl(\tfrac{16}{z}\bigr)} + z^{5/12}\,(1-z)^{1/24}\,\mathrm{NN}(z)\,,
\label{eq:ising-ansatz}
\end{equation}
with $2\Delta_{\rm gap}+1/6 = 5/12$ and $(2-3c)/12 = 1/24$ in the correction-term exponents.  Across $100$ initialization seeds with the training schedule of Section~\ref{sec:nn}, the mean square (MS) training loss is $(5.94\pm 10.5)\times 10^{-10}$, and the NN prediction at $z=0.5$ reads $\widetilde G^{\rm pred}(0.5)=0.86612\pm 0.00099$ against $\widetilde G^{\rm exact}(0.5)=0.86298$.  Results are shown in Fig.~\ref{fig:m3_summary}.

\subsection{Non-Unitary Example: Lee--Yang CFT}\label{sec:lee-yang}
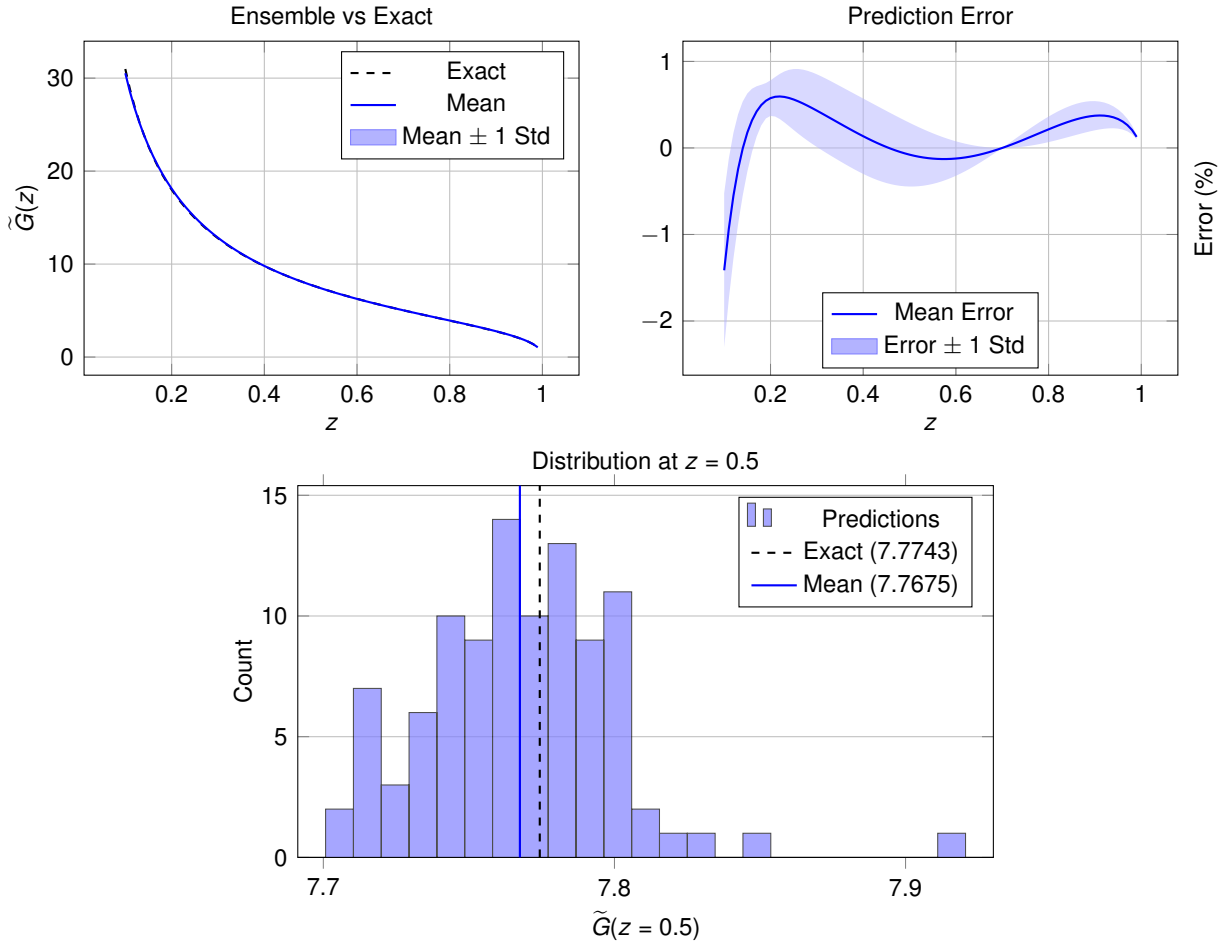
\begin{figure}[H]
    \centering
    \begin{subfigure}[b]{0.49\textwidth}
        \centering
        \begin{tikzpicture}
            \begin{axis}[
                width=\linewidth, height=6cm,
                xlabel={$z$},
                ylabel={$\widetilde{G}(z)$},
                title={Ensemble vs Exact},
                grid=major,
                legend pos=north east,
            ]
                \addplot [black, dashed, thick] table [x=z, y=Exact, col sep=space] {plots/ly_ensemble_comparison.dat};
                \addlegendentry{Exact}
                \addplot [blue, thick] table [x=z, y=Mean, col sep=space] {plots/ly_ensemble_comparison.dat};
                \addlegendentry{Mean}
                \addplot [forget plot, name path=upper, draw=none] table [x=z, y=Mean_plus_Std, col sep=space] {plots/ly_ensemble_comparison.dat};
                \addplot [forget plot, name path=lower, draw=none] table [x=z, y=Mean_minus_Std, col sep=space] {plots/ly_ensemble_comparison.dat};
                \addplot [forget plot, fill=blue!30, fill opacity=0.5, draw=none] fill between [of=upper and lower];
                \addlegendimage{legend image code/.code={\fill[blue!30, draw=blue!50] (0cm,-0.1cm) rectangle (0.6cm,0.1cm);}}
                \addlegendentry{Mean $\pm$ 1 Std}
            \end{axis}
        \end{tikzpicture}
    \end{subfigure}
    \hfill
    \begin{subfigure}[b]{0.49\textwidth}
        \centering
        \begin{tikzpicture}
            \begin{axis}[
                ylabel style={at={(axis description cs:1.10,0.5)}, anchor=south},
                width=\linewidth, height=6cm,
                xlabel={$z$},
                ylabel={Error (\%)},
                title={Prediction Error},
                grid=major,
                legend style={at={(0.5,0.02)}, anchor=south},
            ]
                \addplot [blue, thick] table [x=z, y=PctError, col sep=space] {plots/ly_percentage_error.dat};
                \addlegendentry{Mean Error}
                \addplot [forget plot, name path=upper, draw=none] table [x=z, y=PctError_plus_PctStd, col sep=space] {plots/ly_percentage_error.dat};
                \addplot [forget plot, name path=lower, draw=none] table [x=z, y=PctError_minus_PctStd, col sep=space] {plots/ly_percentage_error.dat};
                \addplot [forget plot, fill=blue!30, fill opacity=0.5, draw=none] fill between [of=upper and lower];
                \addlegendimage{legend image code/.code={\fill[blue!30, draw=blue!50] (0cm,-0.1cm) rectangle (0.6cm,0.1cm);}}
                \addlegendentry{Error $\pm$ 1 Std}
            \end{axis}
        \end{tikzpicture}
    \end{subfigure}

    \begin{subfigure}[b]{0.65\textwidth}
        \centering
        \begin{tikzpicture}
            \begin{axis}[
                width=\linewidth, height=6.5cm,
                xlabel={$\widetilde{G}(z=0.5)$},
                ylabel={Count},
                title={Distribution at $z=0.5$},
                ybar,
                bar width=0.009557,
                xmin=7.691228, xmax=7.930155,
                xtick={7.7,7.8,7.9},
                ymin=0,
                ymajorgrids=true,
                xmajorgrids=false,
                legend pos=north east,
                scaled ticks=false,
                xticklabel style={
                    /pgf/number format/fixed,
                    /pgf/number format/precision=4,
                    font=\sansmath\sffamily\footnotesize,
                }
            ]
                \addplot [fill=blue!50, draw=black, opacity=0.7] table [x=BinCenter, y=Count, col sep=space] {plots/ly_histogram_z0.500.dat};
                \addlegendentry{Predictions}
                \draw [black, dashed, thick] (axis cs:7.77435,\pgfkeysvalueof{/pgfplots/ymin}) -- (axis cs:7.77435,\pgfkeysvalueof{/pgfplots/ymax});
                \addlegendimage{legend image code/.code={\draw[black, dashed, thick] (0cm,0cm) -- (0.6cm,0cm);}}
                \addlegendentry{Exact ($7.7743$)}
                \draw [blue, thick] (axis cs:7.76752,\pgfkeysvalueof{/pgfplots/ymin}) -- (axis cs:7.76752,\pgfkeysvalueof{/pgfplots/ymax});
                \addlegendimage{legend image code/.code={\draw[blue, thick] (0cm,0cm) -- (0.6cm,0cm);}}
                \addlegendentry{Mean ($7.7675$)}
            \end{axis}
        \end{tikzpicture}
    \end{subfigure}

    \caption{NN-predicted reduced correlator $\widetilde G(z)$ for the Lee--Yang minimal model $\mathcal M(2,5)$ ($c=-22/5$, $h_\phi=-1/5$) with the modified ansatz~\eqref{eq:ly-ansatz} and anchor $z_0=0.7$.  The NN prediction at $z=0.5$ is $\widetilde G^{\rm pred}(0.5)=7.7675\pm 0.0323$ against $\widetilde G^{\rm exact}(0.5)=7.7743$.}
    \label{fig:ly_summary}
\end{figure}

The Lee--Yang model $\mathcal M(2,5)$ has $c=-22/5$ and one non-identity primary $\phi=\phi_{1,2}$ with $h_\phi=-1/5$.  Because this primary lies below the identity, the $z\to0$ endpoint of $\widetilde{G}$ is divergent. We therefore build the ansatz around $z\to1$,
\begin{equation}
\widetilde{G}(z) = c_1\lsp(1-z)^{7/15}\sqrt{\log\bigl(\tfrac{16}{1-z}\bigr)} + z^{-19/30}\,(1-z)^{19/15}\,\mathrm{NN}(z)\,,
\label{eq:ly-ansatz}
\end{equation}
The leading coefficient is
\begin{equation}
c_1 = \frac{2^{38/15}}{\sqrt{\pi}} \approx 3.2661\,.
\label{eq:ly-c1}
\end{equation}
We train on $z\in[0.1,0.99]$ with anchor $z_0=0.7$.  Over $100$ seeds, the MS training loss is $(9.50\pm 11.8)\times 10^{-6}$, and $\widetilde G^{\rm pred}(0.5)=7.7675\pm 0.0323$ versus $\widetilde G^{\rm exact}(0.5)=7.7743$; see Fig.~\ref{fig:ly_summary}.

\subsection{Generic ADE-Series Minimal Models}\label{sec:ADE-series}
The unitary minimal models $\mathcal{M}(m,m+1)$ have central charge
\begin{equation}
    c(m)=1-\frac{6}{m(m+1)}\,,\qquad m\ge 3\,,
    \label{eq:c-ADE}
\end{equation}
with primary fields labelled by the Kac table
\begin{equation}
    \mathcal{P}_{m}=\bigl\{(r,s)\in\mathbb{Z}^{2}\,\big|\,1\le r\le m-1,\ 1\le s\le m,\ (r,s)\sim(m-r,m+1-s)\bigr\}\,,
    \label{eq:kac-table}
\end{equation}
and conformal weights
\begin{equation}
    h_{r,s}=\bar h_{r,s}=\frac{\bigl((m+1)r-ms\bigr)^{2}-1}{4m(m+1)}\,.
\end{equation}
The chiral characters are given by the Rocha--Caridi formula
\begin{equation}
    \chi_{r,s}(q)=\frac{1}{\eta(q)}\sum_{n\in\mathbb Z}\!\left[q^{\frac{(2m(m+1)n+(m+1)r-ms)^{2}}{4m(m+1)}}-q^{\frac{(2m(m+1)n+(m+1)r+ms)^{2}}{4m(m+1)}}\right],
\end{equation}
from which the exact target correlator is computed numerically throughout this subsection.

The CIZ classification~\cite{Cappelli:1987xt} groups the modular-invariant partition functions of $\mathcal M(m,m+1)$ by pairs of simply-laced Dynkin diagrams $(G,G')$ with Coxeter numbers $h(G)=m$, $h(G')=m+1$. The invariant reads
\begin{equation}
    \mathcal Z(\tau,\bar\tau)=\sum_{(r,s),(r',s')\in\mathcal P_{m}}M_{(r,s),(r',s')}\,\chi_{r,s}(\tau)\,\overline{\chi_{r',s'}(\tau)}\,,
\end{equation}
with $M$ a non-negative integer matrix commuting with the Virasoro $S$ and $T$ matrices.  Three families arise: the diagonal A-series $(A_{m-1},A_m)$; the D-series, in which one Dynkin factor is a $D_n$-diagram of matching Coxeter number; and the six exceptional E-series invariants $(A_{10},E_6)$, $(E_6,A_{12})$, $(A_{16},E_7)$, $(E_7,A_{18})$, $(A_{28},E_8)$, $(E_8,A_{30})$.  The thermal-line crossing equation is unchanged across all three families, only the spectrum running in the sum is modified, and the relevant gap is the dimension of the lightest non-vacuum primary surviving the projection by $M$,
\begin{equation}
    \Delta_{\rm gap}=\min_{\substack{M_{(r,s),(r',s')}\ne 0\\((r,s),(r',s'))\ne((1,1),(1,1))}}\bigl(h_{r,s}+h_{r',s'}\bigr)\,.
    \label{eq:ADEgap}
\end{equation}
For the A- and D-series this gives the closed forms
\begin{equation}
\Delta^{A}_{\rm gap} = \frac{3}{2m(m+1)}\,,\qquad \Delta^{D}_{\rm gap} = \frac{4}{m(m+1)}\,,
\label{eq:gap-AD}
\end{equation}
while the E-series gaps are read off from the exceptional matrix $M^E$ case by case.

In the reduced-correlator parametrisation of Section~\ref{sec:nn}, we adopt the ansatz~\eqref{eq:line-split} with $\Delta_{\rm gap}$ given by~\eqref{eq:gap-AD} for the A- and D-series (case-by-case for the E-series) and central charge from~\eqref{eq:c-ADE}.

We test the three families on the two representative cases of tricritical Ising (A-series) and the three-state Potts $D_4$ block. Three further examples illustrating higher-$m$ A-series and the $D$/$E$ modular invariants, together with eight additional modular invariants collected in a summary table, are relegated to Appendix~\ref{app:torus-extra}.  The full set of unitary minimal models $\mathcal M(m,m+1)$ is accessible through a \href{https://www.python.org/}{\texttt{Python}} program found in the companion \href{https://github.com}{\texttt{GitHub}} repository \href{https://github.com/andstergiou/nn-cft}{\tt andstergiou/nn-cft}.

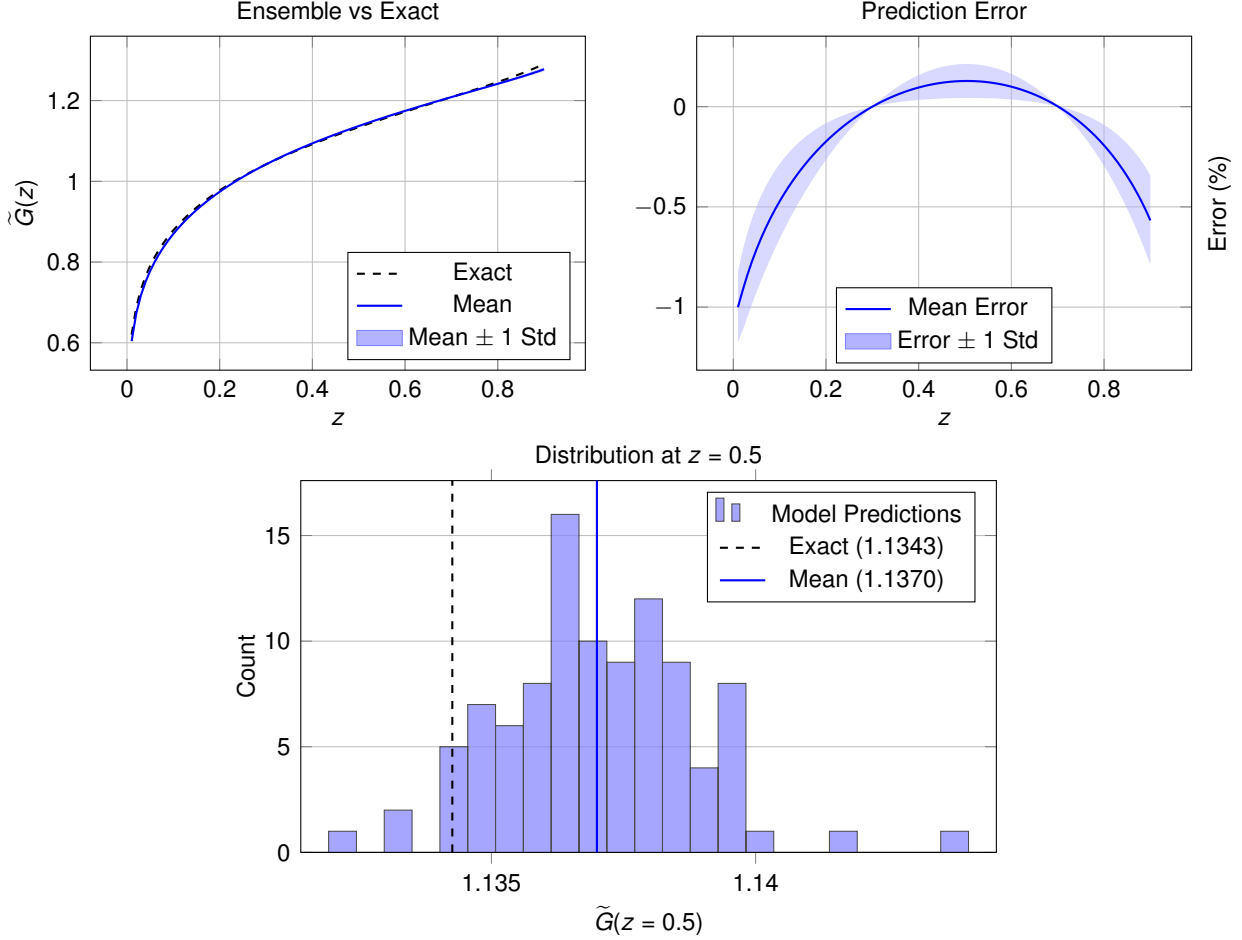
\begin{figure}[H]
    \centering
    \begin{subfigure}[b]{0.49\textwidth}
        \centering
        \begin{tikzpicture}
            \begin{axis}[
                width=\linewidth, height=6cm,
                xlabel={$z$},
                ylabel={$\widetilde{G}(z)$},
                title={Ensemble vs Exact},
                grid=major,
                legend pos=south east,
            ]
                \addplot [black, dashed, thick] table [x=z, y=Exact, col sep=space] {plots/m4_ensemble_comparison.dat};
                \addlegendentry{Exact}
                \addplot [blue, thick] table [x=z, y=Mean, col sep=space] {plots/m4_ensemble_comparison.dat};
                \addlegendentry{Mean}
                \addplot [forget plot, name path=upper, draw=none] table [x=z, y=Mean_plus_Std, col sep=space] {plots/m4_ensemble_comparison.dat};
                \addplot [forget plot, name path=lower, draw=none] table [x=z, y=Mean_minus_Std, col sep=space] {plots/m4_ensemble_comparison.dat};
                \addplot [forget plot, fill=blue!30, fill opacity=0.5, draw=none] fill between [of=upper and lower];
                \addlegendimage{legend image code/.code={\fill[blue!30, draw=blue!50] (0cm,-0.1cm) rectangle (0.6cm,0.1cm);}}
                \addlegendentry{Mean $\pm$ 1 Std}
            \end{axis}
        \end{tikzpicture}
    \end{subfigure}
    \hfill
    \begin{subfigure}[b]{0.49\textwidth}
        \centering
        \begin{tikzpicture}
            \begin{axis}[
                ylabel style={at={(axis description cs:1.10,0.5)}, anchor=south},
                width=\linewidth, height=6cm,
                xlabel={$z$},
                ylabel={Error (\%)},
                title={Prediction Error},
                grid=major,
                legend style={at={(0.5,0.02)}, anchor=south},
            ]
                \addplot [blue, thick] table [x=z, y=PctError, col sep=space] {plots/m4_percentage_error.dat};
                \addlegendentry{Mean Error}
                \addplot [forget plot, name path=upper, draw=none] table [x=z, y=PctError_plus_PctStd, col sep=space] {plots/m4_percentage_error.dat};
                \addplot [forget plot, name path=lower, draw=none] table [x=z, y=PctError_minus_PctStd, col sep=space] {plots/m4_percentage_error.dat};
                \addplot [forget plot, fill=blue!30, fill opacity=0.5, draw=none] fill between [of=upper and lower];
                \addlegendimage{legend image code/.code={\fill[blue!30, draw=blue!50] (0cm,-0.1cm) rectangle (0.6cm,0.1cm);}}
                \addlegendentry{Error $\pm$ 1 Std}
            \end{axis}
        \end{tikzpicture}
    \end{subfigure}

    \begin{subfigure}[b]{0.65\textwidth}
        \centering
        \begin{tikzpicture}
            \begin{axis}[
                width=\linewidth, height=6.5cm,
                xlabel={$\widetilde{G}(z=0.5)$},
                ylabel={Count},
                title={Distribution at $z=0.5$},
                ybar,
                bar width=0.000526,
                xmin=1.131394, xmax=1.144554,
                ymin=0,
                ymajorgrids=true,
                xmajorgrids=false,
                legend pos=north east,
                scaled ticks=false,
                xtick={1.135,1.14},
                xticklabel style={
                    /pgf/number format/fixed,
                    /pgf/number format/precision=4,
                    font=\sansmath\sffamily\footnotesize,
                }
            ]
                \addplot [fill=blue!50, draw=black, opacity=0.7] table [x=BinCenter, y=Count, col sep=space] {plots/m4_histogram_z0.500.dat};
                \addlegendentry{Model Predictions}
                \draw [black, dashed, thick] (axis cs:1.134261,\pgfkeysvalueof{/pgfplots/ymin}) -- (axis cs:1.134261,\pgfkeysvalueof{/pgfplots/ymax});
                \addlegendimage{legend image code/.code={\draw[black, dashed, thick] (0cm,0cm) -- (0.6cm,0cm);}}
                \addlegendentry{Exact ($1.1343$)}
                \draw [blue, thick] (axis cs:1.136997,\pgfkeysvalueof{/pgfplots/ymin}) -- (axis cs:1.136997,\pgfkeysvalueof{/pgfplots/ymax});
                \addlegendimage{legend image code/.code={\draw[blue, thick] (0cm,0cm) -- (0.6cm,0cm);}}
                \addlegendentry{Mean ($1.1370$)}
            \end{axis}
        \end{tikzpicture}
    \end{subfigure}

    \caption{NN-predicted reduced correlator $\widetilde G(z)$ for the $\mathcal M(4,5)$ tricritical Ising model ($c=7/10$, $\Delta_{\rm gap}=3/40$).  The NN prediction at $z=0.5$ is $\widetilde G^{\rm pred}(0.5)=1.13700\pm 0.00181$ against $\widetilde G^{\rm exact}(0.5)=1.13426$.}
    \label{fig:m4_summary}
\end{figure}

\paragraph{$\mathcal M(4,5)$ tricritical Ising: $c=7/10$, $\Delta_{\rm gap}=3/40$.}
This is the simplest non-Ising $A$-series case.  At $c=7/10>2/3$ the $(1-z)^{(2-3c)/12}$ factor in~\eqref{eq:line-split} already diverges weakly at $z=1$.  Over $100$ seeds, the MS training loss is $(1.39\pm 2.45)\times 10^{-9}$, and the NN prediction at $z=0.5$ reads $\widetilde G^{\rm pred}(0.5)=1.13700\pm 0.00181$ against $\widetilde G^{\rm exact}(0.5)=1.13426$.  Results are shown in Fig.~\ref{fig:m4_summary}.

\begin{figure}[H]
    \centering
    \begin{subfigure}[b]{0.49\textwidth}
        \centering
        \begin{tikzpicture}
            \begin{axis}[
                width=\linewidth, height=6cm,
                xlabel={$z$},
                ylabel={$\widetilde{G}(z)$},
                title={Ensemble vs Exact},
                grid=major,
                legend pos=south east,
            ]
                \addplot [black, dashed, thick] table [x=z, y=Exact, col sep=space] {plots/potts_ensemble_comparison.dat};
                \addlegendentry{Exact}
                \addplot [blue, thick] table [x=z, y=Mean, col sep=space] {plots/potts_ensemble_comparison.dat};
                \addlegendentry{Mean}
                \addplot [forget plot, name path=upper, draw=none] table [x=z, y=Mean_plus_Std, col sep=space] {plots/potts_ensemble_comparison.dat};
                \addplot [forget plot, name path=lower, draw=none] table [x=z, y=Mean_minus_Std, col sep=space] {plots/potts_ensemble_comparison.dat};
                \addplot [forget plot, fill=blue!30, fill opacity=0.5, draw=none] fill between [of=upper and lower];
                \addlegendimage{legend image code/.code={\fill[blue!30, draw=blue!50] (0cm,-0.1cm) rectangle (0.6cm,0.1cm);}}
                \addlegendentry{Mean $\pm$ 1 Std}
            \end{axis}
        \end{tikzpicture}
    \end{subfigure}
    \hfill
    \begin{subfigure}[b]{0.49\textwidth}
        \centering
        \begin{tikzpicture}
            \begin{axis}[
                ylabel style={at={(axis description cs:1.10,0.5)}, anchor=south},
                width=\linewidth, height=6cm,
                xlabel={$z$},
                ylabel={Error (\%)},
                title={Prediction Error},
                grid=major,
                legend style={at={(0.5,0.02)}, anchor=south},
            ]
                \addplot [blue, thick] table [x=z, y=PctError, col sep=space] {plots/potts_percentage_error.dat};
                \addlegendentry{Mean Error}
                \addplot [forget plot, name path=upper, draw=none] table [x=z, y=PctError_plus_PctStd, col sep=space] {plots/potts_percentage_error.dat};
                \addplot [forget plot, name path=lower, draw=none] table [x=z, y=PctError_minus_PctStd, col sep=space] {plots/potts_percentage_error.dat};
                \addplot [forget plot, fill=blue!30, fill opacity=0.5, draw=none] fill between [of=upper and lower];
                \addlegendimage{legend image code/.code={\fill[blue!30, draw=blue!50] (0cm,-0.1cm) rectangle (0.6cm,0.1cm);}}
                \addlegendentry{Error $\pm$ 1 Std}
            \end{axis}
        \end{tikzpicture}
    \end{subfigure}

    \begin{subfigure}[b]{0.65\textwidth}
        \centering
        \begin{tikzpicture}
            \begin{axis}[
                width=\linewidth, height=6.5cm,
                xlabel={$\widetilde{G}(z=0.5)$},
                ylabel={Count},
                title={Distribution at $z=0.5$},
                ybar,
                bar width=0.000374,
                xmin=1.109645, xmax=1.118990,
                ymin=0,
                ymajorgrids=true,
                xmajorgrids=false,
                legend pos=north east,
                scaled ticks=false,
                xtick={1.11,1.112,1.114,1.116,1.118},
                xticklabel style={
                    /pgf/number format/fixed,
                    /pgf/number format/precision=4,
                    font=\sansmath\sffamily\footnotesize,
                }
            ]
                \addplot [fill=blue!50, draw=black, opacity=0.7] table [x=BinCenter, y=Count, col sep=space] {plots/potts_histogram_z0.500.dat};
                \addlegendentry{Model Predictions}
                \draw [black, dashed, thick] (axis cs:1.110289,\pgfkeysvalueof{/pgfplots/ymin}) -- (axis cs:1.110289,\pgfkeysvalueof{/pgfplots/ymax});
                \addlegendimage{legend image code/.code={\draw[black, dashed, thick] (0cm,0cm) -- (0.6cm,0cm);}}
                \addlegendentry{Exact ($1.1103$)}
                \draw [blue, thick] (axis cs:1.113423,\pgfkeysvalueof{/pgfplots/ymin}) -- (axis cs:1.113423,\pgfkeysvalueof{/pgfplots/ymax});
                \addlegendimage{legend image code/.code={\draw[blue, thick] (0cm,0cm) -- (0.6cm,0cm);}}
                \addlegendentry{Mean ($1.1134$)}
            \end{axis}
        \end{tikzpicture}
    \end{subfigure}

    \caption{NN-predicted reduced correlator $\widetilde G(z)$ for the three-state Potts CFT ($\mathcal M(5,6)_D$, $c=4/5$, $\Delta_{\rm gap}^{D}=2/15$).  The NN prediction at $z=0.5$ is $\widetilde G^{\rm pred}(0.5)=1.11342\pm 0.00091$ against $\widetilde G^{\rm exact}(0.5)=1.11029$.}
    \label{fig:potts_summary}
\end{figure}
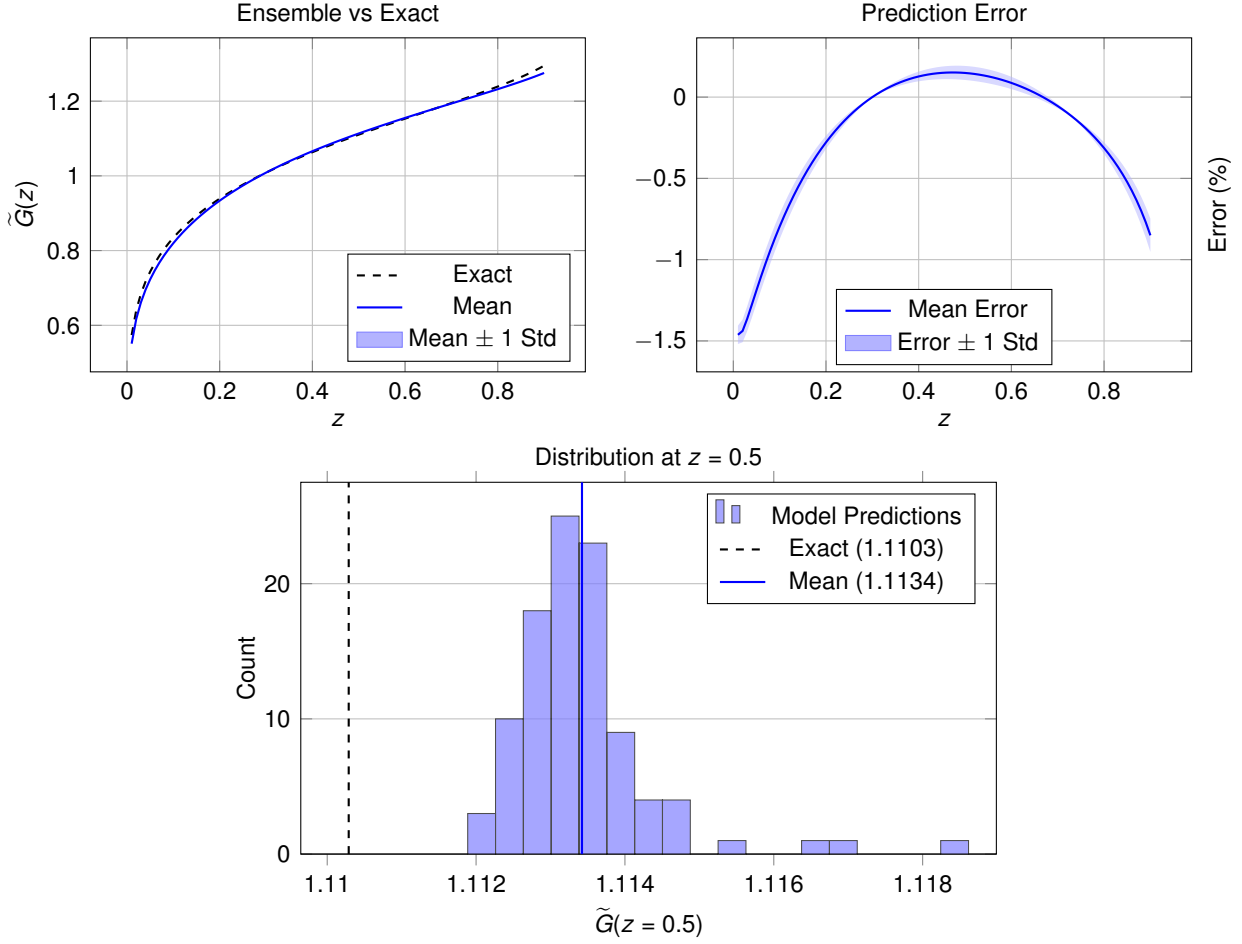

\paragraph{Three-state Potts model: $\mathcal M(5,6)_D$, $c=4/5$, $\Delta_{\rm gap}^{D}=2/15$.}
The canonical $D$-series example is the three-state Potts CFT, obtained from $\mathcal M(5,6)$ by the $(A_4,D_4)$ modular invariant.  Its torus partition function is
\begin{equation}
    Z_{\rm 3\text{-}Potts}(\tau,\bar\tau)
    = |\chi_{1,1}+\chi_{1,5}|^{2} + |\chi_{2,1}+\chi_{2,5}|^{2}
      + 2|\chi_{1,3}|^{2} + 2|\chi_{2,3}|^{2}.
    \label{eq:potts-Z}
\end{equation}
The $(2,2)$ primary of the diagonal $A$-series theory, with chiral weight $h=1/40$, is absent from the $D_4$ modular invariant, and the lightest non-vacuum primary entering the partition function is the spin doublet $\sigma$ with $h_\sigma = h_{3,3} = 1/15$, giving $\Delta_{\rm gap}^{D}=h_\sigma+\bar h_\sigma = 2/15$.  The MS training loss is $(4.32\pm 0.620)\times 10^{-7}$, and the NN prediction at $z=0.5$ reads $\widetilde G^{\rm pred}(0.5)=1.11342\pm 0.000911$ against $\widetilde G^{\rm exact}(0.5)=1.11029$.  Results are shown in Fig.~\ref{fig:potts_summary}.

\subsection{Wess--Zumino--Witten Models}\label{sec:wzw}
WZW models are rational CFTs whose chiral algebra is the affine Kac--Moody algebra $\widehat{\mathfrak g}_k$, with $\mathfrak g$ simple and $k\in\mathbb Z_{>0}$.  The basic input for the torus reconstruction is fixed by the representation theory of highest-weight modules over $\widehat{\mathfrak g}_k$,
\begin{equation}
c=\frac{k\,\dim\mathfrak g}{k+h^\vee}\,,\qquad \Delta_{\rm gap}=2h_{\lambda_\star}=\frac{(\lambda_\star,\lambda_\star+2\rho)}{k+h^\vee}\,,
\end{equation}
where $h^\vee$ is the dual Coxeter number, $\rho$ the Weyl vector, and $\lambda_\star$ the lightest non-vacuum integrable weight.  The integrable weights
\[
P_k^+=\{\lambda\in P^+\,|\,(\lambda,\theta)\le k\}
\]
label affine characters $\chi_\lambda(\tau)=\mathrm{Tr}_{\mathcal H_\lambda}(q^{L_0-c/24})$ that form a finite-dimensional unitary representation of the modular group~\cite{Kac:1984mq,DiFrancesco:1997nk}.  In this section we use the diagonal invariant
\begin{equation}
\mathcal Z_{\mathfrak g,k}(\tau,\bar\tau)=\sum_{\lambda\in P_k^+}\chi_\lambda(\tau)\bar\chi_\lambda(\bar\tau)\,.
\end{equation}
The Weyl--Kac character formula gives
\begin{equation}
\chi_\lambda(\tau)=\lim_{\mathbf u\to0}\frac{\displaystyle\sum_{w\in S_N}\varepsilon(w)\,\Theta^{(k+N)}_{w(\lambda+\rho)}(\tau,\mathbf u)}{\displaystyle\sum_{w\in S_N}\varepsilon(w)\,\Theta^{(N)}_{w(\rho)}(\tau,\mathbf u)}\,,\qquad \Theta^{(m)}_{\mu}(\tau,\mathbf u)=\sum_{\gamma\in mQ^\vee+\mu}q^{(\gamma,\gamma)/(2m)}e^{2\pi i(\gamma,\mathbf u)}\,.
\label{eq:kac-weyl-char}
\end{equation}
Here $\mathbf u$ is a Cartan fugacity, $Q^\vee$ the coroot lattice, $\rho$ the Weyl vector, $S_N$ the Weyl group, and $\varepsilon(w)$ its sign.

\paragraph{$\widehat{\mathfrak{su}}(2)_2$: $c=3/2$, $\Delta_{\rm gap}=3/8$.}
For $\widehat{\mathfrak{su}}(2)_k$, this gives
\begin{equation}
\chi_\ell^{(k)}(\tau)=\frac{1}{\eta(\tau)^{3}}\sum_{n\in\mathbb Z}\bigl[2(k+2)n+\ell+1\bigr]\,q^{[2(k+2)n+\ell+1]^{2}/[4(k+2)]}\,,
\label{eq:su2k-char}
\end{equation}
with integrable representations labelled by $\ell=0,\ldots,k$ (spin $j=\ell/2$) and
\begin{equation}
h_\ell=\frac{\ell(\ell+2)}{4(k+2)}\,,\qquad c=\frac{3k}{k+2}\,,\qquad \Delta_{\rm gap}=2h_1=\frac{3}{2(k+2)}\,.
\end{equation}

\begin{figure}[H]
    \centering
    \begin{subfigure}[b]{0.49\textwidth}
        \centering
        \begin{tikzpicture}
            \begin{axis}[
                width=\linewidth, height=6cm,
                xlabel={$z$},
                ylabel={$\widetilde{G}(z)$},
                title={Ensemble vs Exact},
                grid=major,
                legend pos=north west,
            ]
                \addplot [black, dashed, thick] table [x=z, y=Exact, col sep=space] {plots/su2_k2_ensemble_comparison.dat};
                \addlegendentry{Exact}
                \addplot [blue, thick] table [x=z, y=Mean, col sep=space] {plots/su2_k2_ensemble_comparison.dat};
                \addlegendentry{Mean}
                \addplot [forget plot, name path=upper, draw=none] table [x=z, y=Mean_plus_Std, col sep=space] {plots/su2_k2_ensemble_comparison.dat};
                \addplot [forget plot, name path=lower, draw=none] table [x=z, y=Mean_minus_Std, col sep=space] {plots/su2_k2_ensemble_comparison.dat};
                \addplot [forget plot, fill=blue!30, fill opacity=0.5, draw=none] fill between [of=upper and lower];
                \addlegendimage{legend image code/.code={\fill[blue!30, draw=blue!50] (0cm,-0.1cm) rectangle (0.6cm,0.1cm);}}
                \addlegendentry{Mean $\pm$ 1 Std}
            \end{axis}
        \end{tikzpicture}
    \end{subfigure}
    \hfill
    \begin{subfigure}[b]{0.49\textwidth}
        \centering
        \begin{tikzpicture}
            \begin{axis}[
                ylabel style={at={(axis description cs:1.10,0.5)}, anchor=south},
                width=\linewidth, height=6cm,
                xlabel={$z$},
                ylabel={Error (\%)},
                title={Prediction Error},
                grid=major,
                legend style={at={(0.5,0.02)}, anchor=south},
            ]
                \addplot [blue, thick] table [x=z, y=PctError, col sep=space] {plots/su2_k2_percentage_error.dat};
                \addlegendentry{Mean Error}
                \addplot [forget plot, name path=upper, draw=none] table [x=z, y=PctError_plus_PctStd, col sep=space] {plots/su2_k2_percentage_error.dat};
                \addplot [forget plot, name path=lower, draw=none] table [x=z, y=PctError_minus_PctStd, col sep=space] {plots/su2_k2_percentage_error.dat};
                \addplot [forget plot, fill=blue!30, fill opacity=0.5, draw=none] fill between [of=upper and lower];
                \addlegendimage{legend image code/.code={\fill[blue!30, draw=blue!50] (0cm,-0.1cm) rectangle (0.6cm,0.1cm);}}
                \addlegendentry{Error $\pm$ 1 Std}
            \end{axis}
        \end{tikzpicture}
    \end{subfigure}

    \begin{subfigure}[b]{0.65\textwidth}
        \centering
        \begin{tikzpicture}
            \begin{axis}[
                width=\linewidth, height=6.5cm,
                xlabel={$\widetilde{G}(z=0.5)$},
                ylabel={Count},
                title={Distribution at $z=0.5$},
                ybar,
                bar width=0.003431,
                xmin=0.773249, xmax=0.859029,
                xtick={0.78,0.8,0.82,0.84},
                ymin=0,
                ymajorgrids=true,
                xmajorgrids=false,
                legend pos=north west,
                scaled ticks=false,
                xticklabel style={
                    /pgf/number format/fixed,
                    /pgf/number format/precision=4,
                    font=\sansmath\sffamily\footnotesize,
                }
            ]
                \addplot [fill=blue!50, draw=black, opacity=0.7] table [x=BinCenter, y=Count, col sep=space] {plots/su2_k2_histogram_z0.500.dat};
                \addlegendentry{Predictions}
                \draw [black, dashed, thick] (axis cs:0.83776,\pgfkeysvalueof{/pgfplots/ymin}) -- (axis cs:0.83776,\pgfkeysvalueof{/pgfplots/ymax});
                \addlegendimage{legend image code/.code={\draw[black, dashed, thick] (0cm,0cm) -- (0.6cm,0cm);}}
                \addlegendentry{Exact ($0.8378$)}
                \draw [blue, thick] (axis cs:0.84119,\pgfkeysvalueof{/pgfplots/ymin}) -- (axis cs:0.84119,\pgfkeysvalueof{/pgfplots/ymax});
                \addlegendimage{legend image code/.code={\draw[blue, thick] (0cm,0cm) -- (0.6cm,0cm);}}
                \addlegendentry{Mean ($0.8412$)}
            \end{axis}
        \end{tikzpicture}
    \end{subfigure}

    \caption{NN-predicted reduced correlator $\widetilde G(z)$ for the level-two $\widehat{\mathfrak{su}}(2)_2$ WZW model, on the filtered ensemble ($92/1000$ seeds after the early-stopping cut).  The NN prediction at $z=0.5$ is $\widetilde G^{\rm pred}(0.5)=0.84119\pm 0.01430$ against $\widetilde G^{\rm exact}(0.5)=0.83776$.}
    \label{fig:su2_k2_summary}
\end{figure}
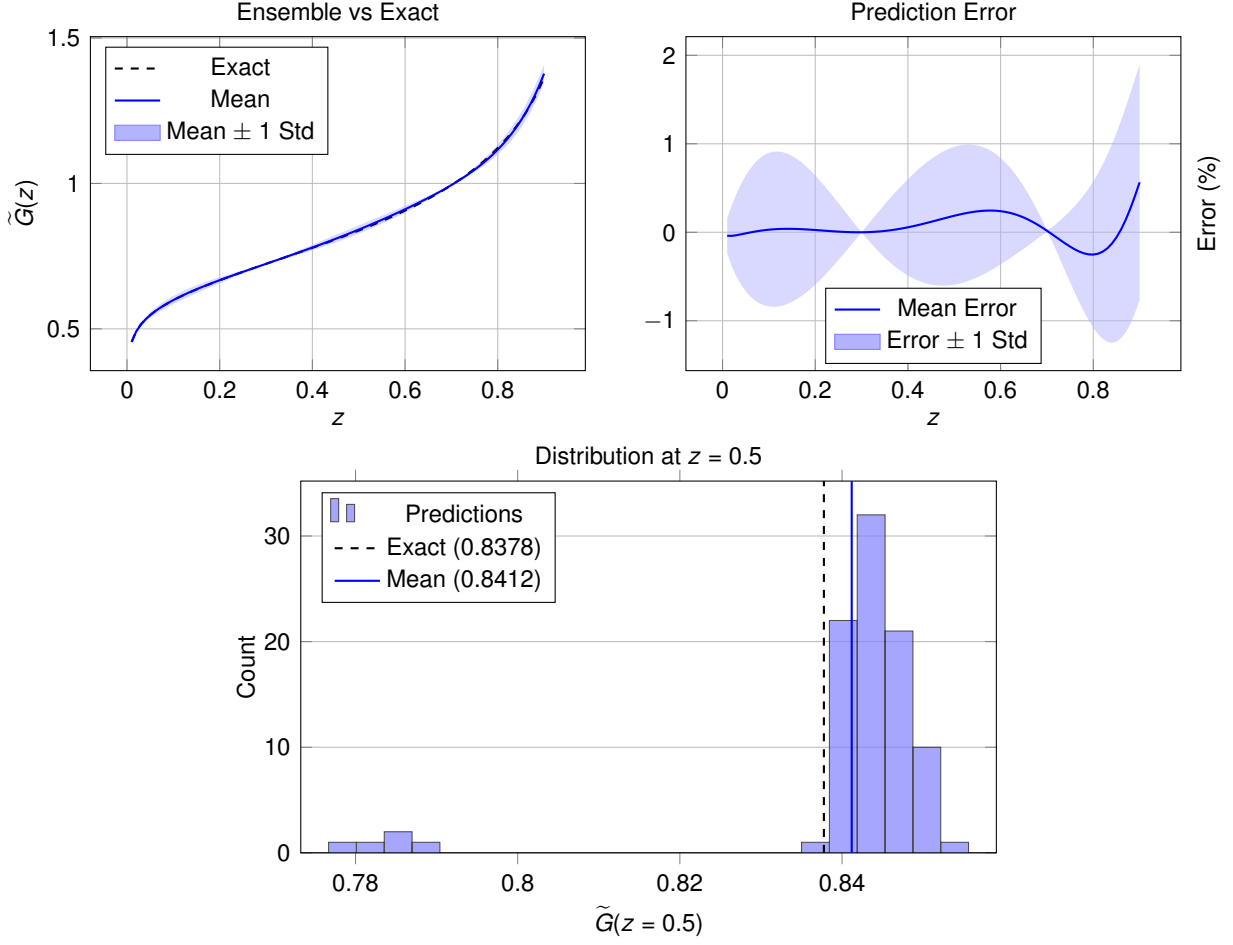

The $k=1$ case is treated in Appendix~\ref{app:torus-extra} (Fig.~\ref{fig:su2_k1_summary}), alongside $\widehat{\mathfrak{su}}(3)_1$ (Fig.~\ref{fig:su3_k1_summary}) and $\widehat{\mathfrak{su}}(4)_1$ (Table~\ref{tab:torus-extra}). For $k=2$ the affine primaries are $\ell=0,1,2$, with $h_\ell=\ell(\ell+2)/16$. The lightest non-vacuum affine primary is $\ell=1$, giving $\Delta_{\rm gap}=3/8$ and correction exponent $2\Delta_{\rm gap}=3/4$. In this case, we observed that the full ensemble of runs is bimodal, exhibiting two peaks at $z=0.5$ cleanly separated. Among the seed-level diagnostics we surveyed, the only feature that correlated with the peak a run landed on was whether the run triggered early stopping.  Runs that terminated via the $5\times10^3$-epoch stagnation criterion before hitting the $2\times 10^5$-epoch cap fell almost exclusively into the peak tracking the exact answer, while runs that ran out the full epoch budget clustered around the spurious peak.  It is unclear why this particular example exhibits a bimodal pattern of configurations, but the positive role of the early stopping criterion in identifying the physical correlator as a low-loss crossing symmetric configuration in the lazy-training regime is consistent with our previous observations in the companion work~\cite{GKNS:1}. Retaining only the early-stopped runs selected $92$ of $1000$ seeds. On this filtered ensemble, the MS training loss is $(5.21\pm 0.413)\times 10^{-6}$, and $\widetilde G^{\rm pred}(0.5)=0.84119\pm 0.0143$ against $\widetilde G^{\rm exact}(0.5)=0.83776$ (Fig.~\ref{fig:su2_k2_summary}). The same type of bimodality appeared also in the $\widehat{\mathfrak{su}}(2)_1$ case in Appendix~\ref{app:torus-extra}, where the same filter selected $184$ of $1000$ seeds. 

\section{Annulus Reconstructions in Compact CFTs}
\label{sec:annulus-results}

\subsection{Minimal Model Boundary Conditions}\label{sec:annulus-minimal}
The diagonal minimal model $\mathcal M(m,m+1)$, with Kac table~\eqref{eq:kac-table}, has modular $S$-matrix~\cite{DiFrancesco:1997nk}
\begin{equation}
S_{(r,s)(r',s')}=2\sqrt{\tfrac{2}{m(m+1)}}\,(-1)^{1+sr'+rs'}\sin\!\Big(\tfrac{\pi(m+1)\,rr'}{m}\Big)\sin\!\Big(\tfrac{\pi m\,ss'}{m+1}\Big)\,.
\label{eq:mm-S}
\end{equation}
Substituting~\eqref{eq:mm-S} into~\eqref{eq:verlinde} and~\eqref{eq:open-closed-decomp} makes the open/closed decompositions explicit, and the leading and next-lightest dimensions read off from each channel feed the annulus ans\"atze~\eqref{eq:annulus-ansatz}.

\subsubsection{Ising}\label{sec:annulus-ising}
The Ising model ($c=1/2$) has Cardy states \(\{\mathbf 1,\boldsymbol\varepsilon,\boldsymbol\sigma\}\), giving four inequivalent boundary pairs. The MS training loss over $100$ seeds, the open- and closed-channel prediction errors, and figure references are collected in Table~\ref{tab:ising-annulus}.  The pairs $(\boldsymbol\varepsilon,\boldsymbol\varepsilon)=(\mathbf 1,\mathbf 1)$ and $(\boldsymbol\sigma,\boldsymbol\varepsilon)=(\mathbf 1,\boldsymbol\sigma)$ are equivalent to their listed representatives. The $(\boldsymbol\sigma,\boldsymbol\sigma)$ annulus has coincident channels.

\begin{table}[H]
\centering
\setlength{\tabcolsep}{5pt}
\begin{tabular}{lcccc}
\hline
$(\alpha,\beta)$ & MS training loss & MRPE$^{(\mathrm o)}$ (\%) & MRPE$^{(\mathrm c)}$ (\%) & Fig.\ \\
\hline
$(\mathbf 1,\mathbf 1)$          & $(9.97\pm 3.60)\times 10^{-8}$ & $0.172\pm 0.096$ & $0.210\pm 0.076$ & \ref{fig:ising_annulus_summary} \\
$(\mathbf 1,\boldsymbol\sigma)$  & $(2.48\pm 0.064)\times 10^{-6}$ & $0.466\pm 0.034$ & $0.393\pm 0.005$ & \ref{fig:ising_annulus_1sig} \\
$(\boldsymbol\sigma,\boldsymbol\sigma)$ & $(1.46\pm 0.012)\times 10^{-6}$ & $0.387\pm 0.012$ & $0.387\pm 0.012$ & \ref{fig:ising_ann_ss} \\
$(\mathbf 1,\boldsymbol\varepsilon)$ & $(5.54\pm 1.11)\times 10^{-6}$ & $0.654\pm 0.336$ & $0.619\pm 0.138$ & \ref{fig:ising_ann_1e} \\
\hline
\end{tabular}
\caption{NN-predicted Ising annulus partition functions.  MRPE denotes the ensemble mean of the per-seed maximum of the relative prediction error~\eqref{eq:reporting-relerr}, expressed as a percentage.}
\label{tab:ising-annulus}
\end{table}

\subsubsection{Tricritical Ising}\label{sec:annulus-tricrit}
\begin{figure}[H]
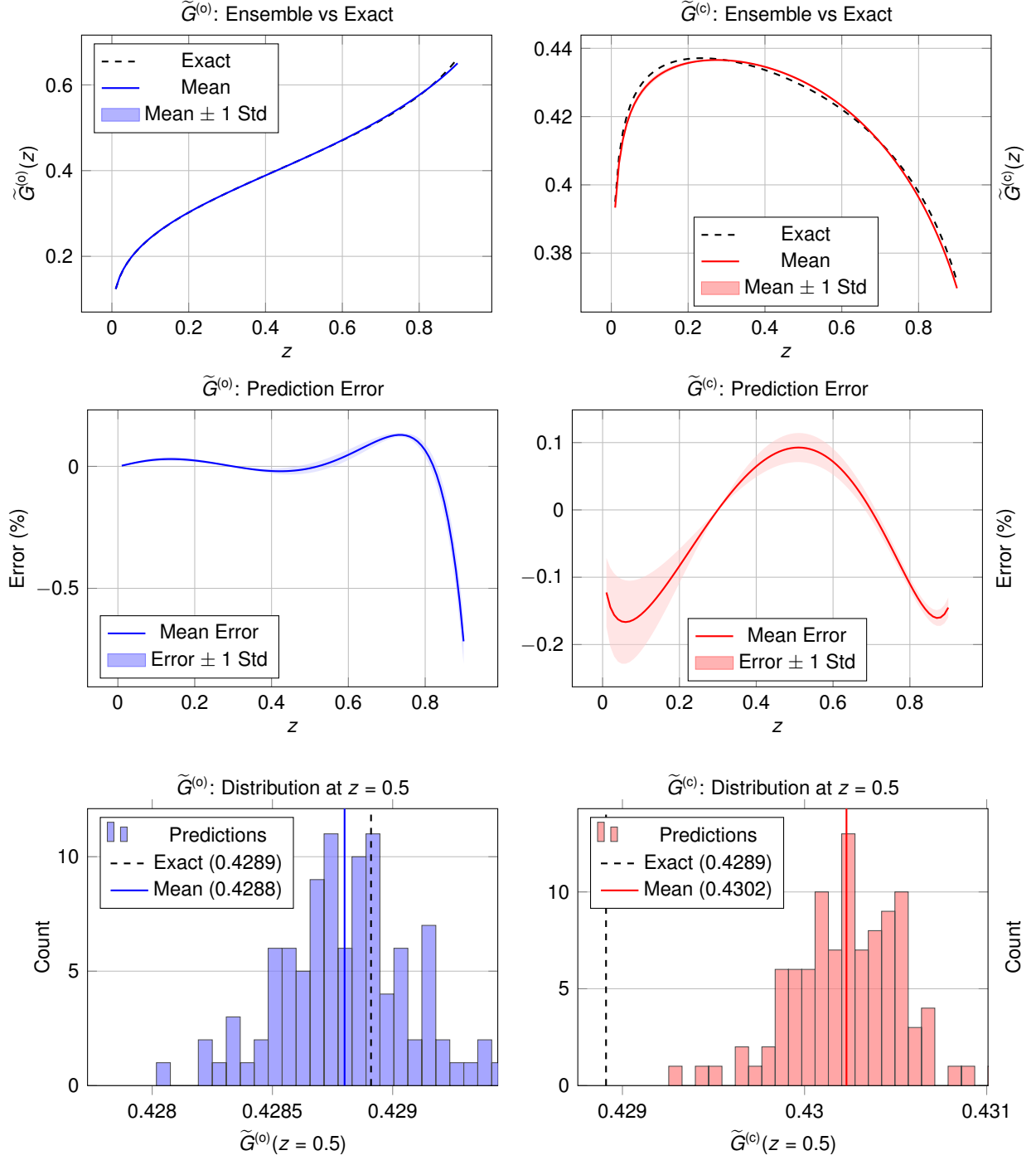

    \centering
    \begin{subfigure}[b]{0.49\textwidth}
        \centering
        %
        \hspace*{-2.9mm}
    \end{subfigure}
    \caption{NN-predicted reduced annulus correlators $\widetilde G^{(\mathrm o)}(z)$ (open, blue) and $\widetilde G^{(\mathrm c)}(z)$ (closed, red) for the tricritical Ising $(\alpha,\beta)=(\boldsymbol\sigma,\boldsymbol\sigma')$ annulus over $100$ seeds.  At $z=0.5$, $\widetilde G^{\rm exact}=0.4289$ vs ensemble means $0.4288$ (open) and $0.4302$ (closed).}
    \label{fig:tricrit_annulus_ssp}
\end{figure}

\begin{figure}[H]
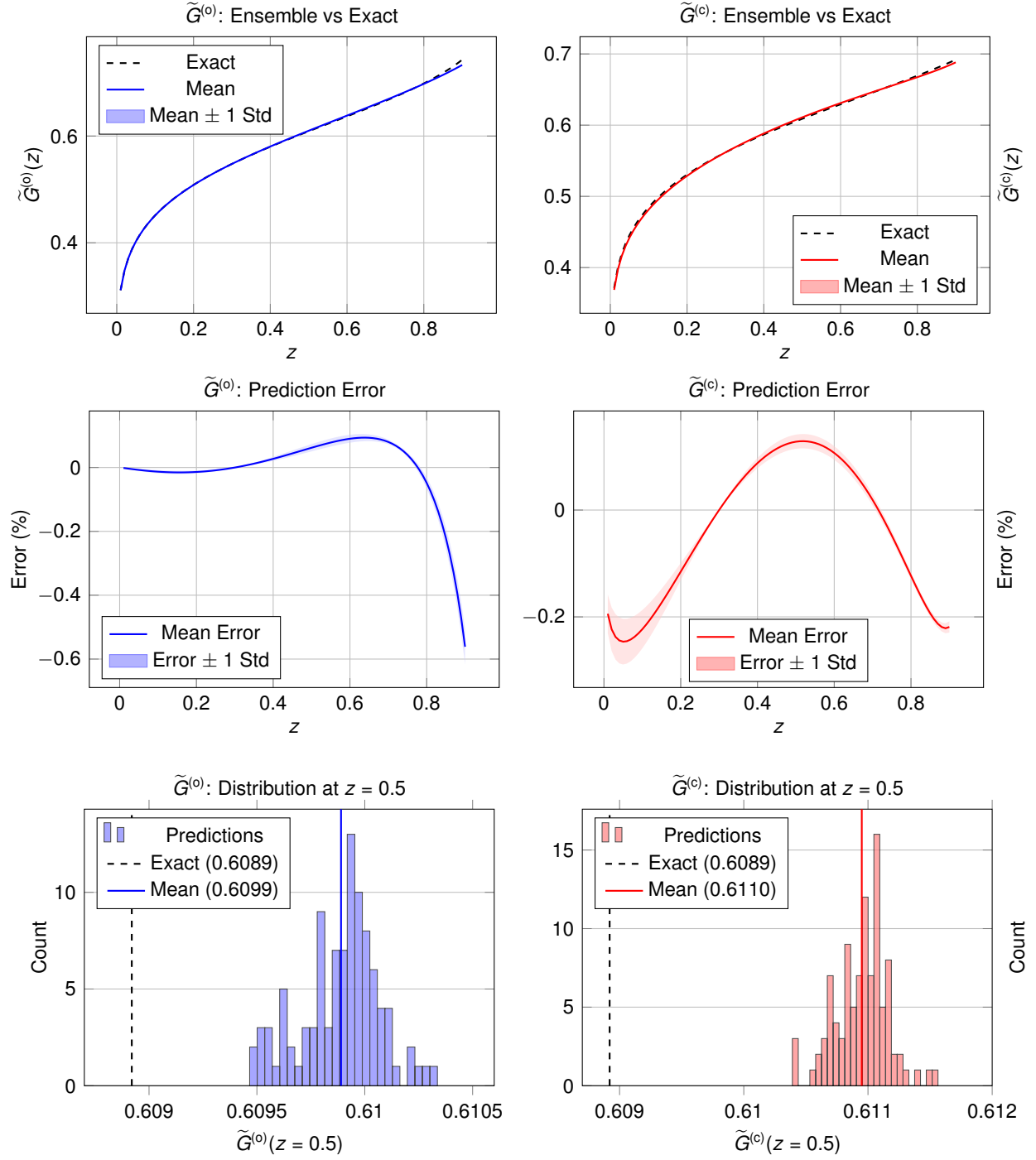

    \centering
    \begin{subfigure}[b]{0.49\textwidth}
        \centering
        %
        \hspace*{-2.9mm}
    \end{subfigure}
    \caption{NN-predicted reduced annulus correlators $\widetilde G^{(\mathrm o)}(z)$ (open, blue) and $\widetilde G^{(\mathrm c)}(z)$ (closed, red) for the tricritical Ising $(\alpha,\beta)=(\boldsymbol\sigma',\boldsymbol\varepsilon')$ annulus over $100$ seeds (equivalent to $(\mathbf 1,\boldsymbol\sigma)$).  At $z=0.5$, $\widetilde G^{\rm exact}=0.6089$ vs ensemble means $0.6099$ (open) and $0.6110$ (closed).}
    \label{fig:tricrit_annulus_spep}
\end{figure}

For tricritical Ising $\mathcal M(4,5)$ ($c=7/10$), we reconstruct the annulus partition functions for the Cardy pairs summarised in Table~\ref{tab:tricrit-annulus}. Two representative pairs with $\widetilde G^{(\mathrm o)}\ne\widetilde G^{(\mathrm c)}$ — $(\boldsymbol\sigma,\boldsymbol\sigma')$ and $(\boldsymbol\sigma',\boldsymbol\varepsilon')$ — are shown in Figs.~\ref{fig:tricrit_annulus_ssp} and~\ref{fig:tricrit_annulus_spep}. The remaining pairs are marked ``--'' in the Fig.\ column and their six-panel plots are available in the companion \href{https://github.com}{\texttt{GitHub}} repository \href{https://github.com/andstergiou/nn-cft}{\tt andstergiou/nn-cft}.

\begin{table}[H]
\centering
\setlength{\tabcolsep}{5pt}
\begin{tabular}{lcccc}
\hline
$(\alpha,\beta)$ & MS training loss & MRPE$^{(\mathrm o)}$ (\%) & MRPE$^{(\mathrm c)}$ (\%) & Fig.\ \\
\hline
$(\boldsymbol\sigma,\boldsymbol\sigma')$ & $(1.26\pm 0.06)\times 10^{-6}$ & $0.717\pm 0.092$ & $0.183\pm 0.039$ & \ref{fig:tricrit_annulus_ssp} \\
$(\boldsymbol\sigma',\boldsymbol\varepsilon')$ & $(8.78\pm 0.19)\times 10^{-7}$ & $0.561\pm 0.055$ & $0.252\pm 0.035$ & \ref{fig:tricrit_annulus_spep} \\
$(\boldsymbol\sigma,\boldsymbol\varepsilon)$ & $(2.58\pm 0.50)\times 10^{-7}$ & $0.231\pm 0.028$ & $0.228\pm 0.040$ & $-$ \\
$(\mathbf 1,\boldsymbol\sigma')$ & $(2.71\pm 0.21)\times 10^{-6}$ & $0.483\pm 0.119$ & $0.428\pm 0.047$ & $-$ \\
$(\mathbf 1,\mathbf 1)$ & $(4.06\pm 1.06)\times 10^{-7}$ & $0.815\pm 0.181$ & $1.812\pm 0.161$ & $-$ \\
$(\boldsymbol\sigma,\boldsymbol\sigma)$ & $(6.54\pm 2.29)\times 10^{-7}$ & $0.505\pm 0.119$ & $0.614\pm 0.120$ & $-$ \\
$(\boldsymbol\sigma',\boldsymbol\sigma')$ & $(3.10\pm 0.25)\times 10^{-6}$ & $0.654\pm 0.092$ & $0.419\pm 0.074$ & $-$ \\
$(\boldsymbol\varepsilon,\boldsymbol\varepsilon)$ & $(2.24\pm 0.08)\times 10^{-6}$ & $1.519\pm 0.146$ & $1.835\pm 0.140$ & $-$ \\
$(\mathbf 1,\boldsymbol\varepsilon)$ & $(5.13\pm 0.15)\times 10^{-6}$ & $1.943\pm 0.177$ & $1.967\pm 0.156$ & $-$ \\
$(\mathbf 1,\boldsymbol\varepsilon')$ & $(3.88\pm 0.78)\times 10^{-6}$ & $0.491\pm 0.394$ & $0.833\pm 0.385$ & $-$ \\
$(\mathbf 1,\boldsymbol\varepsilon'')$ & $(2.16\pm 0.64)\times 10^{-6}$ & $0.902\pm 0.506$ & $1.782\pm 0.395$ & $-$ \\
$(\boldsymbol\varepsilon,\boldsymbol\varepsilon')$ & $(9.71\pm 0.42)\times 10^{-7}$ & $0.848\pm 0.149$ & $0.737\pm 0.120$ & $-$ \\
\hline
\end{tabular}
\caption{NN-predicted Tricritical Ising annulus partition functions.  MRPE denotes the ensemble mean of the per-seed maximum of the relative prediction error~\eqref{eq:reporting-relerr}, expressed as a percentage.  Pairs marked ``--'' in the Fig.\ column are not shown here; their $6$-panel plots are available in the companion \href{https://github.com}{\texttt{GitHub}} repository \href{https://github.com/andstergiou/nn-cft}{\tt andstergiou/nn-cft}.}
\label{tab:tricrit-annulus}
\end{table}

\begin{figure}[H]
    \centering
    \begin{subfigure}[b]{0.49\textwidth}
        \centering

        \hspace*{-2.9mm}
    \end{subfigure}
    \caption{NN-predicted reduced annulus correlators $\widetilde G^{(\mathrm o)}(z)$ (open, blue) and $\widetilde G^{(\mathrm c)}(z)$ (closed, red) for the Ising $(\alpha,\beta)=(\mathbf 1,\mathbf 1)$ annulus over $100$ seeds. At $z=0.5$, $\widetilde G^{\rm exact}=0.7689$ vs ensemble means $0.7685$ (open) and $0.7681$ (closed).}
    \label{fig:ising_annulus_summary}
\end{figure}

\begin{figure}[H]
    \centering
    \begin{subfigure}[b]{0.49\textwidth}
        \centering
        %
        \hspace*{-2.9mm}
    \end{subfigure}
    \caption{NN-predicted reduced annulus correlators $\widetilde G^{(\mathrm o)}(z)$ (open, blue) and $\widetilde G^{(\mathrm c)}(z)$ (closed, red) for the Ising $(\alpha,\beta)=(\mathbf 1,\boldsymbol\sigma)$ (fixed-free) annulus over $100$ seeds.  At $z=0.5$, $\widetilde G^{\rm exact}=0.5202$ vs ensemble means $0.5214$ (open) and $0.5228$ (closed).}
    \label{fig:ising_annulus_1sig}
\end{figure}

\subsection{WZW Boundary States}\label{sec:annulus-wzw}
The annulus reconstruction extends to WZW models with no change to the line-correlator setup, only the spectrum that the affine chiral algebra forces.  Cardy boundary states of $\widehat{\mathfrak g}_k$ are labelled by integrable weights $\lambda\in P^+_k$ and take the form~\eqref{eq:cardy-state} with the modular $S$-matrix, for $\widehat{\mathfrak{su}}(N)_k$, given by the Kac--Peterson formula~\cite{Kac:1984mq,DiFrancesco:1997nk}
\begin{equation}
S_{\lambda\mu}=\frac{i^{N(N-1)/2}}{\sqrt{N}\,(k+N)^{(N-1)/2}}\sum_{w\in S_N}\varepsilon(w)\,\exp\!\Big(\!-\tfrac{2\pi i}{k+N}\,\bigl(w(\lambda+\rho),\,\mu+\rho\bigr)\Big)\,,
\label{eq:kac-peterson}
\end{equation}
where $S_N$ is the Weyl group of $\mathfrak{su}(N)$ and $\rho$ is the Weyl vector.  Setting $N=2$ collapses the Weyl sum to
\begin{equation}
S^{\widehat{\mathfrak{su}}(2)_k}_{\ell\ell'} = \sqrt{\tfrac{2}{k+2}}\,\sin\!\Big(\tfrac{\pi(\ell+1)(\ell'+1)}{k+2}\Big)\,,\qquad \ell,\ell'\in\{0,1,\ldots,k\}\,,
\label{eq:su2k-S}
\end{equation}
while setting $N=3$, $k=1$ gives constant entries $|S_{\lambda\mu}|=1/\sqrt 3$.  The open/closed Verlinde and Cardy decompositions follow from~\eqref{eq:open-closed-decomp}, and the leading and next-lightest dimensions read off from each channel feed the ans\"atze. The training losses and the open- and closed-channel prediction errors for every Cardy pair are collected in Table~\ref{tab:wzw-annulus}. We display the two $\widehat{\mathfrak{su}}(2)_1$ pairs in the main text (Figs.~\ref{fig:wzw_su2_k1_00} and~\ref{fig:wzw_su2_k1_0h}). The four $\widehat{\mathfrak{su}}(2)_2$ and two $\widehat{\mathfrak{su}}(3)_1$ two-channel reconstructions are collected in Appendix~\ref{app:annulus-extra}.

\begin{table}[H]
\centering
\setlength{\tabcolsep}{5pt}
\begin{tabular}{llcccc}
\hline
Model & $(\lambda_\alpha,\lambda_\beta)$ & MS training loss & MRPE$^{(\mathrm o)}$ (\%) & MRPE$^{(\mathrm c)}$ (\%) & Fig.\ \\
\hline
$\widehat{\mathfrak{su}}(2)_1$ & $(\lambda_0,\lambda_0)$ & $(3.28\pm 0.12)\times 10^{-6}$ & $0.744\pm 0.091$ & $0.724\pm 0.021$ & \ref{fig:wzw_su2_k1_00} \\
$\widehat{\mathfrak{su}}(2)_1$ & $(\lambda_0,\lambda_1)$ & $(4.61\pm 0.42)\times 10^{-7}$ & $0.410\pm 0.053$ & $0.084\pm 0.011$ & \ref{fig:wzw_su2_k1_0h} \\
$\widehat{\mathfrak{su}}(2)_2$ & $(\lambda_0,\lambda_0)$ & $(2.48\pm 0.08)\times 10^{-6}$ & $1.083\pm 0.108$ & $0.746\pm 0.051$ & \ref{fig:wzw_su2_k2_00} \\
$\widehat{\mathfrak{su}}(2)_2$ & $(\lambda_0,\lambda_1)$ & $(3.37\pm 0.16)\times 10^{-6}$ & $0.795\pm 0.041$ & $0.420\pm 0.008$ & \ref{fig:wzw_su2_k2_0h} \\
$\widehat{\mathfrak{su}}(2)_2$ & $(\lambda_0,\lambda_2)$ & $(2.53\pm 0.11)\times 10^{-6}$ & $0.615\pm 0.059$ & $0.461\pm 0.019$ & \ref{fig:wzw_su2_k2_01} \\
$\widehat{\mathfrak{su}}(2)_2$ & $(\lambda_1,\lambda_1)$ & $(8.87\pm 1.78)\times 10^{-6}$ & $1.031\pm 0.188$ & $1.031\pm 0.188$ & \ref{fig:wzw_su2_k2_hh} \\
$\widehat{\mathfrak{su}}(3)_1$ & $(\lambda_0,\lambda_0)$ & $(3.10\pm 0.12)\times 10^{-6}$ & $0.653\pm 0.063$ & $0.847\pm 0.024$ & \ref{fig:wzw_su3_k1_00} \\
$\widehat{\mathfrak{su}}(3)_1$ & $(\lambda_0,\lambda_1)$ & $(4.60\pm 0.41)\times 10^{-7}$ & $0.429\pm 0.040$ & $0.088\pm 0.013$ & \ref{fig:wzw_su3_k1_01} \\
\hline
\end{tabular}
\caption{NN-predicted WZW annulus partition functions across $\widehat{\mathfrak{su}}(2)_1$, $\widehat{\mathfrak{su}}(2)_2$, and $\widehat{\mathfrak{su}}(3)_1$.  MRPE denotes the ensemble mean of the per-seed maximum of the relative prediction error~\eqref{eq:reporting-relerr}, expressed as a percentage.}
\label{tab:wzw-annulus}
\end{table}

\begin{figure}[H]
    \centering
    \begin{subfigure}[b]{0.49\textwidth}
        \centering
        \begin{tikzpicture}
            \begin{axis}[width=\linewidth, height=6cm, xlabel={$z$},
                ylabel={$\widetilde G^{(\mathrm o)}(z)$},
                title={$\widetilde G^{(\mathrm o)}$: Ensemble vs Exact},
                grid=major, legend pos=north west]
                \addplot [black, dashed, thick] table [x=z, y=Exact, col sep=space] {plots/wzw_su2_k1_00_ensemble_comparison_g.dat};
                \addlegendentry{Exact}
                \addplot [blue, thick] table [x=z, y=Mean, col sep=space] {plots/wzw_su2_k1_00_ensemble_comparison_g.dat};
                \addlegendentry{Mean}
                \addplot [forget plot, name path=upper, draw=none] table [x=z, y=Mean_plus_Std, col sep=space] {plots/wzw_su2_k1_00_ensemble_comparison_g.dat};
                \addplot [forget plot, name path=lower, draw=none] table [x=z, y=Mean_minus_Std, col sep=space] {plots/wzw_su2_k1_00_ensemble_comparison_g.dat};
                \addplot [forget plot, fill=blue!30, fill opacity=0.5, draw=none] fill between [of=upper and lower];
                \addlegendimage{legend image code/.code={\fill[blue!30, draw=blue!50] (0cm,-0.1cm) rectangle (0.6cm,0.1cm);}}
                \addlegendentry{Mean $\pm$ 1 Std}
            \end{axis}
        \end{tikzpicture}
    \end{subfigure}
    \hfill
    \begin{subfigure}[b]{0.49\textwidth}
        \centering
        \begin{tikzpicture}
            \begin{axis}[
                ylabel style={at={(axis description cs:1.10,0.5)}, anchor=south},
                width=\linewidth, height=6cm, xlabel={$z$},
                ylabel={$\widetilde G^{(\mathrm c)}(z)$},
                title={$\widetilde G^{(\mathrm c)}$: Ensemble vs Exact},
                grid=major, legend pos=north west]
                \addplot [black, dashed, thick] table [x=z, y=Exact, col sep=space] {plots/wzw_su2_k1_00_ensemble_comparison_f.dat};
                \addlegendentry{Exact}
                \addplot [red, thick] table [x=z, y=Mean, col sep=space] {plots/wzw_su2_k1_00_ensemble_comparison_f.dat};
                \addlegendentry{Mean}
                \addplot [forget plot, name path=upper, draw=none] table [x=z, y=Mean_plus_Std, col sep=space] {plots/wzw_su2_k1_00_ensemble_comparison_f.dat};
                \addplot [forget plot, name path=lower, draw=none] table [x=z, y=Mean_minus_Std, col sep=space] {plots/wzw_su2_k1_00_ensemble_comparison_f.dat};
                \addplot [forget plot, fill=red!30, fill opacity=0.5, draw=none] fill between [of=upper and lower];
                \addlegendimage{legend image code/.code={\fill[red!30, draw=red!50] (0cm,-0.1cm) rectangle (0.6cm,0.1cm);}}
                \addlegendentry{Mean $\pm$ 1 Std}
            \end{axis}
        \end{tikzpicture}
    \end{subfigure}
    \vspace{1em}
    \hspace*{-4mm}
    \begin{subfigure}[b]{0.49\textwidth}
        \centering
        \begin{tikzpicture}
            \begin{axis}[width=\linewidth, height=6cm, xlabel={$z$},
                ylabel={Error (\%)}, title={$\widetilde G^{(\mathrm o)}$: Prediction Error},
                grid=major, legend pos=north west]
                \addplot [blue, thick] table [x=z, y=PctError, col sep=space] {plots/wzw_su2_k1_00_percentage_error_g.dat};
                \addlegendentry{Mean Error}
                \addplot [forget plot, name path=upper, draw=none] table [x=z, y=PctError_plus_PctStd, col sep=space] {plots/wzw_su2_k1_00_percentage_error_g.dat};
                \addplot [forget plot, name path=lower, draw=none] table [x=z, y=PctError_minus_PctStd, col sep=space] {plots/wzw_su2_k1_00_percentage_error_g.dat};
                \addplot [forget plot, fill=blue!20, fill opacity=0.5, draw=none] fill between [of=upper and lower];
                \addlegendimage{legend image code/.code={\fill[blue!30, draw=blue!50] (0cm,-0.1cm) rectangle (0.6cm,0.1cm);}}
                \addlegendentry{Error $\pm$ 1 Std}
            \end{axis}
        \end{tikzpicture}
    \end{subfigure}
    \begin{subfigure}[b]{0.49\textwidth}
        \centering
        \begin{tikzpicture}
            \begin{axis}[
                ylabel style={at={(axis description cs:1.10,0.5)}, anchor=south},
                width=\linewidth, height=6cm, xlabel={$z$},
                ylabel={Error (\%)}, title={$\widetilde G^{(\mathrm c)}$: Prediction Error},
                grid=major, legend style={at={(0.5,0.98)}, anchor=north}]
                \addplot [red, thick] table [x=z, y=PctError, col sep=space] {plots/wzw_su2_k1_00_percentage_error_f.dat};
                \addlegendentry{Mean Error}
                \addplot [forget plot, name path=upper, draw=none] table [x=z, y=PctError_plus_PctStd, col sep=space] {plots/wzw_su2_k1_00_percentage_error_f.dat};
                \addplot [forget plot, name path=lower, draw=none] table [x=z, y=PctError_minus_PctStd, col sep=space] {plots/wzw_su2_k1_00_percentage_error_f.dat};
                \addplot [forget plot, fill=red!20, fill opacity=0.5, draw=none] fill between [of=upper and lower];
                \addlegendimage{legend image code/.code={\fill[red!30, draw=red!50] (0cm,-0.1cm) rectangle (0.6cm,0.1cm);}}
                \addlegendentry{Error $\pm$ 1 Std}
            \end{axis}
        \end{tikzpicture}
    \end{subfigure}
    \vspace{1em}
    \hspace*{1mm}
    \begin{subfigure}[b]{0.49\textwidth}
        \centering
        \begin{tikzpicture}
            \begin{axis}[width=\linewidth, height=6cm,
                xlabel={$\widetilde G^{(\mathrm o)}(z=0.5)$}, ylabel={Count},
                title={$\widetilde G^{(\mathrm o)}$: Distribution at $z=0.5$},
                ybar, bar width=0.000318, xmin=0.766586, xmax=0.774533,
                xtick={0.768,0.77,0.772,0.774}, ymin=0, ymajorgrids=true, xmajorgrids=false,
                ytick distance=5, legend pos=north east, scaled ticks=false,
                xticklabel style={ /pgf/number format/fixed,
                    /pgf/number format/precision=4, font=\sansmath\sffamily\footnotesize }]
                \addplot [fill=blue!50, draw=black, opacity=0.7] table [x=BinCenter, y=Count, col sep=space] {plots/wzw_su2_k1_00_histogram_g_z0.500.dat};
                \addlegendentry{Predictions}
                \draw [black, dashed, thick] (axis cs:0.77399,\pgfkeysvalueof{/pgfplots/ymin}) -- (axis cs:0.77399,\pgfkeysvalueof{/pgfplots/ymax});
                \addlegendimage{legend image code/.code={\draw[black, dashed, thick] (0cm,0cm) -- (0.6cm,0cm);}}
                \addlegendentry{Exact ($0.7740$)}
                \draw [blue, thick] (axis cs:0.76835,\pgfkeysvalueof{/pgfplots/ymin}) -- (axis cs:0.76835,\pgfkeysvalueof{/pgfplots/ymax});
                \addlegendimage{legend image code/.code={\draw[blue, thick] (0cm,0cm) -- (0.6cm,0cm);}}
                \addlegendentry{Mean ($0.7683$)}
            \end{axis}
        \end{tikzpicture}
    \end{subfigure}
    \begin{subfigure}[b]{0.49\textwidth}
        \centering
        \begin{tikzpicture}
            \begin{axis}[
                ylabel style={at={(axis description cs:1.10,0.5)}, anchor=south},
                width=\linewidth, height=6cm,
                xlabel={$\widetilde G^{(\mathrm c)}(z=0.5)$}, ylabel={Count},
                title={$\widetilde G^{(\mathrm c)}$: Distribution at $z=0.5$},
                ybar, bar width=0.000390, xmin=0.764905, xmax=0.774658,
                xtick={0.766,0.768,0.77,0.772,0.774}, ymin=0, ymajorgrids=true, xmajorgrids=false,
                ytick distance=5,
                legend pos=north east, scaled ticks=false,
                xticklabel style={ /pgf/number format/fixed,
                    /pgf/number format/precision=4, font=\sansmath\sffamily\footnotesize }]
                \addplot [fill=red!50, draw=black, opacity=0.7] table [x=BinCenter, y=Count, col sep=space] {plots/wzw_su2_k1_00_histogram_f_z0.500.dat};
                \addlegendentry{Predictions}
                \draw [black, dashed, thick] (axis cs:0.77399,\pgfkeysvalueof{/pgfplots/ymin}) -- (axis cs:0.77399,\pgfkeysvalueof{/pgfplots/ymax});
                \addlegendimage{legend image code/.code={\draw[black, dashed, thick] (0cm,0cm) -- (0.6cm,0cm);}}
                \addlegendentry{Exact ($0.7740$)}
                \draw [red, thick] (axis cs:0.76656,\pgfkeysvalueof{/pgfplots/ymin}) -- (axis cs:0.76656,\pgfkeysvalueof{/pgfplots/ymax});
                \addlegendimage{legend image code/.code={\draw[red, thick] (0cm,0cm) -- (0.6cm,0cm);}}
                \addlegendentry{Mean ($0.7666$)}
            \end{axis}
        \end{tikzpicture}
        \hspace*{-2.9mm}
    \end{subfigure}
    \caption{NN-predicted reduced annulus correlators $\widetilde G^{(\mathrm o)}(z)$ (open, blue) and $\widetilde G^{(\mathrm c)}(z)$ (closed, red) for the $\widehat{\mathfrak{su}}(2)_1$ $(\lambda_0,\lambda_0)$ annulus over $100$ seeds.  At $z=0.5$, $\widetilde G^{\rm exact}=0.7740$ vs ensemble means $0.7683$ (open) and $0.7666$ (closed).}
    \label{fig:wzw_su2_k1_00}
\end{figure}
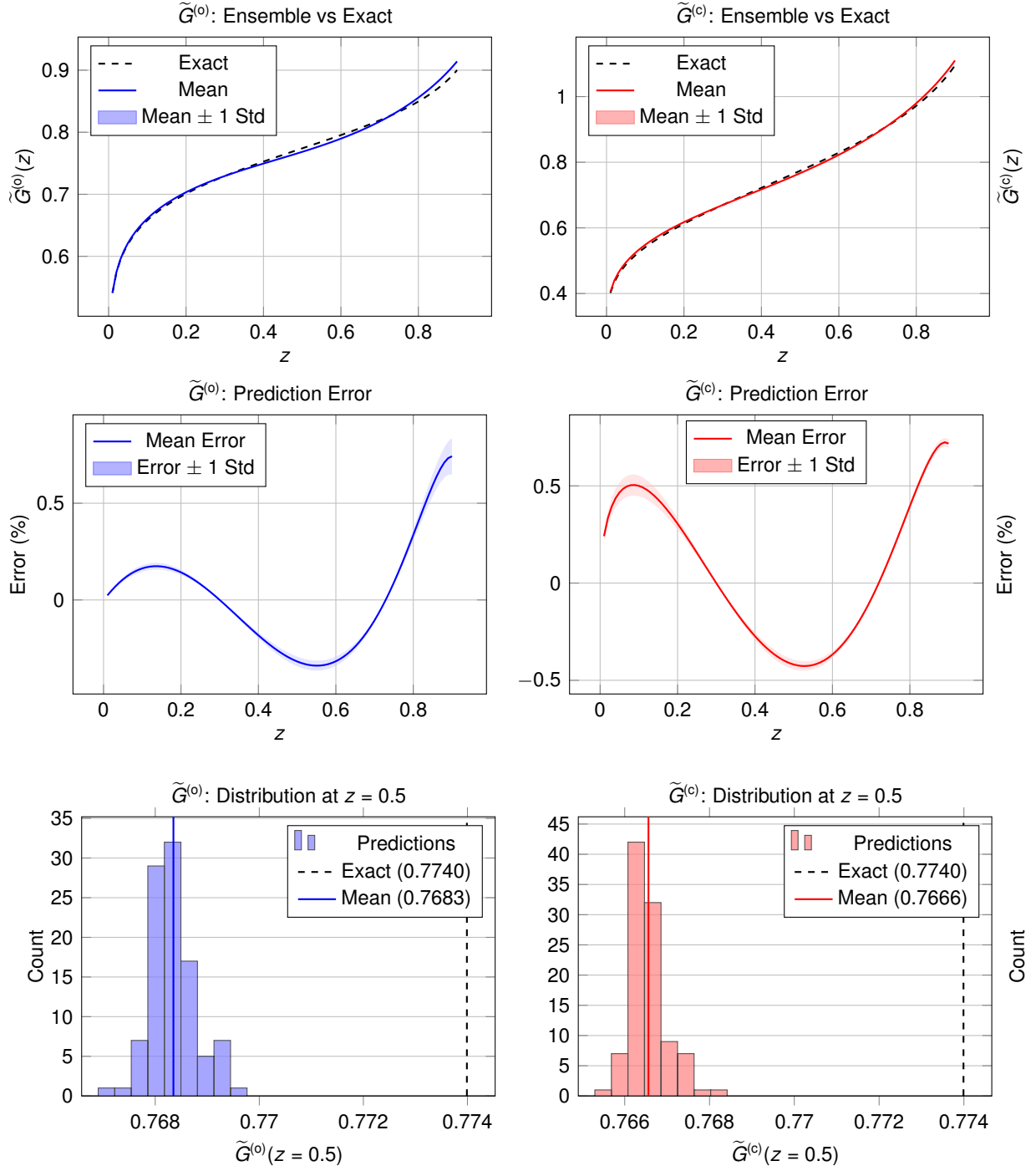
\begin{figure}[H]
    \centering
    \begin{subfigure}[b]{0.49\textwidth}
        \centering
        \begin{tikzpicture}
            \begin{axis}[width=\linewidth, height=6cm, xlabel={$z$},
                ylabel={$\widetilde G^{(\mathrm o)}(z)$},
                title={$\widetilde G^{(\mathrm o)}$: Ensemble vs Exact},
                grid=major, legend pos=north west]
                \addplot [black, dashed, thick] table [x=z, y=Exact, col sep=space] {plots/wzw_su2_k1_0h_ensemble_comparison_g.dat};
                \addlegendentry{Exact}
                \addplot [blue, thick] table [x=z, y=Mean, col sep=space] {plots/wzw_su2_k1_0h_ensemble_comparison_g.dat};
                \addlegendentry{Mean}
                \addplot [forget plot, name path=upper, draw=none] table [x=z, y=Mean_plus_Std, col sep=space] {plots/wzw_su2_k1_0h_ensemble_comparison_g.dat};
                \addplot [forget plot, name path=lower, draw=none] table [x=z, y=Mean_minus_Std, col sep=space] {plots/wzw_su2_k1_0h_ensemble_comparison_g.dat};
                \addplot [forget plot, fill=blue!30, fill opacity=0.5, draw=none] fill between [of=upper and lower];
                \addlegendimage{legend image code/.code={\fill[blue!30, draw=blue!50] (0cm,-0.1cm) rectangle (0.6cm,0.1cm);}}
                \addlegendentry{Mean $\pm$ 1 Std}
            \end{axis}
        \end{tikzpicture}
    \end{subfigure}
    \hfill
    \begin{subfigure}[b]{0.49\textwidth}
        \centering
        \begin{tikzpicture}
            \begin{axis}[
                ylabel style={at={(axis description cs:1.10,0.5)}, anchor=south},
                width=\linewidth, height=6cm, xlabel={$z$},
                ylabel={$\widetilde G^{(\mathrm c)}(z)$},
                title={$\widetilde G^{(\mathrm c)}$: Ensemble vs Exact},
                grid=major, legend pos=south west]
                \addplot [black, dashed, thick] table [x=z, y=Exact, col sep=space] {plots/wzw_su2_k1_0h_ensemble_comparison_f.dat};
                \addlegendentry{Exact}
                \addplot [red, thick] table [x=z, y=Mean, col sep=space] {plots/wzw_su2_k1_0h_ensemble_comparison_f.dat};
                \addlegendentry{Mean}
                \addplot [forget plot, name path=upper, draw=none] table [x=z, y=Mean_plus_Std, col sep=space] {plots/wzw_su2_k1_0h_ensemble_comparison_f.dat};
                \addplot [forget plot, name path=lower, draw=none] table [x=z, y=Mean_minus_Std, col sep=space] {plots/wzw_su2_k1_0h_ensemble_comparison_f.dat};
                \addplot [forget plot, fill=red!30, fill opacity=0.5, draw=none] fill between [of=upper and lower];
                \addlegendimage{legend image code/.code={\fill[red!30, draw=red!50] (0cm,-0.1cm) rectangle (0.6cm,0.1cm);}}
                \addlegendentry{Mean $\pm$ 1 Std}
            \end{axis}
        \end{tikzpicture}
    \end{subfigure}
    \vspace{1em}
    \hspace*{-4mm}
    \begin{subfigure}[b]{0.49\textwidth}
        \centering
        \begin{tikzpicture}
            \begin{axis}[width=\linewidth, height=6cm, xlabel={$z$},
                ylabel={Error (\%)}, title={$\widetilde G^{(\mathrm o)}$: Prediction Error},
                grid=major, legend pos=south west]
                \addplot [blue, thick] table [x=z, y=PctError, col sep=space] {plots/wzw_su2_k1_0h_percentage_error_g.dat};
                \addlegendentry{Mean Error}
                \addplot [forget plot, name path=upper, draw=none] table [x=z, y=PctError_plus_PctStd, col sep=space] {plots/wzw_su2_k1_0h_percentage_error_g.dat};
                \addplot [forget plot, name path=lower, draw=none] table [x=z, y=PctError_minus_PctStd, col sep=space] {plots/wzw_su2_k1_0h_percentage_error_g.dat};
                \addplot [forget plot, fill=blue!20, fill opacity=0.5, draw=none] fill between [of=upper and lower];
                \addlegendimage{legend image code/.code={\fill[blue!30, draw=blue!50] (0cm,-0.1cm) rectangle (0.6cm,0.1cm);}}
                \addlegendentry{Error $\pm$ 1 Std}
            \end{axis}
        \end{tikzpicture}
    \end{subfigure}
    \begin{subfigure}[b]{0.49\textwidth}
        \centering
        \begin{tikzpicture}
            \begin{axis}[
                ylabel style={at={(axis description cs:1.10,0.5)}, anchor=south},
                width=\linewidth, height=6cm, xlabel={$z$},
                ylabel={Error (\%)}, title={$\widetilde G^{(\mathrm c)}$: Prediction Error},
                grid=major, legend pos=north west,
                scaled y ticks=false, yticklabel style={/pgf/number format/fixed, /pgf/number format/precision=2}]
                \addplot [red, thick] table [x=z, y=PctError, col sep=space] {plots/wzw_su2_k1_0h_percentage_error_f.dat};
                \addlegendentry{Mean Error}
                \addplot [forget plot, name path=upper, draw=none] table [x=z, y=PctError_plus_PctStd, col sep=space] {plots/wzw_su2_k1_0h_percentage_error_f.dat};
                \addplot [forget plot, name path=lower, draw=none] table [x=z, y=PctError_minus_PctStd, col sep=space] {plots/wzw_su2_k1_0h_percentage_error_f.dat};
                \addplot [forget plot, fill=red!20, fill opacity=0.5, draw=none] fill between [of=upper and lower];
                \addlegendimage{legend image code/.code={\fill[red!30, draw=red!50] (0cm,-0.1cm) rectangle (0.6cm,0.1cm);}}
                \addlegendentry{Error $\pm$ 1 Std}
            \end{axis}
        \end{tikzpicture}
    \end{subfigure}
    \vspace{1em}
    \hspace*{1mm}
    \begin{subfigure}[b]{0.49\textwidth}
        \centering
        \begin{tikzpicture}
            \begin{axis}[width=\linewidth, height=6cm,
                xlabel={$\widetilde G^{(\mathrm o)}(z=0.5)$}, ylabel={Count},
                title={$\widetilde G^{(\mathrm o)}$: Distribution at $z=0.5$},
                ybar, bar width=0.000073, xmin=0.318905, xmax=0.320720,
                xtick={0.319,0.3195,0.320,0.3205}, ymin=0, ymajorgrids=true, xmajorgrids=false,
                ytick distance=5, legend pos=north east, scaled ticks=false,
                xticklabel style={ /pgf/number format/fixed,
                    /pgf/number format/precision=4, font=\sansmath\sffamily\footnotesize }]
                \addplot [fill=blue!50, draw=black, opacity=0.7] table [x=BinCenter, y=Count, col sep=space] {plots/wzw_su2_k1_0h_histogram_g_z0.500.dat};
                \addlegendentry{Predictions}
                \draw [black, dashed, thick] (axis cs:0.32060,\pgfkeysvalueof{/pgfplots/ymin}) -- (axis cs:0.32060,\pgfkeysvalueof{/pgfplots/ymax});
                \addlegendimage{legend image code/.code={\draw[black, dashed, thick] (0cm,0cm) -- (0.6cm,0cm);}}
                \addlegendentry{Exact ($0.3206$)}
                \draw [blue, thick] (axis cs:0.31948,\pgfkeysvalueof{/pgfplots/ymin}) -- (axis cs:0.31948,\pgfkeysvalueof{/pgfplots/ymax});
                \addlegendimage{legend image code/.code={\draw[blue, thick] (0cm,0cm) -- (0.6cm,0cm);}}
                \addlegendentry{Mean ($0.3195$)}
            \end{axis}
        \end{tikzpicture}
    \end{subfigure}
    \begin{subfigure}[b]{0.49\textwidth}
        \centering
        \begin{tikzpicture}
            \begin{axis}[
                ylabel style={at={(axis description cs:1.10,0.5)}, anchor=south},
                width=\linewidth, height=6cm,
                xlabel={$\widetilde G^{(\mathrm c)}(z=0.5)$}, ylabel={Count},
                title={$\widetilde G^{(\mathrm c)}$: Distribution at $z=0.5$},
                ybar, bar width=0.000056, xmin=0.319574, xmax=0.320973,
                xtick={0.3196,0.3200,0.3204,0.3208}, ymin=0, ymajorgrids=true, xmajorgrids=false,
                ytick distance=5, legend pos=north west, scaled ticks=false,
                xticklabel style={ /pgf/number format/fixed,
                    /pgf/number format/precision=4, font=\sansmath\sffamily\footnotesize }]
                \addplot [fill=red!50, draw=black, opacity=0.7] table [x=BinCenter, y=Count, col sep=space] {plots/wzw_su2_k1_0h_histogram_f_z0.500.dat};
                \addlegendentry{Predictions}
                \draw [black, dashed, thick] (axis cs:0.32060,\pgfkeysvalueof{/pgfplots/ymin}) -- (axis cs:0.32060,\pgfkeysvalueof{/pgfplots/ymax});
                \addlegendimage{legend image code/.code={\draw[black, dashed, thick] (0cm,0cm) -- (0.6cm,0cm);}}
                \addlegendentry{Exact ($0.3206$)}
                \draw [red, thick] (axis cs:0.32036,\pgfkeysvalueof{/pgfplots/ymin}) -- (axis cs:0.32036,\pgfkeysvalueof{/pgfplots/ymax});
                \addlegendimage{legend image code/.code={\draw[red, thick] (0cm,0cm) -- (0.6cm,0cm);}}
                \addlegendentry{Mean ($0.3204$)}
            \end{axis}
        \end{tikzpicture}
        \hspace*{-2.9mm}
    \end{subfigure}
    \caption{NN-predicted reduced annulus correlators $\widetilde G^{(\mathrm o)}(z)$ (open, blue) and $\widetilde G^{(\mathrm c)}(z)$ (closed, red) for the $\widehat{\mathfrak{su}}(2)_1$ $(\lambda_0,\lambda_1)$ annulus over $100$ seeds.  At $z=0.5$, $\widetilde G^{\rm exact}=0.3206$ vs ensemble means $0.3195$ (open) and $0.3204$ (closed).}
    \label{fig:wzw_su2_k1_0h}
\end{figure}
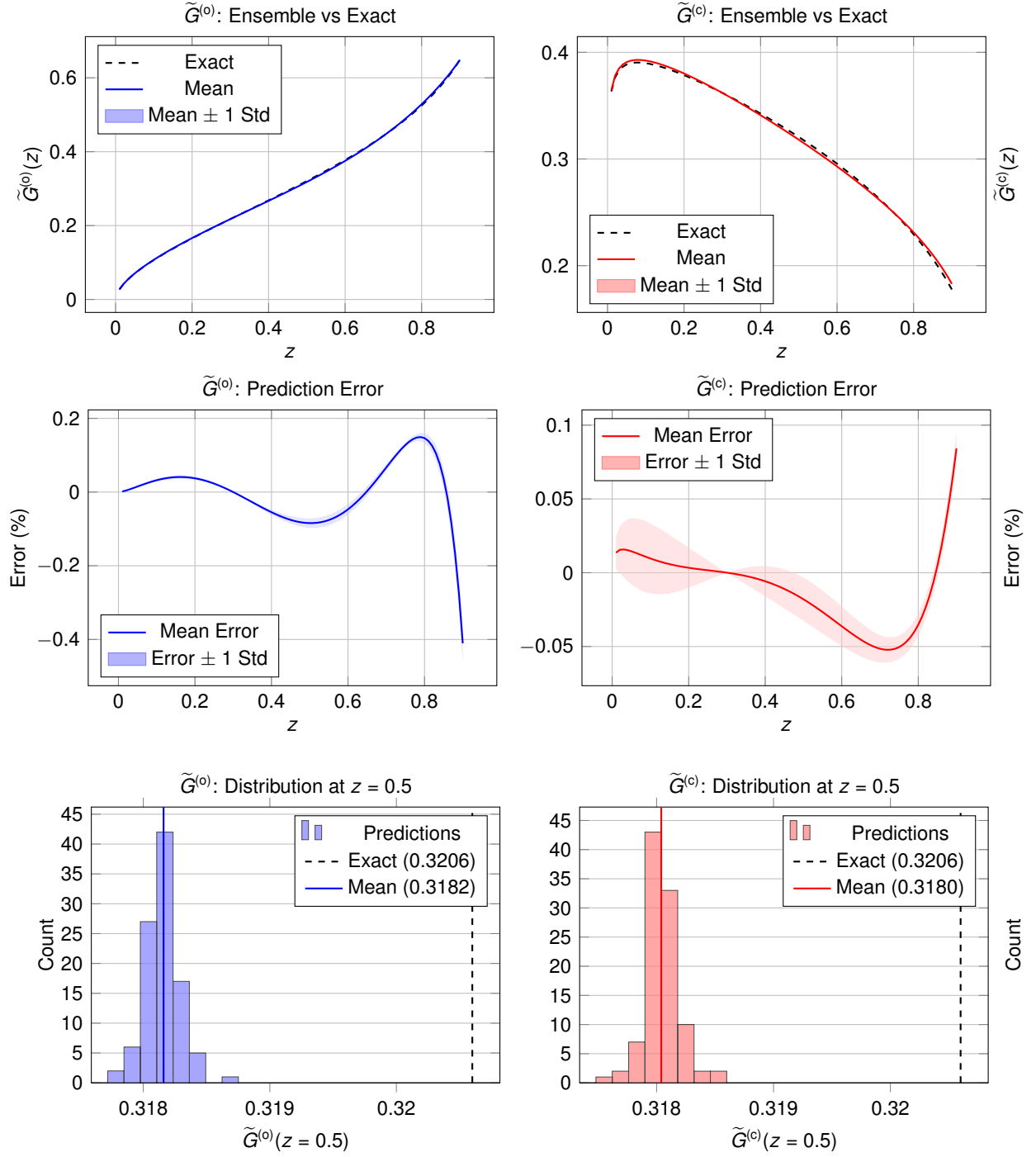

\section{Partition Functions in Non-compact CFTs}
\label{noncompact}

In this section we consider the case of irrational non-compact CFTs. One of the simplest examples is the CFT of a non-compact scalar whose torus partition function is (up to a constant that includes the infinite target volume of the scalar)
\begin{equation}
    \label{ncaa}
    \ZZ(\tau,\bar \tau) = \frac{1}{\sqrt{\Ima\tau} |\eta(\tau)|^2}\,.
\end{equation}
The reduced partition function $\sqrt{\Ima\tau} |\eta(\tau)|^2 \ZZ(\tau,\bar \tau)$ is therefore a trivial constant in this case and the corresponding reduced four-point correlator (on a line) is
\begin{equation}
    \label{ncab}
    \widetilde{G}(z)=B_{\rm t}(z)\lsp G(z)=
    \left(\frac{z^2}{2^8(1-z)}\right)^{1/12}.
\end{equation}

Exactly the same partition function \eqref{ncaa} appears also in Liouville theory for any real value of the central charge $c$ (including the cases of time-like and space-like Liouville theories). In this larger family of non-compact CFTs, the central charge appears in the reduced four-point correlator
\begin{equation}
    \label{ncac}
    \widetilde{G}(z)=B_{\rm t}(z)\lsp G(z)=
    \left(
    \frac{z^2}{2^8(1-z)}
    \right)^{c/12},
\end{equation}
which allows us to probe a variable functional dependence on the level of the four-point correlator.

We have applied the anchored NN bootstrap for moderate values of the central charge of order 1 (negative and positive) recovering the physical correlator at the same sub-percent-level accuracy as in all the other examples explored in this paper. Since the application is straightforward, we will refrain from an explicit demonstration of these results here. We should note, however, that as we increase $c$ the target function \eqref{ncac} takes values over an exponentially increasing range. This makes the numerics increasingly harder. This difficulty is generic and requires appropriate treatment. In typical machine learning applications it is addressed by a suitable rescaling of the data. In our context, one can try similar implementations by composing the NN function with a smooth non-linear function and/or by suitably modifying the loss function to make it sensitive to the correlators across the whole range of the $z$ interval. One has to deal with this problem on a case-by-case basis. For concreteness, let us examine here how to deal with such difficulties in a more interesting (and less trivial) application of annulus partition functions in spacelike Liouville theory with $c\in (25,\infty)$ involving the ZZ and FZZ boundary conditions.

There are two well-known independent classes of boundary conditions we can consider in spacelike Liouville theory. The first class involves the discrete ZZ boundary conditions $[n,m]$ labelled by two positive integers $(n,m=1,2,\ldots)$ in one-to-one correspondence with the degenerate Virasoro representations \cite{Zamolodchikov:2001ah}. The second class involves the FZZ boundary conditions, in one-to-one correspondence with the principal-series representations, which are labelled by a continuous parameter $s\in [0,\infty)$ \cite{Fateev:2000ik,Teschner:2000md}. 

In what follows, we consider for illustration annulus partition functions involving the basic ZZ boundary condition $[1,1]$: the $\ZZL[1,1]$-$\FZZ[s=0.2]$ annulus partition function and the $\ZZL[1,1]$-$\ZZL[1,1]$ annulus partition function. The parameter $s=0.2$ was chosen as an arbitrary example. For concreteness, we also chose $c=31.5$. Other examples can be treated similarly.

For the ZZ$(1,1)$-FZZ$(s)$ annulus, the open-channel correlator is
\begin{equation}
    G^{\mathrm{open}}(z)
    =
    B(z)\,\frac{q(z)^{s^2/4}}{\eta(it(z))}\,,\qquad q(z) = e^{2\pi i \tau(z)} = e^{-2\pi t(z)}\,,
\end{equation}
where $B(z)$ denotes the kinematic prefactor isolating the Virasoro character as defined in \eqref{eq:annulus-partition-four-point-map}, and the closed-channel correlator is obtained from crossing,
\begin{equation}
    G^{\mathrm{closed}}(z)
    =
    \left(\frac{z}{1-z}\right)^{c/8}
    G^{\mathrm{open}}(1-z)\,.
\end{equation}
The reduced correlators are defined by
\begin{equation}
    G^{\text{(o)}}_{\text{red}}(z)=\eta(it(z))\,G^{\mathrm{open}}(z)\,,
    \qquad
    G^{\text{(c)}}_{\text{red}}(z)=\eta(it(z))\,G^{\mathrm{closed}}(z)\,,
\end{equation}
and accordingly the corresponding crossing equation becomes, as in the previous section,
\begin{equation}
    G^{\text{(o)}}_{\mathrm{red}}(z)
    =
    \left(\frac{z}{1-z}\right)^{c/8}
    \frac{1}{\sqrt{t(z)}}
    \,G^{\text{(c)}}_{\mathrm{red}}(1-z)\,.
\end{equation}
We parametrise these correlation functions using NNs as
\begin{equation}
    G^{\text{(o)}}_{\mathrm{red}}(z)= z^{\frac{c}{12}+\frac{s^2}{2}}
    (1-z)^{-\frac{c}{24}}\,e^{\text{NN}^{\rm (o)}(z)},
    \qquad  G^{\text{(c)}}_{\mathrm{red}}(z)=z^{\frac{c}{12}}
    (1-z)^{\frac{s^2}{2}-\frac{c}{24}}
    \sqrt{\frac{\log\bigl(\frac{16}{1-z}\bigr)}{\log\bigl(\frac{16}{z}\bigr)}}\,e^{\text{NN}^{\rm (c)}(z)}.
\end{equation}
Here it is not necessary to split the contribution into $L$ and $H$ pieces, although doing so can further improve the reconstruction. Since the correlators take very small values, we reconstruct them directly. We also parametrise the neural-network output exponentially, which prevents the prediction from becoming negative and is consistent with the positivity of the correlator.

With this prescription, we reconstruct the $\ZZL[1,1]$-$\FZZ[s=0.2]$ annulus at $c=31.5$, now training with the interior anchor $z_0=0.7$ over $100$ seeds on the wider grid of $N=150$ points $z\in[0.05,0.95]$. Across the ensemble the MS training loss is $(6.18\pm 3.34)\times 10^{-8}$, and at $z=0.5$ the exact open reduced correlator $G^{(\mathrm o),\rm exact}_{\rm red}=2.63\times 10^{-4}$ is recovered as $(3.08\pm 0.62)\times 10^{-4}$, while the closed value $G^{(\mathrm c),\rm exact}_{\rm red}=2.64\times 10^{-4}$ is recovered as $(2.45\pm 0.18)\times 10^{-4}$. The corresponding six-panel diagnostic is shown in Fig.~\ref{fig:liouville_zz11_fzz02}.

For the $\ZZL[1,1]$-$\ZZL[1,1]$ annulus, a similar prescription struggles to reconstruct the correlator accurately close to the edge in the open channel due to the significant disparity in the magnitude of the correlators in the two channels. This is another example where a suitable reformulation of the approach can help the numerical implementation. To make the values of the correlators comparable in the two channels, we redefined the reduced correlators as
\begin{equation}
    \widetilde{G}^{(\mathrm o/\mathrm c)}(z)=(1-z)^{c/8} G^{(\mathrm o/\mathrm c)}(z)\,.
\end{equation}
With this ansatz, the crossing equation takes the form
\begin{equation}
    \widetilde{G}^{(\mathrm o)}(z)=\widetilde{G}^{(\mathrm c)}(1-z)\,.
\end{equation}
Then, we set $\widetilde{G}^{(\mathrm o)}(z)=L(z)+H(z)$ with
\begin{equation}
    L(z)=(1-z)^{c/8},\qquad H(z)= z^2 (1-z)^{(c-1)/12} \log(16/(1-z))^{-3/2} \lsp \mathrm{NN}^{\rm (o)}(z)
\end{equation}
in the open channel and
\begin{equation}
    \widetilde{G}^{(\mathrm c)}(z)=z^{(c-1)/12}  \log(16/z)^{-3/2}\lsp\mathrm{NN}^{\rm (c)}(z)
\end{equation}
in the closed channel.

\begin{figure}[H]
    \centering
    \begin{subfigure}[b]{0.49\textwidth}
        \centering
        \begin{tikzpicture}
            \begin{axis}[width=\linewidth, height=6cm, xlabel={$z$},
                ylabel={$\widetilde G^{(\mathrm o)}(z)$},
                title={$\widetilde G^{(\mathrm o)}$: Ensemble vs Exact},
                grid=major, legend pos=north east]
                \addplot [black, dashed, thick] table [x=z, y=Exact, col sep=space] {plots/liouville_zz11_ensemble_g.dat};
                \addlegendentry{Exact}
                \addplot [blue, thick] table [x=z, y=Mean, col sep=space] {plots/liouville_zz11_ensemble_g.dat};
                \addlegendentry{Mean}
                \addplot [forget plot, name path=upZZo, draw=none] table [x=z, y=Mean_plus_Std, col sep=space] {plots/liouville_zz11_ensemble_g.dat};
                \addplot [forget plot, name path=loZZo, draw=none] table [x=z, y=Mean_minus_Std, col sep=space] {plots/liouville_zz11_ensemble_g.dat};
                \addplot [forget plot, fill=blue!30, fill opacity=0.5, draw=none] fill between [of=upZZo and loZZo];
                \addlegendimage{legend image code/.code={\fill[blue!30, draw=blue!50] (0cm,-0.1cm) rectangle (0.6cm,0.1cm);}}
                \addlegendentry{Mean $\pm$ 1 Std}
            \end{axis}
        \end{tikzpicture}
    \end{subfigure}
    \hfill
    \begin{subfigure}[b]{0.49\textwidth}
        \centering
        \begin{tikzpicture}
            \begin{axis}[
                ylabel style={at={(axis description cs:1.10,0.5)}, anchor=south},
                width=\linewidth, height=6cm, xlabel={$z$},
                ylabel={$\widetilde G^{(\mathrm c)}(z)$},
                title={$\widetilde G^{(\mathrm c)}$: Ensemble vs Exact},
                grid=major, legend pos=north west]
                \addplot [black, dashed, thick] table [x=z, y=Exact, col sep=space] {plots/liouville_zz11_ensemble_f.dat};
                \addlegendentry{Exact}
                \addplot [red, thick] table [x=z, y=Mean, col sep=space] {plots/liouville_zz11_ensemble_f.dat};
                \addlegendentry{Mean}
                \addplot [forget plot, name path=upZZc, draw=none] table [x=z, y=Mean_plus_Std, col sep=space] {plots/liouville_zz11_ensemble_f.dat};
                \addplot [forget plot, name path=loZZc, draw=none] table [x=z, y=Mean_minus_Std, col sep=space] {plots/liouville_zz11_ensemble_f.dat};
                \addplot [forget plot, fill=red!30, fill opacity=0.5, draw=none] fill between [of=upZZc and loZZc];
                \addlegendimage{legend image code/.code={\fill[red!30, draw=red!50] (0cm,-0.1cm) rectangle (0.6cm,0.1cm);}}
                \addlegendentry{Mean $\pm$ 1 Std}
            \end{axis}
        \end{tikzpicture}
    \end{subfigure}
    \vspace{1em}
    \hspace*{-4mm}
    \begin{subfigure}[b]{0.49\textwidth}
        \centering
        \begin{tikzpicture}
            \begin{axis}[width=\linewidth, height=6cm, xlabel={$z$},
                ylabel={Error (\%)}, title={$\widetilde G^{(\mathrm o)}$: Prediction Error},
                grid=major, legend pos=north west,
                scaled y ticks=false, yticklabel style={/pgf/number format/fixed, /pgf/number format/precision=3}]
                \addplot [blue, thick] table [x=z, y=PctError, col sep=space] {plots/liouville_zz11_pct_g.dat};
                \addlegendentry{Mean Error}
                \addplot [forget plot, name path=upEZZo, draw=none] table [x=z, y=PctError_plus_PctStd, col sep=space] {plots/liouville_zz11_pct_g.dat};
                \addplot [forget plot, name path=loEZZo, draw=none] table [x=z, y=PctError_minus_PctStd, col sep=space] {plots/liouville_zz11_pct_g.dat};
                \addplot [forget plot, fill=blue!20, fill opacity=0.5, draw=none] fill between [of=upEZZo and loEZZo];
                \addlegendimage{legend image code/.code={\fill[blue!30, draw=blue!50] (0cm,-0.1cm) rectangle (0.6cm,0.1cm);}}
                \addlegendentry{Error $\pm$ 1 Std}
            \end{axis}
        \end{tikzpicture}
    \end{subfigure}
    \begin{subfigure}[b]{0.49\textwidth}
        \centering
        \begin{tikzpicture}
            \begin{axis}[
                ylabel style={at={(axis description cs:1.10,0.5)}, anchor=south},
                width=\linewidth, height=6cm, xlabel={$z$},
                ylabel={Error (\%)}, title={$\widetilde G^{(\mathrm c)}$: Prediction Error},
                grid=major, legend pos=north west,
                scaled y ticks=false, yticklabel style={/pgf/number format/fixed, /pgf/number format/precision=2}]
                \addplot [red, thick] table [x=z, y=PctError, col sep=space] {plots/liouville_zz11_pct_f.dat};
                \addlegendentry{Mean Error}
                \addplot [forget plot, name path=upEZZc, draw=none] table [x=z, y=PctError_plus_PctStd, col sep=space] {plots/liouville_zz11_pct_f.dat};
                \addplot [forget plot, name path=loEZZc, draw=none] table [x=z, y=PctError_minus_PctStd, col sep=space] {plots/liouville_zz11_pct_f.dat};
                \addplot [forget plot, fill=red!20, fill opacity=0.5, draw=none] fill between [of=upEZZc and loEZZc];
                \addlegendimage{legend image code/.code={\fill[red!30, draw=red!50] (0cm,-0.1cm) rectangle (0.6cm,0.1cm);}}
                \addlegendentry{Error $\pm$ 1 Std}
            \end{axis}
        \end{tikzpicture}
    \end{subfigure}
    \vspace{1em}
    \hspace*{1mm}
    \begin{subfigure}[b]{0.49\textwidth}
        \centering
        \begin{tikzpicture}
            \begin{axis}[width=\linewidth, height=6cm,
                xlabel={$\widetilde G^{(\mathrm o)}(z=0.5)$}, ylabel={Count},
                title={$\widetilde G^{(\mathrm o)}$: Distribution at $z=0.5$},
                ybar, bar width=0.000027, xmin=0.06845, xmax=0.06925,
                xtick={0.0685,0.0687,0.0689,0.0691}, ymin=0, ymajorgrids=true, xmajorgrids=false,
                ytick distance=5, legend pos=north west, scaled ticks=false,
                xticklabel style={ /pgf/number format/fixed, /pgf/number format/precision=4, font=\sansmath\sffamily\footnotesize }]
                \addplot [fill=blue!50, draw=black, opacity=0.7] table [x=BinCenter, y=Count, col sep=space] {plots/liouville_zz11_hist_g.dat};
                \addlegendentry{Predictions}
                \draw [black, dashed, thick] (axis cs:0.068847,\pgfkeysvalueof{/pgfplots/ymin}) -- (axis cs:0.068847,\pgfkeysvalueof{/pgfplots/ymax});
                \addlegendimage{legend image code/.code={\draw[black, dashed, thick] (0cm,0cm) -- (0.6cm,0cm);}}
                \addlegendentry{Exact ($0.0688$)}
                \draw [blue, thick] (axis cs:0.068862,\pgfkeysvalueof{/pgfplots/ymin}) -- (axis cs:0.068862,\pgfkeysvalueof{/pgfplots/ymax});
                \addlegendimage{legend image code/.code={\draw[blue, thick] (0cm,0cm) -- (0.6cm,0cm);}}
                \addlegendentry{Mean ($0.0689$)}
            \end{axis}
        \end{tikzpicture}
    \end{subfigure}
    \begin{subfigure}[b]{0.49\textwidth}
        \centering
        \begin{tikzpicture}
            \begin{axis}[
                ylabel style={at={(axis description cs:1.10,0.5)}, anchor=south},
                width=\linewidth, height=6cm,
                xlabel={$\widetilde G^{(\mathrm c)}(z=0.5)$}, ylabel={Count},
                title={$\widetilde G^{(\mathrm c)}$: Distribution at $z=0.5$},
                ybar, bar width=0.000035, xmin=0.06915, xmax=0.07010,
                xtick={0.0692,0.0694,0.0696,0.0698,0.0700}, ymin=0, ymajorgrids=true, xmajorgrids=false,
                ytick distance=5, legend pos=north west, scaled ticks=false,
                xticklabel style={ /pgf/number format/fixed, /pgf/number format/precision=4, font=\sansmath\sffamily\footnotesize }]
                \addplot [fill=red!50, draw=black, opacity=0.7] table [x=BinCenter, y=Count, col sep=space] {plots/liouville_zz11_hist_f.dat};
                \addlegendentry{Predictions}
                \draw [black, dashed, thick] (axis cs:0.069606,\pgfkeysvalueof{/pgfplots/ymin}) -- (axis cs:0.069606,\pgfkeysvalueof{/pgfplots/ymax});
                \addlegendimage{legend image code/.code={\draw[black, dashed, thick] (0cm,0cm) -- (0.6cm,0cm);}}
                \addlegendentry{Exact ($0.0696$)}
                \draw [red, thick] (axis cs:0.069699,\pgfkeysvalueof{/pgfplots/ymin}) -- (axis cs:0.069699,\pgfkeysvalueof{/pgfplots/ymax});
                \addlegendimage{legend image code/.code={\draw[red, thick] (0cm,0cm) -- (0.6cm,0cm);}}
                \addlegendentry{Mean ($0.0697$)}
            \end{axis}
        \end{tikzpicture}
        \hspace*{-2.9mm}
    \end{subfigure}
    \caption{NN-predicted reduced annulus correlators $\widetilde G^{(\mathrm o)}(z)$ (open, blue) and $\widetilde G^{(\mathrm c)}(z)$ (closed, red) for the Liouville $\ZZL[1,1]$--$\ZZL[1,1]$ annulus at $c=31.5$ over $100$ seeds.  At $z=0.5$, $\widetilde G^{(\mathrm o),\rm exact}=0.0688$ vs ensemble mean $0.0689$; $\widetilde G^{(\mathrm c),\rm exact}=0.0696$ vs ensemble mean $0.0697$.}
    \label{fig:liouville_zz11_zz11}
\end{figure}
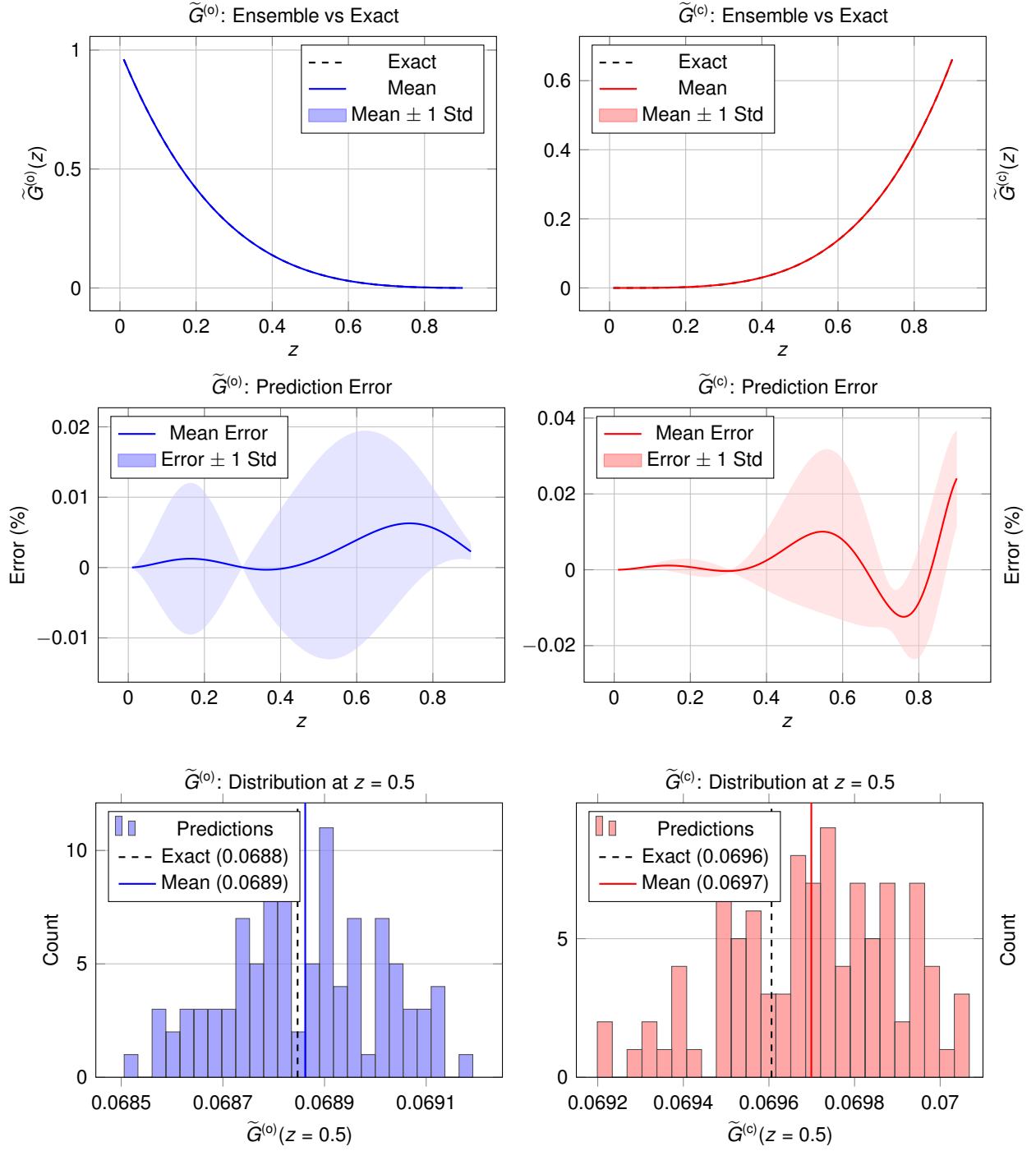

A numerical computation was again performed for illustration at $c=31.5$. The network follows the setup of Section~\ref{sec:annulus-results}, trained with the single interior anchor $z_0=0.3$ over $100$ initialisation seeds on a uniform grid of $N=150$ points $z\in[0.01,0.9]$.  Across the ensemble, the MS training loss was $(4.96\pm 2.76)\times 10^{-8}$, and at $z=0.5$ the exact reduced open correlator $\widetilde G^{(\mathrm{o}),\rm{exact}}=0.0688$ was recovered as $0.0689\pm 0.0002$, while the closed value $\widetilde G^{(\mathrm{c}),\rm{exact}}=0.0696$ was recovered as $0.0697\pm 0.0002$. The maximum relative prediction error over the ensemble, in the sense of \eqref{eq:reporting-relerr}, sits at $0.015\%$ in the open channel and $0.030\%$ in the closed channel, so the reconstruction is accurate to within a few parts in $10^{4}$ across the full $z$ range even though $c$ is well outside the rational-model regime of the earlier sections.  The corresponding six-panel diagnostic is shown in Fig.~\ref{fig:liouville_zz11_zz11}.

\begin{figure}[H]
    \centering
    \begin{subfigure}[b]{0.49\textwidth}
        \centering
        \begin{tikzpicture}
            \begin{axis}[width=\linewidth, height=6cm, xlabel={$z$},
                ylabel={$G^{(\mathrm o)}_{\rm red}(z)$},
                title={$G^{(\mathrm o)}_{\rm red}$: Ensemble vs Exact},
                grid=major, legend pos=north west,
                scaled y ticks=false, yticklabel style={/pgf/number format/fixed, /pgf/number format/precision=2}]
                \addplot [black, dashed, thick] table [x=z, y=Exact, col sep=space] {plots/liouville_zz11_fzz02_ensemble_g.dat};
                \addlegendentry{Exact}
                \addplot [blue, thick] table [x=z, y=Mean, col sep=space] {plots/liouville_zz11_fzz02_ensemble_g.dat};
                \addlegendentry{Mean}
                \addplot [forget plot, name path=upFZo, draw=none] table [x=z, y=Mean_plus_Std, col sep=space] {plots/liouville_zz11_fzz02_ensemble_g.dat};
                \addplot [forget plot, name path=loFZo, draw=none] table [x=z, y=Mean_minus_Std, col sep=space] {plots/liouville_zz11_fzz02_ensemble_g.dat};
                \addplot [forget plot, fill=blue!30, fill opacity=0.5, draw=none] fill between [of=upFZo and loFZo];
                \addlegendimage{legend image code/.code={\fill[blue!30, draw=blue!50] (0cm,-0.1cm) rectangle (0.6cm,0.1cm);}}
                \addlegendentry{Mean $\pm$ 1 Std}
            \end{axis}
        \end{tikzpicture}
    \end{subfigure}
    \hfill
    \begin{subfigure}[b]{0.49\textwidth}
        \centering
        \begin{tikzpicture}
            \begin{axis}[
                ylabel style={at={(axis description cs:1.10,0.5)}, anchor=south},
                width=\linewidth, height=6cm, xlabel={$z$},
                ylabel={$G^{(\mathrm c)}_{\rm red}(z)$},
                title={$G^{(\mathrm c)}_{\rm red}$: Ensemble vs Exact},
                grid=major, legend pos=north west,
                scaled y ticks=false, yticklabel style={/pgf/number format/fixed, /pgf/number format/precision=2}]
                \addplot [black, dashed, thick] table [x=z, y=Exact, col sep=space] {plots/liouville_zz11_fzz02_ensemble_f.dat};
                \addlegendentry{Exact}
                \addplot [red, thick] table [x=z, y=Mean, col sep=space] {plots/liouville_zz11_fzz02_ensemble_f.dat};
                \addlegendentry{Mean}
                \addplot [forget plot, name path=upFZc, draw=none] table [x=z, y=Mean_plus_Std, col sep=space] {plots/liouville_zz11_fzz02_ensemble_f.dat};
                \addplot [forget plot, name path=loFZc, draw=none] table [x=z, y=Mean_minus_Std, col sep=space] {plots/liouville_zz11_fzz02_ensemble_f.dat};
                \addplot [forget plot, fill=red!30, fill opacity=0.5, draw=none] fill between [of=upFZc and loFZc];
                \addlegendimage{legend image code/.code={\fill[red!30, draw=red!50] (0cm,-0.1cm) rectangle (0.6cm,0.1cm);}}
                \addlegendentry{Mean $\pm$ 1 Std}
            \end{axis}
        \end{tikzpicture}
    \end{subfigure}
    \vspace{1em}
    \hspace*{-3mm}
    \begin{subfigure}[b]{0.49\textwidth}
        \centering
        \begin{tikzpicture}
            \begin{axis}[width=\linewidth, height=6cm, xlabel={$z$},
                ylabel={Error (\%)}, title={$G^{(\mathrm o)}_{\rm red}$: Prediction Error},
                grid=major, legend pos=north west,
                scaled y ticks=false, yticklabel style={/pgf/number format/fixed, /pgf/number format/precision=2}]
                \addplot [blue, thick] table [x=z, y=PctError, col sep=space] {plots/liouville_zz11_fzz02_pct_g.dat};
                \addlegendentry{Mean Error}
                \addplot [forget plot, name path=upEFZo, draw=none] table [x=z, y=PctError_plus_PctStd, col sep=space] {plots/liouville_zz11_fzz02_pct_g.dat};
                \addplot [forget plot, name path=loEFZo, draw=none] table [x=z, y=PctError_minus_PctStd, col sep=space] {plots/liouville_zz11_fzz02_pct_g.dat};
                \addplot [forget plot, fill=blue!20, fill opacity=0.5, draw=none] fill between [of=upEFZo and loEFZo];
                \addlegendimage{legend image code/.code={\fill[blue!30, draw=blue!50] (0cm,-0.1cm) rectangle (0.6cm,0.1cm);}}
                \addlegendentry{Error $\pm$ 1 Std}
            \end{axis}
        \end{tikzpicture}
    \end{subfigure}
    \hspace*{1mm}
    \begin{subfigure}[b]{0.49\textwidth}
        \centering
        \begin{tikzpicture}
            \begin{axis}[
                ylabel style={at={(axis description cs:1.10,0.5)}, anchor=south},
                width=\linewidth, height=6cm, xlabel={$z$},
                ylabel={Error (\%)}, title={$G^{(\mathrm c)}_{\rm red}$: Prediction Error},
                grid=major, legend pos=north west]
                \addplot [red, thick] table [x=z, y=PctError, col sep=space] {plots/liouville_zz11_fzz02_pct_f.dat};
                \addlegendentry{Mean Error}
                \addplot [forget plot, name path=upEFZc, draw=none] table [x=z, y=PctError_plus_PctStd, col sep=space] {plots/liouville_zz11_fzz02_pct_f.dat};
                \addplot [forget plot, name path=loEFZc, draw=none] table [x=z, y=PctError_minus_PctStd, col sep=space] {plots/liouville_zz11_fzz02_pct_f.dat};
                \addplot [forget plot, fill=red!20, fill opacity=0.5, draw=none] fill between [of=upEFZc and loEFZc];
                \addlegendimage{legend image code/.code={\fill[red!30, draw=red!50] (0cm,-0.1cm) rectangle (0.6cm,0.1cm);}}
                \addlegendentry{Error $\pm$ 1 Std}
            \end{axis}
        \end{tikzpicture}
    \end{subfigure}
    \vspace{1em}
    \hspace*{1mm}
    \begin{subfigure}[b]{0.49\textwidth}
        \centering
        \begin{tikzpicture}
            \begin{axis}[width=\linewidth, height=6cm,
                xlabel={$G^{(\mathrm o)}_{\rm red}(z=0.5)$}, ylabel={Count},
                title={$G^{(\mathrm o)}_{\rm red}$: Distribution at $z=0.5$},
                ybar, bar width=0.000013, xmin=0.00016, xmax=0.00052,
                xtick={0.0002,0.0003,0.0004,0.0005}, ymin=0, ymajorgrids=true, xmajorgrids=false,
                ytick distance=5, legend pos=north east, scaled ticks=false,
                xticklabel style={ /pgf/number format/fixed, /pgf/number format/precision=4, font=\sansmath\sffamily\footnotesize }]
                \addplot [fill=blue!50, draw=black, opacity=0.7] table [x=BinCenter, y=Count, col sep=space] {plots/liouville_zz11_fzz02_hist_g.dat};
                \addlegendentry{Predictions}
                \draw [black, dashed, thick] (axis cs:0.000263,\pgfkeysvalueof{/pgfplots/ymin}) -- (axis cs:0.000263,\pgfkeysvalueof{/pgfplots/ymax});
                \addlegendimage{legend image code/.code={\draw[black, dashed, thick] (0cm,0cm) -- (0.6cm,0cm);}}
                \addlegendentry{Exact ($2.63\!\times\!10^{-4}$)}
                \draw [blue, thick] (axis cs:0.000308,\pgfkeysvalueof{/pgfplots/ymin}) -- (axis cs:0.000308,\pgfkeysvalueof{/pgfplots/ymax});
                \addlegendimage{legend image code/.code={\draw[blue, thick] (0cm,0cm) -- (0.6cm,0cm);}}
                \addlegendentry{Mean ($3.08\!\times\!10^{-4}$)}
            \end{axis}
        \end{tikzpicture}
    \end{subfigure}
    \begin{subfigure}[b]{0.49\textwidth}
        \centering
        \begin{tikzpicture}
            \begin{axis}[
                ylabel style={at={(axis description cs:1.10,0.5)}, anchor=south},
                width=\linewidth, height=6cm,
                xlabel={$G^{(\mathrm c)}_{\rm red}(z=0.5)$}, ylabel={Count},
                title={$G^{(\mathrm c)}_{\rm red}$: Distribution at $z=0.5$},
                ybar, bar width=0.000004, xmin=0.000195, xmax=0.000305,
                xtick={0.0002,0.00025,0.0003}, ymin=0, ymajorgrids=true, xmajorgrids=false,
                ytick distance=5, legend pos=north east, scaled ticks=false,
                xticklabel style={ /pgf/number format/fixed, /pgf/number format/precision=5, font=\sansmath\sffamily\footnotesize }]
                \addplot [fill=red!50, draw=black, opacity=0.7] table [x=BinCenter, y=Count, col sep=space] {plots/liouville_zz11_fzz02_hist_f.dat};
                \addlegendentry{Predictions}
                \draw [black, dashed, thick] (axis cs:0.000264,\pgfkeysvalueof{/pgfplots/ymin}) -- (axis cs:0.000264,\pgfkeysvalueof{/pgfplots/ymax});
                \addlegendimage{legend image code/.code={\draw[black, dashed, thick] (0cm,0cm) -- (0.6cm,0cm);}}
                \addlegendentry{Exact ($2.64\!\times\!10^{-4}$)}
                \draw [red, thick] (axis cs:0.000245,\pgfkeysvalueof{/pgfplots/ymin}) -- (axis cs:0.000245,\pgfkeysvalueof{/pgfplots/ymax});
                \addlegendimage{legend image code/.code={\draw[red, thick] (0cm,0cm) -- (0.6cm,0cm);}}
                \addlegendentry{Mean ($2.45\!\times\!10^{-4}$)}
            \end{axis}
        \end{tikzpicture}
        \hspace*{-2.9mm}
    \end{subfigure}
    \caption{NN-predicted reduced annulus correlators $G^{(\mathrm o)}_{\rm red}(z)$ (open, blue) and $G^{(\mathrm c)}_{\rm red}(z)$ (closed, red) for the Liouville $\ZZL[1,1]$-$\FZZ[s=0.2]$ annulus at $c=31.5$ over $100$ seeds.  At $z=0.5$, $G^{(\mathrm o),\rm exact}_{\rm red}=2.63\times 10^{-4}$ vs ensemble mean $3.08\times 10^{-4}$; $G^{(\mathrm c),\rm exact}_{\rm red}=2.64\times 10^{-4}$ vs ensemble mean $2.45\times 10^{-4}$.}
    \label{fig:liouville_zz11_fzz02}
\end{figure}
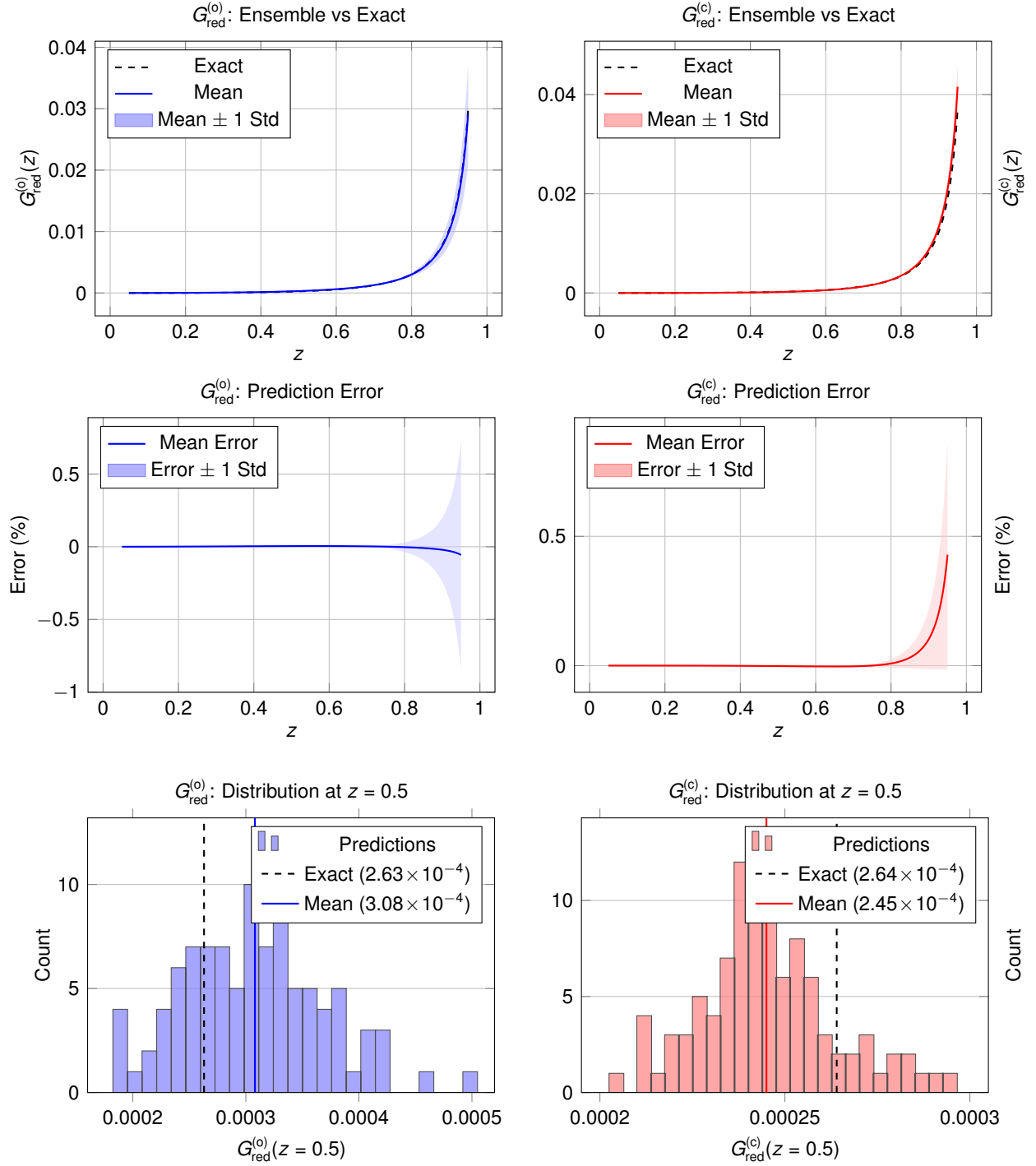

\section{Conclusion}
\label{sec:outlook}
In this work we extended the neural-network approach of Refs.~\cite{GKNS:0, GKNS:1} to torus and annulus partition functions in two-dimensional CFTs. The central observation is that both problems can be brought to the familiar language of four-point crossing, where the techniques of \cite{GKNS:0, GKNS:1} can be immediately applied. Modular $S$-invariance of the torus partition function becomes crossing symmetry of a four-point function of $\mathbb Z_2$ twist fields \cite{Hartman:2019pcd}, while the Cardy condition becomes crossing between the open- and closed-channel descriptions of a mixed four-point function of defect-changing operators \cite{Collier:2021ngi}. The input given to the network is deliberately sparse. We impose the relevant crossing equation, factor out the known endpoint behaviour, specify the leading spectral information and provide the value of the correlator at a single interior point. This information does not determine a unique crossing-symmetric function. Nevertheless, across many examples, the network selects a solution remarkably close to the exact partition function over the full interval. This provides further evidence that the spectral bias of neural networks gives an operational way to select smooth and physically relevant solutions from a much larger space of crossing-symmetric functions.

For torus partition functions, we tested the method on a broad class of compact CFTs, including A, D and E-series minimal models and WZW models. The same architecture and training prescription works across theories with rather different central charges, spectra and modular invariants. We also considered the Lee--Yang model, showing that unitarity is not necessary for the reconstruction. For annulus partition functions, we reconstructed the open and closed channels simultaneously for several Cardy boundary conditions in minimal models and WZW theories. This is an important test because the two channels have different spectra and can have very different magnitudes, while still being related by a single crossing equation.

We also studied non-compact theories. The free non-compact boson and Liouville theory provide more difficult tests because their spectra are continuous. The Liouville annulus examples are particularly challenging since the open- and closed-channel correlators can be very small and can scale differently. In such situations, achieving a small absolute crossing loss does not automatically imply an equally accurate reconstruction of both channels. The results nevertheless show that the same basic method continues to identify the expected functional behaviour with simple modification. At the same time, these examples make clear that a more carefully balanced loss function will be useful when different correlators appear at parametrically different scales.

More broadly, modular invariance and the Cardy condition are only the first examples of consistency conditions beyond ordinary four-point crossing. Higher-genus partition functions, higher-point correlators and general sewing constraints provide a much larger class of functional equations to which the same philosophy may be applied. The main lesson of this work is that one need not always search directly over spectra and OPE coefficients. It is also possible to search directly in the space of functions satisfying the relevant consistency conditions. Neural networks, through their spectral bias, provide a simple and surprisingly effective way to perform this search. Our results suggest that this viewpoint can provide a useful new direction for the modular bootstrap and, more generally, for the non-perturbative reconstruction of quantum field theory observables from sparse physical input.

In the future, it would be interesting to explore further the reconstruction of the torus partition functions beyond the diagonal $\tau=-\bar\tau$ kinematics using the concentric circle approach of \cite{GKNS:0, GKNS:1}. It is also imperative to understand better on general grounds how the smoothness of the four-point reformulation correlates with the spectral information of the partition functions and the extent of the universal success of the neural spectral bias approach in capturing physical CFT partition functions and correlators. Eventually, one would like to combine the presented technology with independent (analytical and numerical) results in CFT to perform efficient bootstrap beyond the current state-of-the-art.

\ack{Research presented in this work was initiated with and supported by an ``International Exchanges 2024 Global Round 1'' grant from the Royal Society (IES\textbackslash{}R1\textbackslash{}241082). Numerical computations in this work have been largely performed on King's College London's CREATE~\cite{CREATE} computing cluster. KG is supported by the Royal Society under grant RF\textbackslash{}ERE\textbackslash{}231142. SK is supported by the UK's Engineering and Physical Sciences Research Council under grant EP/Z535035/1, through an EPSRC Doctoral Landscape Award. AS is supported by the Royal Society under grant URF\textbackslash{}R1\textbackslash211417 and by STFC under grant ST/X000753/1.}

\appendix

\section{Additional Numerical Examples}\label{app:extra}
\subsection{Additional torus reconstructions}\label{app:torus-extra}
This appendix collects reconstructions for torus modular invariants not shown in the main text.  All runs use the anchored ansatz~\eqref{eq:line-split} and its trivial rewriting for D- and E-series and for higher WZW ranks, with the single-anchor training schedule of Section~\ref{sec:nn} at $z_0=0.3$ and $100$ initialisation seeds per model.  We first display individual three-panel reconstructions for a higher-$m$ A-series example ($\mathcal M(13,14)$), the two non-diagonal minimal-model invariants studied in the main text ($\mathcal M(15,16)_{D_9}$ and $\mathcal M(11,12)_{E_6}$), and the two WZW examples $\widehat{\mathfrak{su}}(2)_1$ and $\widehat{\mathfrak{su}}(3)_1$.  Table~\ref{tab:torus-extra} then records the ensemble mean of $\widetilde G^{\rm pred}(0.5)$, its standard deviation, the MS training loss, and MRPE for eight further modular invariants: the A-series entries $\mathcal M(5,6)$--$\mathcal M(8,9)$ interpolate between the Ising warm-up and the higher-$c$ examples of Section~\ref{sec:ADE-series}, and the D-, E-, and $\widehat{\mathfrak{su}}(4)_1$ rows illustrate that the same ansatz handles non-diagonal invariants and higher affine ranks with no structural change.

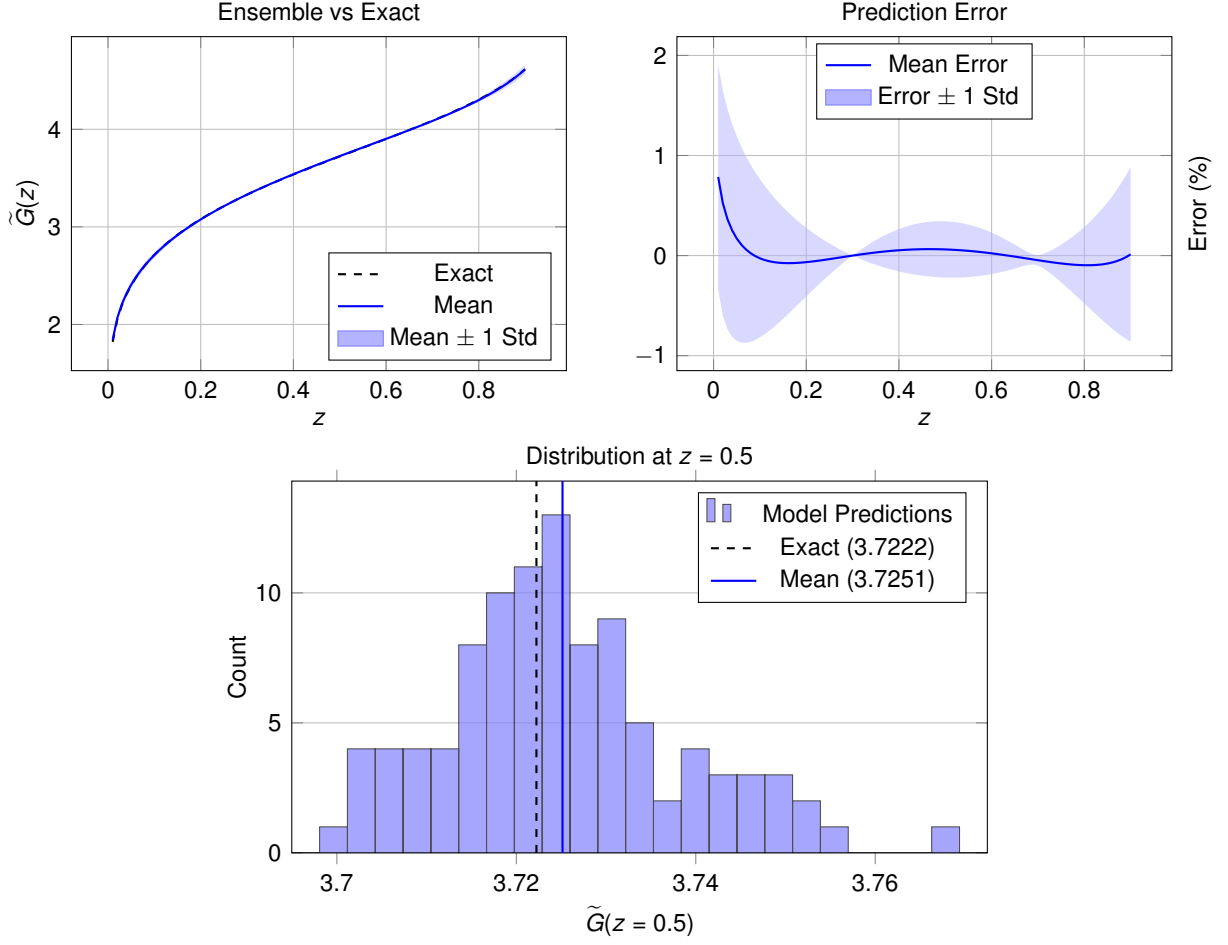
\begin{figure}[H]
    \centering
    \begin{subfigure}[b]{0.49\textwidth}
        \centering
        \begin{tikzpicture}
            \begin{axis}[
                width=\linewidth, height=6cm,
                xlabel={$z$},
                ylabel={$\widetilde{G}(z)$},
                title={Ensemble vs Exact},
                grid=major,
                legend pos=south east,
            ]
                \addplot [black, dashed, thick] table [x=z, y=Exact, col sep=space] {plots/m13_ensemble_comparison.dat};
                \addlegendentry{Exact}
                \addplot [blue, thick] table [x=z, y=Mean, col sep=space] {plots/m13_ensemble_comparison.dat};
                \addlegendentry{Mean}
                \addplot [forget plot, name path=upper, draw=none] table [x=z, y=Mean_plus_Std, col sep=space] {plots/m13_ensemble_comparison.dat};
                \addplot [forget plot, name path=lower, draw=none] table [x=z, y=Mean_minus_Std, col sep=space] {plots/m13_ensemble_comparison.dat};
                \addplot [forget plot, fill=blue!30, fill opacity=0.5, draw=none] fill between [of=upper and lower];
                \addlegendimage{legend image code/.code={\fill[blue!30, draw=blue!50] (0cm,-0.1cm) rectangle (0.6cm,0.1cm);}}
                \addlegendentry{Mean $\pm$ 1 Std}
            \end{axis}
        \end{tikzpicture}
    \end{subfigure}
    \hfill
    \begin{subfigure}[b]{0.49\textwidth}
        \centering
        \begin{tikzpicture}
            \begin{axis}[
                ylabel style={at={(axis description cs:1.10,0.5)}, anchor=south},
                width=\linewidth, height=6cm,
                xlabel={$z$},
                ylabel={Error (\%)},
                title={Prediction Error},
                grid=major,
                legend style={at={(0.5,0.98)}, anchor=north},
            ]
                \addplot [blue, thick] table [x=z, y=PctError, col sep=space] {plots/m13_percentage_error.dat};
                \addlegendentry{Mean Error}
                \addplot [forget plot, name path=upper, draw=none] table [x=z, y=PctError_plus_PctStd, col sep=space] {plots/m13_percentage_error.dat};
                \addplot [forget plot, name path=lower, draw=none] table [x=z, y=PctError_minus_PctStd, col sep=space] {plots/m13_percentage_error.dat};
                \addplot [forget plot, fill=blue!30, fill opacity=0.5, draw=none] fill between [of=upper and lower];
                \addlegendimage{legend image code/.code={\fill[blue!30, draw=blue!50] (0cm,-0.1cm) rectangle (0.6cm,0.1cm);}}
                \addlegendentry{Error $\pm$ 1 Std}
            \end{axis}
        \end{tikzpicture}
    \end{subfigure}

    \begin{subfigure}[b]{0.65\textwidth}
        \centering
        \begin{tikzpicture}
            \begin{axis}[
                width=\linewidth, height=6.5cm,
                xlabel={$\widetilde{G}(z=0.5)$},
                ylabel={Count},
                title={Distribution at $z=0.5$},
                ybar,
                bar width=0.003101,
                xmin=3.694973, xmax=3.772498,
                ymin=0,
                ymajorgrids=true,
                xmajorgrids=false,
                legend pos=north east,
                scaled ticks=false,
                xtick={3.7,3.72,3.74,3.76},
                xticklabel style={
                    /pgf/number format/fixed,
                    /pgf/number format/precision=3,
                    font=\sansmath\sffamily\footnotesize,
                }
            ]
                \addplot [fill=blue!50, draw=black, opacity=0.7] table [x=BinCenter, y=Count, col sep=space] {plots/m13_histogram_z0.500.dat};
                \addlegendentry{Model Predictions}
                \draw [black, dashed, thick] (axis cs:3.722240,\pgfkeysvalueof{/pgfplots/ymin}) -- (axis cs:3.722240,\pgfkeysvalueof{/pgfplots/ymax});
                \addlegendimage{legend image code/.code={\draw[black, dashed, thick] (0cm,0cm) -- (0.6cm,0cm);}}
                \addlegendentry{Exact ($3.7222$)}
                \draw [blue, thick] (axis cs:3.725148,\pgfkeysvalueof{/pgfplots/ymin}) -- (axis cs:3.725148,\pgfkeysvalueof{/pgfplots/ymax});
                \addlegendimage{legend image code/.code={\draw[blue, thick] (0cm,0cm) -- (0.6cm,0cm);}}
                \addlegendentry{Mean ($3.7251$)}
            \end{axis}
        \end{tikzpicture}
    \end{subfigure}

    \caption{NN-predicted reduced correlator $\widetilde G(z)$ for the $\mathcal M(13,14)$ minimal model ($c=88/91$, $\Delta_{\rm gap}=3/364$).  The NN prediction at $z=0.5$ is $\widetilde G^{\rm pred}(0.5)=3.72515\pm 0.01330$ against $\widetilde G^{\rm exact}(0.5)=3.72224$.}
    \label{fig:m13_summary}
\end{figure}

\paragraph{$\mathcal M(13,14)$: $c=88/91$, $\Delta_{\rm gap}=3/364$.}
Over $100$ seeds, the MS training loss is $(1.17\pm 2.32)\times 10^{-7}$, and the NN prediction at $z=0.5$ reads $\widetilde G^{\rm pred}(0.5)=3.72515\pm 0.0133$ against $\widetilde G^{\rm exact}(0.5)=3.72224$.  Results are shown in Fig.~\ref{fig:m13_summary}.

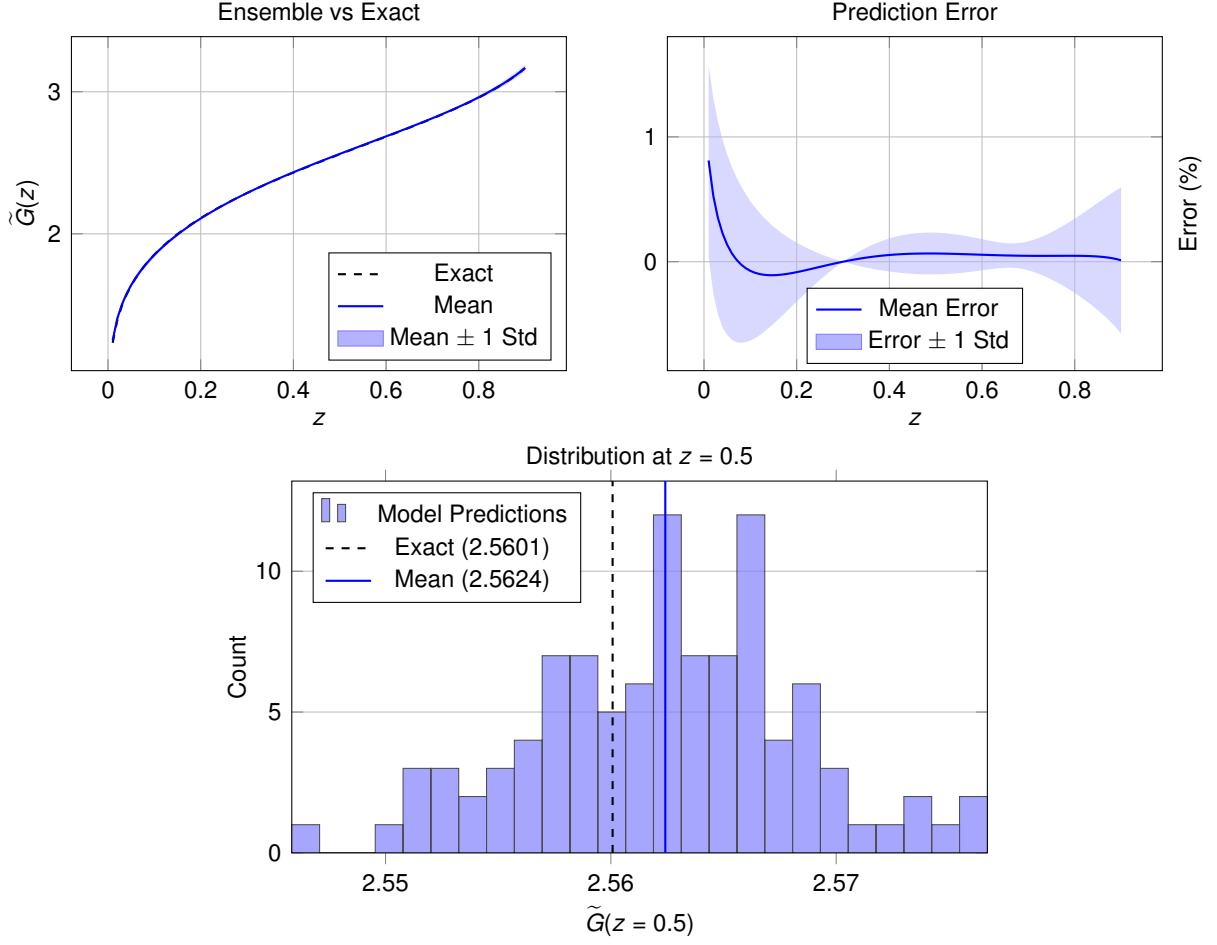
\begin{figure}[H]
    \centering
    \begin{subfigure}[b]{0.49\textwidth}
        \centering
        \begin{tikzpicture}
            \begin{axis}[
                width=\linewidth, height=6cm,
                xlabel={$z$},
                ylabel={$\widetilde{G}(z)$},
                title={Ensemble vs Exact},
                grid=major,
                legend pos=south east,
            ]
                \addplot [black, dashed, thick] table [x=z, y=Exact, col sep=space] {plots/m15_d9_ensemble_comparison.dat};
                \addlegendentry{Exact}
                \addplot [blue, thick] table [x=z, y=Mean, col sep=space] {plots/m15_d9_ensemble_comparison.dat};
                \addlegendentry{Mean}
                \addplot [forget plot, name path=upper, draw=none] table [x=z, y=Mean_plus_Std, col sep=space] {plots/m15_d9_ensemble_comparison.dat};
                \addplot [forget plot, name path=lower, draw=none] table [x=z, y=Mean_minus_Std, col sep=space] {plots/m15_d9_ensemble_comparison.dat};
                \addplot [forget plot, fill=blue!30, fill opacity=0.5, draw=none] fill between [of=upper and lower];
                \addlegendimage{legend image code/.code={\fill[blue!30, draw=blue!50] (0cm,-0.1cm) rectangle (0.6cm,0.1cm);}}
                \addlegendentry{Mean $\pm$ 1 Std}
            \end{axis}
        \end{tikzpicture}
    \end{subfigure}
    \hfill
    \begin{subfigure}[b]{0.49\textwidth}
        \centering
        \begin{tikzpicture}
            \begin{axis}[
                ylabel style={at={(axis description cs:1.10,0.5)}, anchor=south},
                width=\linewidth, height=6cm,
                xlabel={$z$},
                ylabel={Error (\%)},
                title={Prediction Error},
                grid=major,
                legend style={at={(0.5,0.02)}, anchor=south},
            ]
                \addplot [blue, thick] table [x=z, y=PctError, col sep=space] {plots/m15_d9_percentage_error.dat};
                \addlegendentry{Mean Error}
                \addplot [forget plot, name path=upper, draw=none] table [x=z, y=PctError_plus_PctStd, col sep=space] {plots/m15_d9_percentage_error.dat};
                \addplot [forget plot, name path=lower, draw=none] table [x=z, y=PctError_minus_PctStd, col sep=space] {plots/m15_d9_percentage_error.dat};
                \addplot [forget plot, fill=blue!30, fill opacity=0.5, draw=none] fill between [of=upper and lower];
                \addlegendimage{legend image code/.code={\fill[blue!30, draw=blue!50] (0cm,-0.1cm) rectangle (0.6cm,0.1cm);}}
                \addlegendentry{Error $\pm$ 1 Std}
            \end{axis}
        \end{tikzpicture}
    \end{subfigure}

    \begin{subfigure}[b]{0.65\textwidth}
        \centering
        \begin{tikzpicture}
            \begin{axis}[
                width=\linewidth, height=6.5cm,
                xlabel={$\widetilde{G}(z=0.5)$},
                ylabel={Count},
                title={Distribution at $z=0.5$},
                ybar,
                bar width=0.001235,
                xmin=2.545838, xmax=2.576710,
                ymin=0,
                ymajorgrids=true,
                xmajorgrids=false,
                legend pos=north west,
                scaled ticks=false,
                xtick={2.55,2.56,2.57},
                xticklabel style={
                    /pgf/number format/fixed,
                    /pgf/number format/precision=3,
                    font=\sansmath\sffamily\footnotesize,
                }
            ]
                \addplot [fill=blue!50, draw=black, opacity=0.7] table [x=BinCenter, y=Count, col sep=space] {plots/m15_d9_histogram_z0.500.dat};
                \addlegendentry{Model Predictions}
                \draw [black, dashed, thick] (axis cs:2.560077,\pgfkeysvalueof{/pgfplots/ymin}) -- (axis cs:2.560077,\pgfkeysvalueof{/pgfplots/ymax});
                \addlegendimage{legend image code/.code={\draw[black, dashed, thick] (0cm,0cm) -- (0.6cm,0cm);}}
                \addlegendentry{Exact ($2.5601$)}
                \draw [blue, thick] (axis cs:2.562416,\pgfkeysvalueof{/pgfplots/ymin}) -- (axis cs:2.562416,\pgfkeysvalueof{/pgfplots/ymax});
                \addlegendimage{legend image code/.code={\draw[blue, thick] (0cm,0cm) -- (0.6cm,0cm);}}
                \addlegendentry{Mean ($2.5624$)}
            \end{axis}
        \end{tikzpicture}
    \end{subfigure}

    \caption{NN-predicted reduced correlator $\widetilde G(z)$ for the $(A_{14},D_9)$ D-series modular invariant of $\mathcal M(15,16)$ ($c=39/40$, $\Delta_{\rm gap}^{D}=1/60$).  The NN prediction at $z=0.5$ is $\widetilde G^{\rm pred}(0.5)=2.56242\pm 0.00595$ against $\widetilde G^{\rm exact}(0.5)=2.56008$.}
    \label{fig:m15_d9_summary}
\end{figure}

\paragraph{$\mathcal M(15,16)_{D_9}$: $c=39/40$, $\Delta_{\rm gap}^{D}=1/60$.}
The corresponding $D$-series invariant at $m=15$ is $(A_{14},D_9)$ with lightest non-vacuum primary at $\Delta_{\rm gap}^{D}=1/60$.  Over $100$ seeds, the MS training loss was $(1.80\pm 1.01)\times 10^{-6}$, and the NN prediction at $z=0.5$ reads $\widetilde G^{\rm pred}(0.5)=2.56242\pm 0.00595$ against $\widetilde G^{\rm exact}(0.5)=2.56008$.  Results are shown in Fig.~\ref{fig:m15_d9_summary}.

\begin{figure}[H]
    \centering
    \begin{subfigure}[b]{0.49\textwidth}
        \centering
        \begin{tikzpicture}
            \begin{axis}[
                width=\linewidth, height=6cm,
                xlabel={$z$},
                ylabel={$\widetilde{G}(z)$},
                title={Ensemble vs Exact},
                grid=major,
                legend pos=north west,
            ]
                \addplot [black, dashed, thick] table [x=z, y=Exact, col sep=space] {plots/m11_e6_ensemble_comparison.dat};
                \addlegendentry{Exact}
                \addplot [blue, thick] table [x=z, y=Mean, col sep=space] {plots/m11_e6_ensemble_comparison.dat};
                \addlegendentry{Mean}
                \addplot [forget plot, name path=upper, draw=none] table [x=z, y=Mean_plus_Std, col sep=space] {plots/m11_e6_ensemble_comparison.dat};
                \addplot [forget plot, name path=lower, draw=none] table [x=z, y=Mean_minus_Std, col sep=space] {plots/m11_e6_ensemble_comparison.dat};
                \addplot [forget plot, fill=blue!30, fill opacity=0.5, draw=none] fill between [of=upper and lower];
                \addlegendimage{legend image code/.code={\fill[blue!30, draw=blue!50] (0cm,-0.1cm) rectangle (0.6cm,0.1cm);}}
                \addlegendentry{Mean $\pm$ 1 Std}
            \end{axis}
        \end{tikzpicture}
    \end{subfigure}
    \hfill
    \begin{subfigure}[b]{0.49\textwidth}
        \centering
        \begin{tikzpicture}
            \begin{axis}[
                ylabel style={at={(axis description cs:1.10,0.5)}, anchor=south},
                width=\linewidth, height=6cm,
                xlabel={$z$},
                ylabel={Error (\%)},
                title={Prediction Error},
                grid=major,
                legend style={at={(0.5,0.02)}, anchor=south},
            ]
                \addplot [blue, thick] table [x=z, y=PctError, col sep=space] {plots/m11_e6_percentage_error.dat};
                \addlegendentry{Mean Error}
                \addplot [forget plot, name path=upper, draw=none] table [x=z, y=PctError_plus_PctStd, col sep=space] {plots/m11_e6_percentage_error.dat};
                \addplot [forget plot, name path=lower, draw=none] table [x=z, y=PctError_minus_PctStd, col sep=space] {plots/m11_e6_percentage_error.dat};
                \addplot [forget plot, fill=blue!30, fill opacity=0.5, draw=none] fill between [of=upper and lower];
                \addlegendimage{legend image code/.code={\fill[blue!30, draw=blue!50] (0cm,-0.1cm) rectangle (0.6cm,0.1cm);}}
                \addlegendentry{Error $\pm$ 1 Std}
            \end{axis}
        \end{tikzpicture}
    \end{subfigure}

    \begin{subfigure}[b]{0.65\textwidth}
        \centering
        \begin{tikzpicture}
            \begin{axis}[
                width=\linewidth, height=6.5cm,
                xlabel={$\widetilde{G}(z=0.5)$},
                ylabel={Count},
                title={Distribution at $z=0.5$},
                ybar,
                bar width=0.000813,
                xmin=1.672422, xmax=1.692740,
                ymin=0,
                ymajorgrids=true,
                xmajorgrids=false,
                legend pos=north east,
                scaled ticks=false,
                xtick={1.675,1.68,1.685,1.69},
                xticklabel style={
                    /pgf/number format/fixed,
                    /pgf/number format/precision=3,
                    font=\sansmath\sffamily\footnotesize,
                }
            ]
                \addplot [fill=blue!50, draw=black, opacity=0.7] table [x=BinCenter, y=Count, col sep=space] {plots/m11_e6_histogram_z0.500.dat};
                \addlegendentry{Model Predictions}
                \draw [black, dashed, thick] (axis cs:1.679586,\pgfkeysvalueof{/pgfplots/ymin}) -- (axis cs:1.679586,\pgfkeysvalueof{/pgfplots/ymax});
                \addlegendimage{legend image code/.code={\draw[black, dashed, thick] (0cm,0cm) -- (0.6cm,0cm);}}
                \addlegendentry{Exact ($1.6796$)}
                \draw [blue, thick] (axis cs:1.680983,\pgfkeysvalueof{/pgfplots/ymin}) -- (axis cs:1.680983,\pgfkeysvalueof{/pgfplots/ymax});
                \addlegendimage{legend image code/.code={\draw[blue, thick] (0cm,0cm) -- (0.6cm,0cm);}}
                \addlegendentry{Mean ($1.6810$)}
            \end{axis}
        \end{tikzpicture}
    \end{subfigure}

    \caption{NN-predicted reduced correlator $\widetilde G(z)$ for the $(A_{10},E_6)$ E-series modular invariant of $\mathcal M(11,12)$ ($c=21/22$, $\Delta_{\rm gap}^{E}=5/88$).  The NN prediction at $z=0.5$ is $\widetilde G^{\rm pred}(0.5)=1.68098\pm 0.00360$ against $\widetilde G^{\rm exact}(0.5)=1.67959$.}
    \label{fig:m11_e6_summary}
\end{figure}
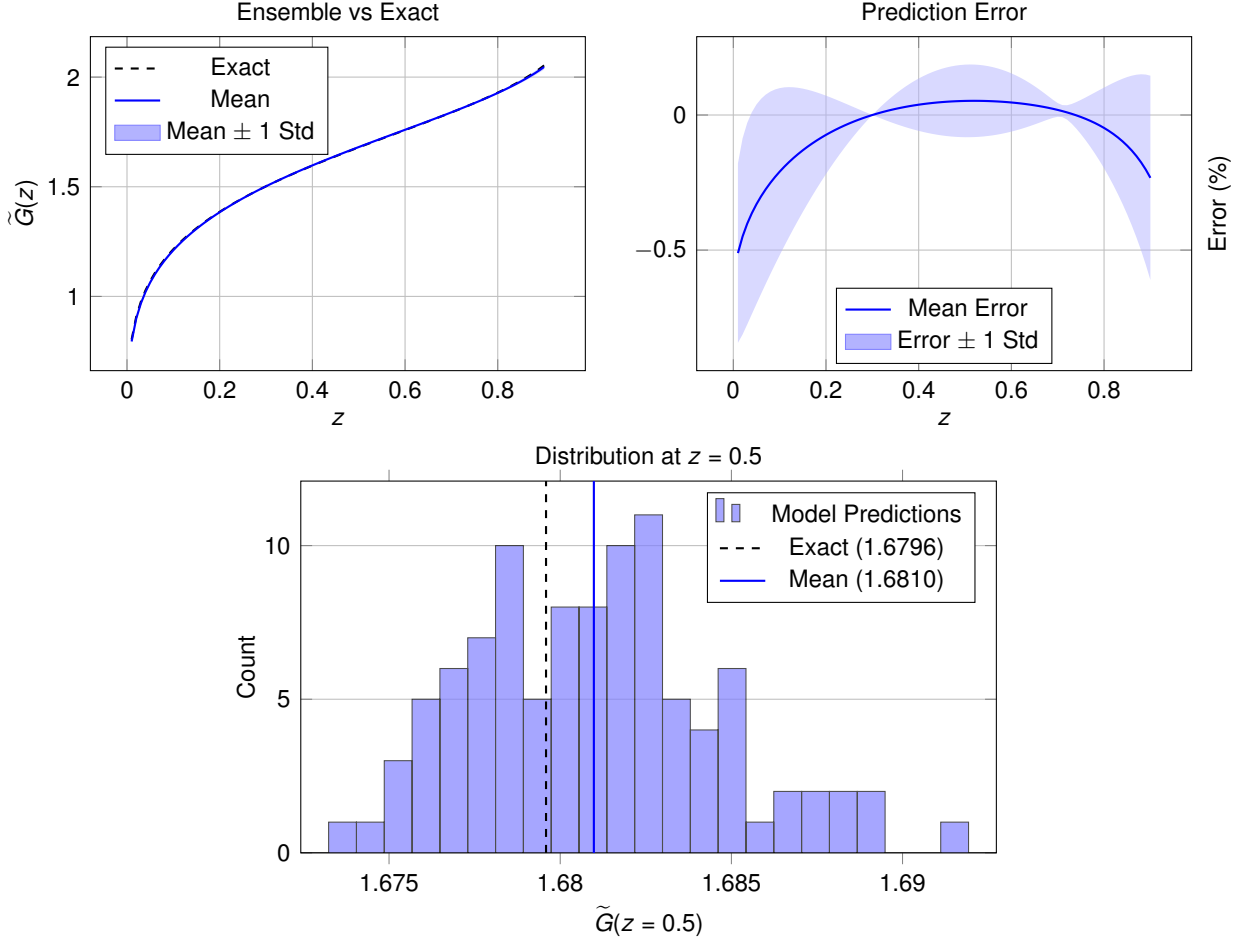

\paragraph{$\mathcal M(11,12)_{E_6}$: $c=21/22$, $\Delta_{\rm gap}^{E}=5/88$.}
The smallest exceptional invariant is $(A_{10},E_6)$ with lightest non-vacuum primary at $\Delta_{\rm gap}^{E}=5/88$. Over $100$ seeds, the MS training loss is $(1.57\pm 1.12)\times 10^{-7}$, and the NN prediction at $z=0.5$ reads $\widetilde G^{\rm pred}(0.5)=1.68098\pm 0.00360$ against $\widetilde G^{\rm exact}(0.5)=1.67959$.  Results are shown in Fig.~\ref{fig:m11_e6_summary}.

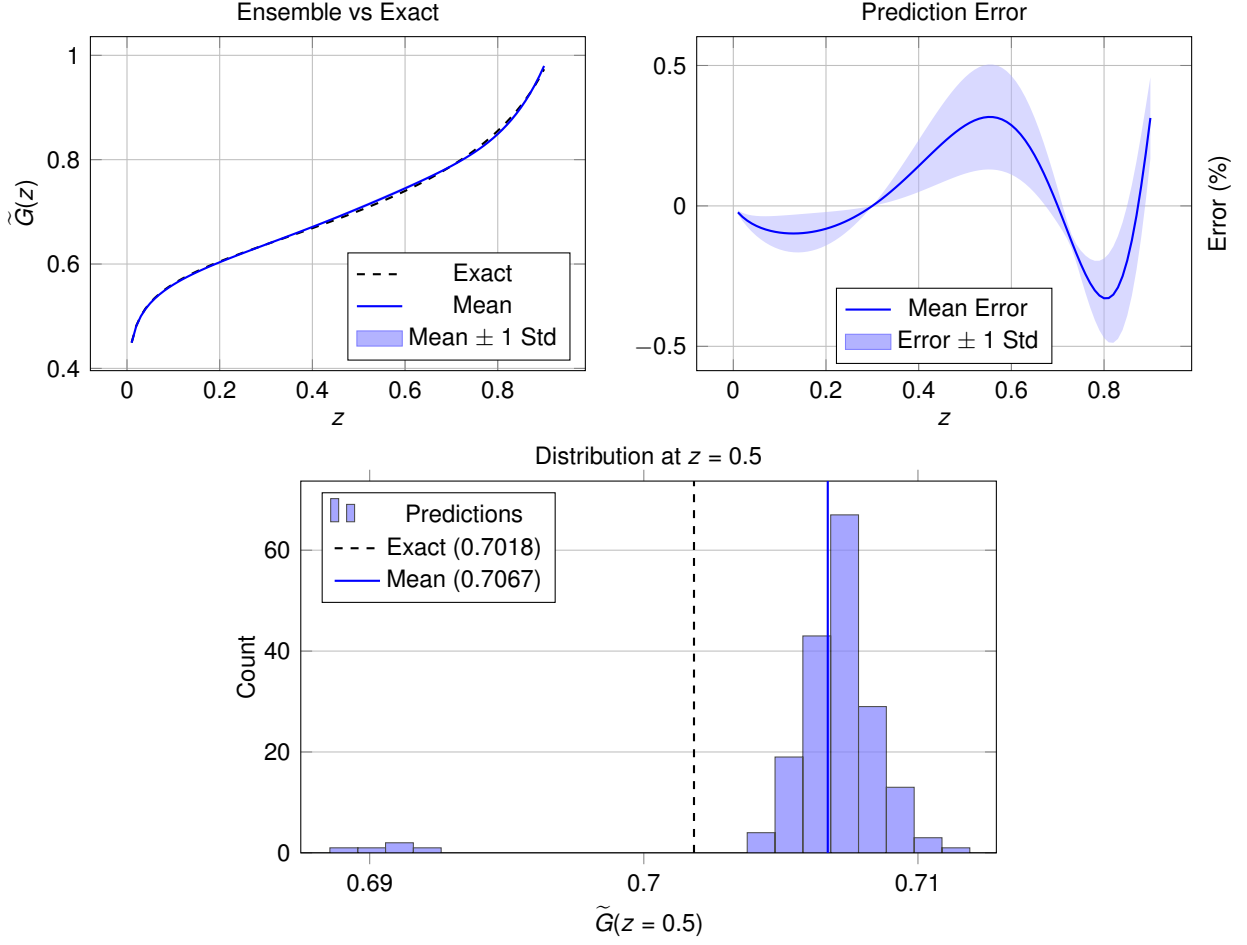
\begin{figure}[H]
    \centering
    \begin{subfigure}[b]{0.49\textwidth}
        \centering
        \begin{tikzpicture}
            \begin{axis}[
                width=\linewidth, height=6cm,
                xlabel={$z$},
                ylabel={$\widetilde{G}(z)$},
                title={Ensemble vs Exact},
                grid=major,
                legend pos=south east,
            ]
                \addplot [black, dashed, thick] table [x=z, y=Exact, col sep=space] {plots/su2_k1_ensemble_comparison.dat};
                \addlegendentry{Exact}
                \addplot [blue, thick] table [x=z, y=Mean, col sep=space] {plots/su2_k1_ensemble_comparison.dat};
                \addlegendentry{Mean}
                \addplot [forget plot, name path=upper, draw=none] table [x=z, y=Mean_plus_Std, col sep=space] {plots/su2_k1_ensemble_comparison.dat};
                \addplot [forget plot, name path=lower, draw=none] table [x=z, y=Mean_minus_Std, col sep=space] {plots/su2_k1_ensemble_comparison.dat};
                \addplot [forget plot, fill=blue!30, fill opacity=0.5, draw=none] fill between [of=upper and lower];
                \addlegendimage{legend image code/.code={\fill[blue!30, draw=blue!50] (0cm,-0.1cm) rectangle (0.6cm,0.1cm);}}
                \addlegendentry{Mean $\pm$ 1 Std}
            \end{axis}
        \end{tikzpicture}
    \end{subfigure}
    \hfill
    \begin{subfigure}[b]{0.49\textwidth}
        \centering
        \begin{tikzpicture}
            \begin{axis}[
                ylabel style={at={(axis description cs:1.10,0.5)}, anchor=south},
                width=\linewidth, height=6cm,
                xlabel={$z$},
                ylabel={Error (\%)},
                title={Prediction Error},
                grid=major,
                legend style={at={(0.5,0.02)}, anchor=south},
            ]
                \addplot [blue, thick] table [x=z, y=PctError, col sep=space] {plots/su2_k1_percentage_error.dat};
                \addlegendentry{Mean Error}
                \addplot [forget plot, name path=upper, draw=none] table [x=z, y=PctError_plus_PctStd, col sep=space] {plots/su2_k1_percentage_error.dat};
                \addplot [forget plot, name path=lower, draw=none] table [x=z, y=PctError_minus_PctStd, col sep=space] {plots/su2_k1_percentage_error.dat};
                \addplot [forget plot, fill=blue!30, fill opacity=0.5, draw=none] fill between [of=upper and lower];
                \addlegendimage{legend image code/.code={\fill[blue!30, draw=blue!50] (0cm,-0.1cm) rectangle (0.6cm,0.1cm);}}
                \addlegendentry{Error $\pm$ 1 Std}
            \end{axis}
        \end{tikzpicture}
    \end{subfigure}

    \begin{subfigure}[b]{0.65\textwidth}
        \centering
        \begin{tikzpicture}
            \begin{axis}[
                width=\linewidth, height=6.5cm,
                xlabel={$\widetilde{G}(z=0.5)$},
                ylabel={Count},
                title={Distribution at $z=0.5$},
                ybar,
                bar width=0.001015,
                xmin=0.687486, xmax=0.712860,
                xtick={0.69,0.7,0.71},
                ymin=0,
                ymajorgrids=true,
                xmajorgrids=false,
                legend pos=north west,
                scaled ticks=false,
                xticklabel style={
                    /pgf/number format/fixed,
                    /pgf/number format/precision=4,
                    font=\sansmath\sffamily\footnotesize,
                }
            ]
                \addplot [fill=blue!50, draw=black, opacity=0.7] table [x=BinCenter, y=Count, col sep=space] {plots/su2_k1_histogram_z0.500.dat};
                \addlegendentry{Predictions}
                \draw [black, dashed, thick] (axis cs:0.70183,\pgfkeysvalueof{/pgfplots/ymin}) -- (axis cs:0.70183,\pgfkeysvalueof{/pgfplots/ymax});
                \addlegendimage{legend image code/.code={\draw[black, dashed, thick] (0cm,0cm) -- (0.6cm,0cm);}}
                \addlegendentry{Exact ($0.7018$)}
                \draw [blue, thick] (axis cs:0.70671,\pgfkeysvalueof{/pgfplots/ymin}) -- (axis cs:0.70671,\pgfkeysvalueof{/pgfplots/ymax});
                \addlegendimage{legend image code/.code={\draw[blue, thick] (0cm,0cm) -- (0.6cm,0cm);}}
                \addlegendentry{Mean ($0.7067$)}
            \end{axis}
        \end{tikzpicture}
    \end{subfigure}

    \caption{NN-predicted reduced correlator $\widetilde G(z)$ for the level-one $\widehat{\mathfrak{su}}(2)_1$ WZW model, on the filtered ensemble ($184/1000$ seeds after the early-stopping cut).  The NN prediction at $z=0.5$ is $\widetilde G^{\rm pred}(0.5)=0.70671\pm 0.00292$ against $\widetilde G^{\rm exact}(0.5)=0.70183$.}
    \label{fig:su2_k1_summary}
\end{figure}

\paragraph{$\widehat{\mathfrak{su}}(2)_1$: $c=1$, $\Delta_{\rm gap}=1/2$.}
At level one the spectrum of affine primaries contains only the vacuum and the $\ell=1$ affine primary, with $h_1=\bar h_1=1/4$ and hence $\Delta_{\rm gap}=1/2$. The unfiltered loss landscape is bimodal, so we train $1000$ seeds and keep the $184$ that satisfy the early-stopping criterion.  On this filtered ensemble, the MS training loss is $(4.89\pm 2.42)\times 10^{-6}$, and $\widetilde G^{\rm pred}(0.5)=0.70671\pm 0.00292$, compared with $\widetilde G^{\rm exact}(0.5)=0.70183$; see Fig.~\ref{fig:su2_k1_summary}.

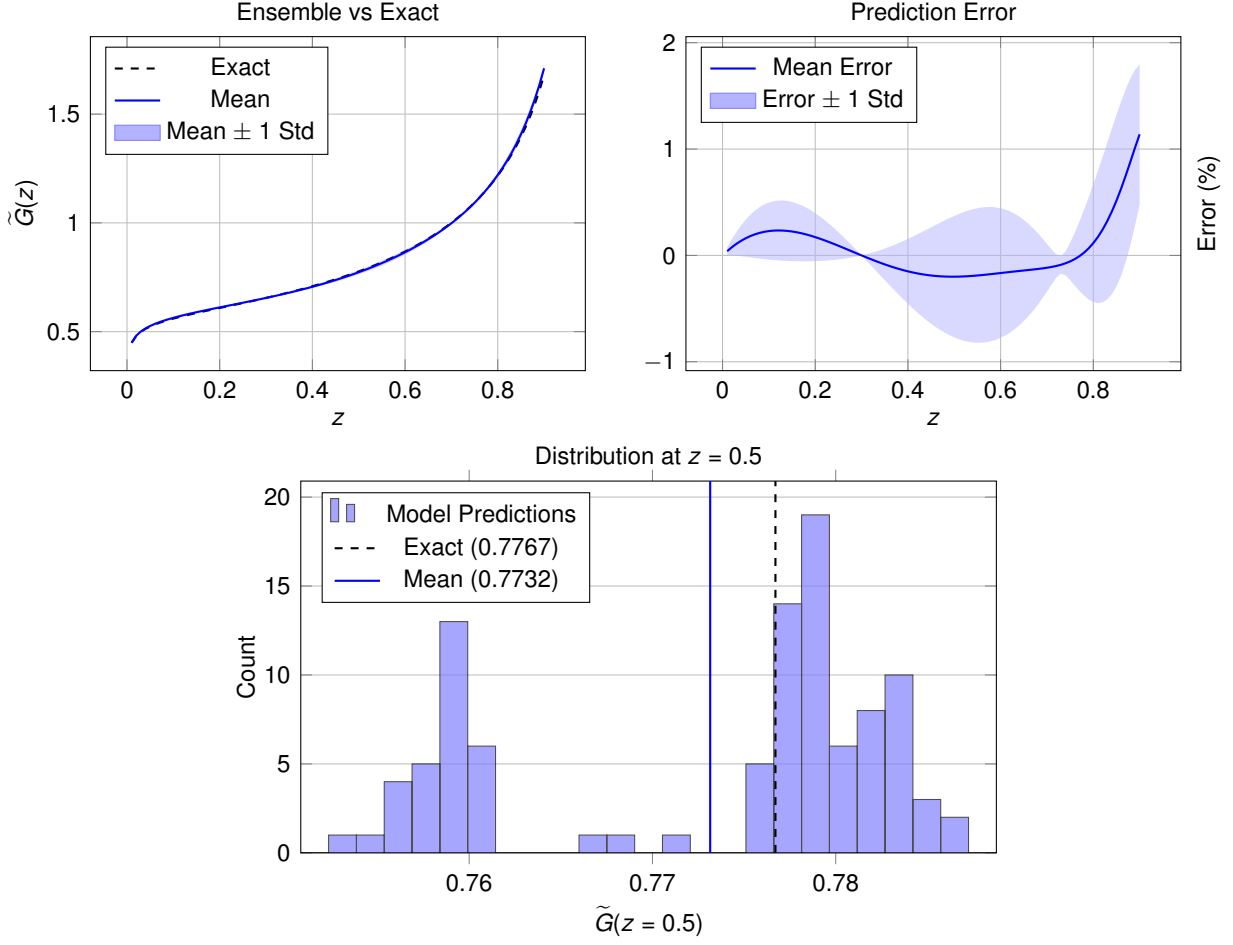
\begin{figure}[H]
    \centering
    \begin{subfigure}[b]{0.49\textwidth}
        \centering
        \begin{tikzpicture}
            \begin{axis}[
                width=\linewidth, height=6cm,
                xlabel={$z$},
                ylabel={$\widetilde{G}(z)$},
                title={Ensemble vs Exact},
                grid=major,
                legend pos=north west,
            ]
                \addplot [black, dashed, thick] table [x=z, y=Exact, col sep=space] {plots/su3_k1_ensemble_comparison.dat};
                \addlegendentry{Exact}
                \addplot [blue, thick] table [x=z, y=Mean, col sep=space] {plots/su3_k1_ensemble_comparison.dat};
                \addlegendentry{Mean}
                \addplot [forget plot, name path=upper, draw=none] table [x=z, y=Mean_plus_Std, col sep=space] {plots/su3_k1_ensemble_comparison.dat};
                \addplot [forget plot, name path=lower, draw=none] table [x=z, y=Mean_minus_Std, col sep=space] {plots/su3_k1_ensemble_comparison.dat};
                \addplot [forget plot, fill=blue!30, fill opacity=0.5, draw=none] fill between [of=upper and lower];
                \addlegendimage{legend image code/.code={\fill[blue!30, draw=blue!50] (0cm,-0.1cm) rectangle (0.6cm,0.1cm);}}
                \addlegendentry{Mean $\pm$ 1 Std}
            \end{axis}
        \end{tikzpicture}
    \end{subfigure}
    \hfill
    \begin{subfigure}[b]{0.49\textwidth}
        \centering
        \begin{tikzpicture}
            \begin{axis}[
                ylabel style={at={(axis description cs:1.10,0.5)}, anchor=south},
                width=\linewidth, height=6cm,
                xlabel={$z$},
                ylabel={Error (\%)},
                title={Prediction Error},
                grid=major,
                legend pos=north west,
            ]
                \addplot [blue, thick] table [x=z, y=PctError, col sep=space] {plots/su3_k1_percentage_error.dat};
                \addlegendentry{Mean Error}
                \addplot [forget plot, name path=upper, draw=none] table [x=z, y=PctError_plus_PctStd, col sep=space] {plots/su3_k1_percentage_error.dat};
                \addplot [forget plot, name path=lower, draw=none] table [x=z, y=PctError_minus_PctStd, col sep=space] {plots/su3_k1_percentage_error.dat};
                \addplot [forget plot, fill=blue!30, fill opacity=0.5, draw=none] fill between [of=upper and lower];
                \addlegendimage{legend image code/.code={\fill[blue!30, draw=blue!50] (0cm,-0.1cm) rectangle (0.6cm,0.1cm);}}
                \addlegendentry{Error $\pm$ 1 Std}
            \end{axis}
        \end{tikzpicture}
    \end{subfigure}

    \begin{subfigure}[b]{0.65\textwidth}
        \centering
        \begin{tikzpicture}
            \begin{axis}[
                width=\linewidth, height=6.5cm,
                xlabel={$\widetilde{G}(z=0.5)$},
                ylabel={Count},
                title={Distribution at $z=0.5$},
                ybar,
                bar width=0.001518,
                xmin=0.750812, xmax=0.788764,
                ymin=0,
                ymajorgrids=true,
                xmajorgrids=false,
                legend pos=north west,
                scaled ticks=false,
                xtick={0.76,0.77,0.78},
                xticklabel style={
                    /pgf/number format/fixed,
                    /pgf/number format/precision=3,
                    font=\sansmath\sffamily\footnotesize,
                }
            ]
                \addplot [fill=blue!50, draw=black, opacity=0.7] table [x=BinCenter, y=Count, col sep=space] {plots/su3_k1_histogram_z0.500.dat};
                \addlegendentry{Model Predictions}
                \draw [black, dashed, thick] (axis cs:0.77671,\pgfkeysvalueof{/pgfplots/ymin}) -- (axis cs:0.77671,\pgfkeysvalueof{/pgfplots/ymax});
                \addlegendimage{legend image code/.code={\draw[black, dashed, thick] (0cm,0cm) -- (0.6cm,0cm);}}
                \addlegendentry{Exact ($0.7767$)}
                \draw [blue, thick] (axis cs:0.77315,\pgfkeysvalueof{/pgfplots/ymin}) -- (axis cs:0.77315,\pgfkeysvalueof{/pgfplots/ymax});
                \addlegendimage{legend image code/.code={\draw[blue, thick] (0cm,0cm) -- (0.6cm,0cm);}}
                \addlegendentry{Mean ($0.7732$)}
            \end{axis}
        \end{tikzpicture}
    \end{subfigure}

    \caption{NN-predicted reduced correlator $\widetilde G(z)$ for the level-one $\widehat{\mathfrak{su}}(3)_1$ WZW model ($c=2$, $\Delta_{\rm gap}=2/3$).  The NN prediction at $z=0.5$ is $\widetilde G^{\rm pred}(0.5)=0.77315\pm 0.01010$ against $\widetilde G^{\rm exact}(0.5)=0.77671$.}
    \label{fig:su3_k1_summary}
\end{figure}

\paragraph{$\widehat{\mathfrak{su}}(3)_1$: $c=2$, $\Delta_{\rm gap}=2/3$.}
For $\widehat{\mathfrak{su}}(3)_1$ there are three affine primaries: the vacuum and the two fundamentals $\mathbf 3,\bar{\mathbf 3}$, both with $h=1/3$.  Therefore, $\Delta_{\rm gap}=2/3$.  Over $100$ seeds, the MS training loss is $(1.09\pm 0.928)\times 10^{-5}$.  Across the full ensemble, $\widetilde G^{\rm pred}(0.5)=0.77315\pm 0.0101$, compared with $\widetilde G^{\rm exact}(0.5)=0.77671$; see Fig.~\ref{fig:su3_k1_summary}.

\begin{table}[H]
\centering
\setlength{\tabcolsep}{5pt}
\begin{tabular}{lccccc}
\hline
Model & $c$ & $\widetilde G^{\rm exact}(0.5)$ & $\widetilde G^{\rm pred}(0.5)$ & $\overline{\mathcal L}$ & MRPE (\%) \\
\hline

A-series $\mathcal M(5,6)$ & 4/5 & $1.4114$ & $1.4130\pm 0.00271$ & $2.85\times 10^{-9}$ & $1.003\pm 0.394$ \\

A-series $\mathcal M(6,7)$ & 6/7 & $1.6931$ & $1.6937\pm 0.00355$ & $6.74\times 10^{-9}$ & $0.576\pm 0.349$ \\

A-series $\mathcal M(7,8)$ & 25/28 & $1.9781$ & $1.9781\pm 0.00457$ & $1.43\times 10^{-8}$ & $0.550\pm 0.420$ \\

A-series $\mathcal M(8,9)$ & 11/12 & $2.2655$ & $2.2651\pm 0.00575$ & $2.58\times 10^{-8}$ & $0.950\pm 0.564$ \\

E-series $(E_6,A_{12})$ of $\mathcal M(12,13)$ & 25/26 & $1.8508$ & $1.8530\pm 0.00420$ & $9.88\times 10^{-8}$ & $0.857\pm 0.484$ \\

D-series $(A_{12},D_8)$ of $\mathcal M(13,14)$ & 88/91 & $2.2651$ & $2.2660\pm 0.00501$ & $1.65\times 10^{-6}$ & $1.239\pm 0.713$ \\

A-series $\mathcal M(15,16)$ & 39/40 & $4.3098$ & $4.3160\pm 0.0172$ & $1.47\times 10^{-7}$ & $1.658\pm 0.914$ \\

WZW $\widehat{\mathfrak{su}}(4)_1$ & 3 & $0.8472$ & $0.8426\pm 0.00988$ & $9.03\times 10^{-6}$ & $1.363\pm 0.586$ \\

\hline
\end{tabular}
\caption{Torus reconstructions for eight additional rational modular invariants beyond those of Section~\ref{sec:minimal} and the individual figures of this appendix, at single anchor $z_0=0.3$ over $100$ seeds each.  Ensemble means at $z=0.5$ track the exact values to sub-$1\%$ in every case, and the MS training loss sits between $10^{-9}$ and $10^{-6}$.  MRPE denotes the ensemble mean of the per-seed maximum of the relative prediction error~\eqref{eq:reporting-relerr}, expressed as a percentage.}
\label{tab:torus-extra}
\end{table}

\subsection{Additional annulus reconstructions}\label{app:annulus-extra}
For completeness we display the two Ising Cardy pairs referenced from Table~\ref{tab:ising-annulus} but not shown in the main text — $(\boldsymbol\sigma,\boldsymbol\sigma)$ and $(\mathbf 1,\boldsymbol\varepsilon)$ — obtained with the ansatz~\eqref{eq:annulus-ansatz} and the single anchor $z_0=0.3$ over $100$ seeds.  Each figure follows the reporting scheme of Section~\ref{sec:nn} duplicated into six panels: rows show the ensemble mean of $\widetilde G^{(\mathrm o,c)}(z)$ vs.\ the exact correlator (top), the per-seed relative error (middle), and the histogram at $z=0.5$ (bottom), with the open channel in blue and the closed channel in red.  The full set of $12$ tricritical Ising Cardy pairs summarised in Table~\ref{tab:tricrit-annulus} is available in the companion \href{https://github.com}{\texttt{GitHub}} repository \href{https://github.com/andstergiou/nn-cft}{\tt andstergiou/nn-cft}.

\begin{figure}[H]
    \centering
    \begin{subfigure}[b]{0.49\textwidth}
        \centering
        \begin{tikzpicture}
            \begin{axis}[width=\linewidth, height=6cm, xlabel={$z$},
                ylabel={$\widetilde G^{(\mathrm o)}(z)$},
                title={$\widetilde G^{(\mathrm o)}$: Ensemble vs Exact},
                grid=major, ytick distance=0.1, legend pos=south east]
                \addplot [black, dashed, thick] table [x=z, y=Exact, col sep=space] {plots/ising_ann_ss_ensemble_g.dat};
                \addlegendentry{Exact}
                \addplot [blue, thick] table [x=z, y=Mean, col sep=space] {plots/ising_ann_ss_ensemble_g.dat};
                \addlegendentry{Mean}
                \addplot [forget plot, name path=up, draw=none] table [x=z, y=Mean_plus_Std, col sep=space] {plots/ising_ann_ss_ensemble_g.dat};
                \addplot [forget plot, name path=lo, draw=none] table [x=z, y=Mean_minus_Std, col sep=space] {plots/ising_ann_ss_ensemble_g.dat};
                \addplot [forget plot, fill=blue!30, fill opacity=0.5, draw=none] fill between [of=up and lo];
                \addlegendimage{legend image code/.code={\fill[blue!30, draw=blue!50] (0cm,-0.1cm) rectangle (0.6cm,0.1cm);}}
                \addlegendentry{Mean $\pm$ 1 Std}
            \end{axis}
        \end{tikzpicture}
    \end{subfigure}
    \hfill
    \begin{subfigure}[b]{0.49\textwidth}
        \centering
        \begin{tikzpicture}
            \begin{axis}[
                ylabel style={at={(axis description cs:1.10,0.5)}, anchor=south},
                width=\linewidth, height=6cm, xlabel={$z$},
                ylabel={$\widetilde G^{(\mathrm c)}(z)$},
                title={$\widetilde G^{(\mathrm c)}$: Ensemble vs Exact},
                grid=major, ytick distance=0.1, legend pos=south east]
                \addplot [black, dashed, thick] table [x=z, y=Exact, col sep=space] {plots/ising_ann_ss_ensemble_g.dat};
                \addlegendentry{Exact}
                \addplot [red, thick] table [x=z, y=Mean, col sep=space] {plots/ising_ann_ss_ensemble_g.dat};
                \addlegendentry{Mean}
                \addplot [forget plot, name path=uF, draw=none] table [x=z, y=Mean_plus_Std, col sep=space] {plots/ising_ann_ss_ensemble_g.dat};
                \addplot [forget plot, name path=lF, draw=none] table [x=z, y=Mean_minus_Std, col sep=space] {plots/ising_ann_ss_ensemble_g.dat};
                \addplot [forget plot, fill=red!30, fill opacity=0.5, draw=none] fill between [of=uF and lF];
                \addlegendimage{legend image code/.code={\fill[red!30, draw=red!50] (0cm,-0.1cm) rectangle (0.6cm,0.1cm);}}
                \addlegendentry{Mean $\pm$ 1 Std}
            \end{axis}
        \end{tikzpicture}
    \end{subfigure}
    \vspace{1em}
    \hspace*{-4mm}
    \begin{subfigure}[b]{0.49\textwidth}
        \centering
        \begin{tikzpicture}
            \begin{axis}[width=\linewidth, height=6cm, xlabel={$z$},
                ylabel={Error (\%)}, title={$\widetilde G^{(\mathrm o)}$: Prediction Error},
                grid=major, legend pos=north west]
                \addplot [blue, thick] table [x=z, y=PctError, col sep=space] {plots/ising_ann_ss_pct_g.dat};
                \addlegendentry{Mean Error}
                \addplot [forget plot, name path=uE, draw=none] table [x=z, y=PctError_plus_PctStd, col sep=space] {plots/ising_ann_ss_pct_g.dat};
                \addplot [forget plot, name path=lE, draw=none] table [x=z, y=PctError_minus_PctStd, col sep=space] {plots/ising_ann_ss_pct_g.dat};
                \addplot [forget plot, fill=blue!20, fill opacity=0.5, draw=none] fill between [of=uE and lE];
                \addlegendimage{legend image code/.code={\fill[blue!30, draw=blue!50] (0cm,-0.1cm) rectangle (0.6cm,0.1cm);}}
                \addlegendentry{Error $\pm$ 1 Std}
            \end{axis}
        \end{tikzpicture}
    \end{subfigure}
    \hspace*{1mm}
    \begin{subfigure}[b]{0.49\textwidth}
        \centering
        \begin{tikzpicture}
            \begin{axis}[
                ylabel style={at={(axis description cs:1.10,0.5)}, anchor=south},
                width=\linewidth, height=6cm, xlabel={$z$},
                ylabel={Error (\%)}, title={$\widetilde G^{(\mathrm c)}$: Prediction Error},
                grid=major, legend pos=north west]
                \addplot [red, thick] table [x=z, y=PctError, col sep=space] {plots/ising_ann_ss_pct_g.dat};
                \addlegendentry{Mean Error}
                \addplot [forget plot, name path=uEF, draw=none] table [x=z, y=PctError_plus_PctStd, col sep=space] {plots/ising_ann_ss_pct_g.dat};
                \addplot [forget plot, name path=lEF, draw=none] table [x=z, y=PctError_minus_PctStd, col sep=space] {plots/ising_ann_ss_pct_g.dat};
                \addplot [forget plot, fill=red!20, fill opacity=0.5, draw=none] fill between [of=uEF and lEF];
                \addlegendimage{legend image code/.code={\fill[red!30, draw=red!50] (0cm,-0.1cm) rectangle (0.6cm,0.1cm);}}
                \addlegendentry{Error $\pm$ 1 Std}
            \end{axis}
        \end{tikzpicture}
    \end{subfigure}
    \vspace{1em}
    \hspace*{1mm}
    \begin{subfigure}[b]{0.49\textwidth}
        \centering
        \begin{tikzpicture}
            \begin{axis}[width=\linewidth, height=6cm,
                xlabel={$\widetilde G^{(\mathrm o)}(z=0.5)$}, ylabel={Count},
                title={$\widetilde G^{(\mathrm o)}$: Distribution at $z=0.5$},
                ybar, bar width=0.000042, xmin=0.797320, xmax=0.802602,
                xtick={0.798,0.8,0.802}, ymin=0, ymajorgrids=true, xmajorgrids=false,
                ytick distance=5, legend pos=north east, scaled ticks=false,
                xticklabel style={ /pgf/number format/fixed, /pgf/number format/precision=4, font=\sansmath\sffamily\footnotesize }]
                \addplot [fill=blue!50, draw=black, opacity=0.7] table [x=BinCenter, y=Count, col sep=space] {plots/ising_ann_ss_hist_g.dat};
                \addlegendentry{Predictions}
                \draw [black, dashed, thick] (axis cs:0.80224,\pgfkeysvalueof{/pgfplots/ymin}) -- (axis cs:0.80224,\pgfkeysvalueof{/pgfplots/ymax});
                \addlegendimage{legend image code/.code={\draw[black, dashed, thick] (0cm,0cm) -- (0.6cm,0cm);}}
                \addlegendentry{Exact ($0.8022$)}
                \draw [blue, thick] (axis cs:0.79814,\pgfkeysvalueof{/pgfplots/ymin}) -- (axis cs:0.79814,\pgfkeysvalueof{/pgfplots/ymax});
                \addlegendimage{legend image code/.code={\draw[blue, thick] (0cm,0cm) -- (0.6cm,0cm);}}
                \addlegendentry{Mean ($0.7981$)}
            \end{axis}
        \end{tikzpicture}
    \end{subfigure}
    \begin{subfigure}[b]{0.49\textwidth}
        \centering
        \begin{tikzpicture}
            \begin{axis}[
                ylabel style={at={(axis description cs:1.10,0.5)}, anchor=south},
                width=\linewidth, height=6cm,
                xlabel={$\widetilde G^{(\mathrm c)}(z=0.5)$}, ylabel={Count},
                title={$\widetilde G^{(\mathrm c)}$: Distribution at $z=0.5$},
                ybar, bar width=0.000042, xmin=0.797320, xmax=0.802602,
                xtick={0.798,0.8,0.802}, ymin=0, ymajorgrids=true, xmajorgrids=false,
                ytick distance=5, legend pos=north east, scaled ticks=false,
                xticklabel style={ /pgf/number format/fixed, /pgf/number format/precision=4, font=\sansmath\sffamily\footnotesize }]
                \addplot [fill=red!50, draw=black, opacity=0.7] table [x=BinCenter, y=Count, col sep=space] {plots/ising_ann_ss_hist_g.dat};
                \addlegendentry{Predictions}
                \draw [black, dashed, thick] (axis cs:0.80224,\pgfkeysvalueof{/pgfplots/ymin}) -- (axis cs:0.80224,\pgfkeysvalueof{/pgfplots/ymax});
                \addlegendimage{legend image code/.code={\draw[black, dashed, thick] (0cm,0cm) -- (0.6cm,0cm);}}
                \addlegendentry{Exact ($0.8022$)}
                \draw [red, thick] (axis cs:0.79814,\pgfkeysvalueof{/pgfplots/ymin}) -- (axis cs:0.79814,\pgfkeysvalueof{/pgfplots/ymax});
                \addlegendimage{legend image code/.code={\draw[red, thick] (0cm,0cm) -- (0.6cm,0cm);}}
                \addlegendentry{Mean ($0.7981$)}
            \end{axis}
        \end{tikzpicture}
        \hspace*{-2.9mm}
    \end{subfigure}
    \caption{NN-predicted reduced annulus correlators $\widetilde G^{(\mathrm o)}(z)$ (open, blue) and $\widetilde G^{(\mathrm c)}(z)$ (closed, red) for the Ising $(\alpha,\beta)=(\boldsymbol\sigma,\boldsymbol\sigma)$ annulus over $100$ seeds.  At $z=0.5$, $\widetilde G^{\rm exact}=0.8022$ vs ensemble means $0.7981$ (open) and $0.7981$ (closed).}
    \label{fig:ising_ann_ss}
\end{figure}

\begin{figure}[H]
    \centering
    \begin{subfigure}[b]{0.49\textwidth}
        \centering
        \begin{tikzpicture}
            \begin{axis}[width=\linewidth, height=6cm, xlabel={$z$},
                ylabel={$\widetilde G^{(\mathrm o)}(z)$},
                title={$\widetilde G^{(\mathrm o)}$: Ensemble vs Exact},
                grid=major, ytick distance=0.1, legend pos=north west]
                \addplot [black, dashed, thick] table [x=z, y=Exact, col sep=space] {plots/ising_ann_1e_ensemble_g.dat};
                \addlegendentry{Exact}
                \addplot [blue, thick] table [x=z, y=Mean, col sep=space] {plots/ising_ann_1e_ensemble_g.dat};
                \addlegendentry{Mean}
                \addplot [forget plot, name path=up, draw=none] table [x=z, y=Mean_plus_Std, col sep=space] {plots/ising_ann_1e_ensemble_g.dat};
                \addplot [forget plot, name path=lo, draw=none] table [x=z, y=Mean_minus_Std, col sep=space] {plots/ising_ann_1e_ensemble_g.dat};
                \addplot [forget plot, fill=blue!30, fill opacity=0.5, draw=none] fill between [of=up and lo];
                \addlegendimage{legend image code/.code={\fill[blue!30, draw=blue!50] (0cm,-0.1cm) rectangle (0.6cm,0.1cm);}}
                \addlegendentry{Mean $\pm$ 1 Std}
            \end{axis}
        \end{tikzpicture}
    \end{subfigure}
    \hfill
    \begin{subfigure}[b]{0.49\textwidth}
        \centering
        \begin{tikzpicture}
            \begin{axis}[
                ylabel style={at={(axis description cs:1.10,0.5)}, anchor=south},
                width=\linewidth, height=6cm, xlabel={$z$},
                ylabel={$\widetilde G^{(\mathrm c)}(z)$},
                title={$\widetilde G^{(\mathrm c)}$: Ensemble vs Exact},
                grid=major, ytick distance=0.1, legend pos=north east]
                \addplot [black, dashed, thick] table [x=z, y=Exact, col sep=space] {plots/ising_ann_1e_ensemble_f.dat};
                \addlegendentry{Exact}
                \addplot [red, thick] table [x=z, y=Mean, col sep=space] {plots/ising_ann_1e_ensemble_f.dat};
                \addlegendentry{Mean}
                \addplot [forget plot, name path=uF, draw=none] table [x=z, y=Mean_plus_Std, col sep=space] {plots/ising_ann_1e_ensemble_f.dat};
                \addplot [forget plot, name path=lF, draw=none] table [x=z, y=Mean_minus_Std, col sep=space] {plots/ising_ann_1e_ensemble_f.dat};
                \addplot [forget plot, fill=red!30, fill opacity=0.5, draw=none] fill between [of=uF and lF];
                \addlegendimage{legend image code/.code={\fill[red!30, draw=red!50] (0cm,-0.1cm) rectangle (0.6cm,0.1cm);}}
                \addlegendentry{Mean $\pm$ 1 Std}
            \end{axis}
        \end{tikzpicture}
    \end{subfigure}
    \vspace{1em}
    \hspace*{-4mm}
    \begin{subfigure}[b]{0.49\textwidth}
        \centering
        \begin{tikzpicture}
            \begin{axis}[width=\linewidth, height=6cm, xlabel={$z$},
                ylabel={Error (\%)}, title={$\widetilde G^{(\mathrm o)}$: Prediction Error},
                grid=major, legend pos=north west]
                \addplot [blue, thick] table [x=z, y=PctError, col sep=space] {plots/ising_ann_1e_pct_g.dat};
                \addlegendentry{Mean Error}
                \addplot [forget plot, name path=uE, draw=none] table [x=z, y=PctError_plus_PctStd, col sep=space] {plots/ising_ann_1e_pct_g.dat};
                \addplot [forget plot, name path=lE, draw=none] table [x=z, y=PctError_minus_PctStd, col sep=space] {plots/ising_ann_1e_pct_g.dat};
                \addplot [forget plot, fill=blue!20, fill opacity=0.5, draw=none] fill between [of=uE and lE];
                \addlegendimage{legend image code/.code={\fill[blue!30, draw=blue!50] (0cm,-0.1cm) rectangle (0.6cm,0.1cm);}}
                \addlegendentry{Error $\pm$ 1 Std}
            \end{axis}
        \end{tikzpicture}
    \end{subfigure}
    \hspace*{1mm}
    \begin{subfigure}[b]{0.49\textwidth}
        \centering
        \begin{tikzpicture}
            \begin{axis}[
                ylabel style={at={(axis description cs:1.10,0.5)}, anchor=south},
                width=\linewidth, height=6cm, xlabel={$z$},
                ylabel={Error (\%)}, title={$\widetilde G^{(\mathrm c)}$: Prediction Error},
                grid=major, legend style={at={(0.5,0.98)}, anchor=north}]
                \addplot [red, thick] table [x=z, y=PctError, col sep=space] {plots/ising_ann_1e_pct_f.dat};
                \addlegendentry{Mean Error}
                \addplot [forget plot, name path=uEF, draw=none] table [x=z, y=PctError_plus_PctStd, col sep=space] {plots/ising_ann_1e_pct_f.dat};
                \addplot [forget plot, name path=lEF, draw=none] table [x=z, y=PctError_minus_PctStd, col sep=space] {plots/ising_ann_1e_pct_f.dat};
                \addplot [forget plot, fill=red!20, fill opacity=0.5, draw=none] fill between [of=uEF and lEF];
                \addlegendimage{legend image code/.code={\fill[red!30, draw=red!50] (0cm,-0.1cm) rectangle (0.6cm,0.1cm);}}
                \addlegendentry{Error $\pm$ 1 Std}
            \end{axis}
        \end{tikzpicture}
    \end{subfigure}
    \vspace{1em}
    \hspace*{1mm}
    \begin{subfigure}[b]{0.49\textwidth}
        \centering
        \begin{tikzpicture}
            \begin{axis}[width=\linewidth, height=6cm,
                xlabel={$\widetilde G^{(\mathrm o)}(z=0.5)$}, ylabel={Count},
                title={$\widetilde G^{(\mathrm o)}$: Distribution at $z=0.5$},
                ybar, bar width=0.000199, xmin=0.028414, xmax=0.034181,
                xtick={0.03,0.032,0.034}, ymin=0, ymajorgrids=true, xmajorgrids=false,
                ytick distance=5, legend pos=north east, scaled ticks=false,
                xticklabel style={ /pgf/number format/fixed, /pgf/number format/precision=4, font=\sansmath\sffamily\footnotesize }]
                \addplot [fill=blue!50, draw=black, opacity=0.7] table [x=BinCenter, y=Count, col sep=space] {plots/ising_ann_1e_hist_g.dat};
                \addlegendentry{Predictions}
                \draw [black, dashed, thick] (axis cs:0.03329,\pgfkeysvalueof{/pgfplots/ymin}) -- (axis cs:0.03329,\pgfkeysvalueof{/pgfplots/ymax});
                \addlegendimage{legend image code/.code={\draw[black, dashed, thick] (0cm,0cm) -- (0.6cm,0cm);}}
                \addlegendentry{Exact ($0.0333$)}
                \draw [blue, thick] (axis cs:0.03168,\pgfkeysvalueof{/pgfplots/ymin}) -- (axis cs:0.03168,\pgfkeysvalueof{/pgfplots/ymax});
                \addlegendimage{legend image code/.code={\draw[blue, thick] (0cm,0cm) -- (0.6cm,0cm);}}
                \addlegendentry{Mean ($0.0317$)}
            \end{axis}
        \end{tikzpicture}
    \end{subfigure}
    \begin{subfigure}[b]{0.49\textwidth}
        \centering
        \begin{tikzpicture}
            \begin{axis}[
                ylabel style={at={(axis description cs:1.10,0.5)}, anchor=south},
                width=\linewidth, height=6cm,
                xlabel={$\widetilde G^{(\mathrm c)}(z=0.5)$}, ylabel={Count},
                title={$\widetilde G^{(\mathrm c)}$: Distribution at $z=0.5$},
                ybar, bar width=0.000249, xmin=0.027835, xmax=0.035053,
                xtick={0.028,0.03,0.032,0.034}, ymin=0, ymajorgrids=true, xmajorgrids=false,
                ytick distance=5, legend pos=north east, scaled ticks=false,
                xticklabel style={ /pgf/number format/fixed, /pgf/number format/precision=4, font=\sansmath\sffamily\footnotesize }]
                \addplot [fill=red!50, draw=black, opacity=0.7] table [x=BinCenter, y=Count, col sep=space] {plots/ising_ann_1e_hist_f.dat};
                \addlegendentry{Predictions}
                \draw [black, dashed, thick] (axis cs:0.03329,\pgfkeysvalueof{/pgfplots/ymin}) -- (axis cs:0.03329,\pgfkeysvalueof{/pgfplots/ymax});
                \addlegendimage{legend image code/.code={\draw[black, dashed, thick] (0cm,0cm) -- (0.6cm,0cm);}}
                \addlegendentry{Exact ($0.0333$)}
                \draw [red, thick] (axis cs:0.03073,\pgfkeysvalueof{/pgfplots/ymin}) -- (axis cs:0.03073,\pgfkeysvalueof{/pgfplots/ymax});
                \addlegendimage{legend image code/.code={\draw[red, thick] (0cm,0cm) -- (0.6cm,0cm);}}
                \addlegendentry{Mean ($0.0307$)}
            \end{axis}
        \end{tikzpicture}
        \hspace*{-2.9mm}
    \end{subfigure}
    \caption{NN-predicted reduced annulus correlators $\widetilde G^{(\mathrm o)}(z)$ (open, blue) and $\widetilde G^{(\mathrm c)}(z)$ (closed, red) for the Ising $(\alpha,\beta)=(\mathbf 1,\boldsymbol\varepsilon)$ annulus over $100$ seeds.  At $z=0.5$, $\widetilde G^{\rm exact}=0.0333$ vs ensemble means $0.0317$ (open) and $0.0307$ (closed).}
    \label{fig:ising_ann_1e}
\end{figure}

We also collect here the remaining six WZW reconstructions of Section~\ref{sec:annulus-wzw} listed in Table~\ref{tab:wzw-annulus}.

\begin{figure}[H]
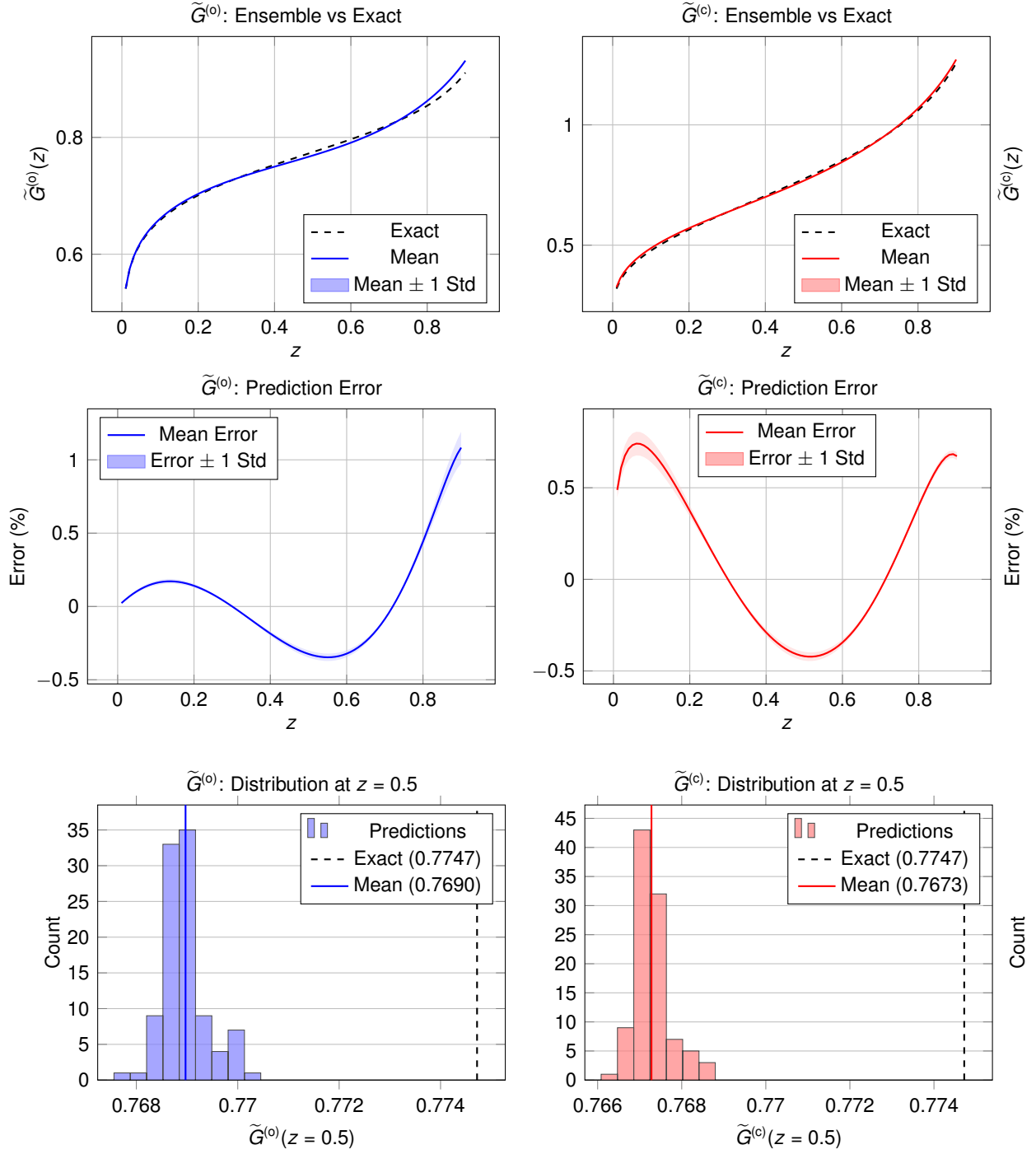

    \centering
    \begin{subfigure}[b]{0.49\textwidth}
        \centering

        \hspace*{-2.9mm}
    \end{subfigure}
    \caption{NN-predicted reduced annulus correlators $\widetilde G^{(\mathrm o)}(z)$ (open, blue) and $\widetilde G^{(\mathrm c)}(z)$ (closed, red) for the $\widehat{\mathfrak{su}}(2)_2$ $(\lambda_0,\lambda_0)$ annulus over $100$ seeds.  At $z=0.5$, $\widetilde G^{\rm exact}=0.7747$ vs ensemble means $0.7690$ (open) and $0.7673$ (closed).}
    \label{fig:wzw_su2_k2_00}
\end{figure}
\begin{figure}[H]
    \centering
    \begin{subfigure}[b]{0.49\textwidth}
        \centering
        %
        \hspace*{-2.9mm}
    \end{subfigure}
    \caption{NN-predicted reduced annulus correlators $\widetilde G^{(\mathrm o)}(z)$ (open, blue) and $\widetilde G^{(\mathrm c)}(z)$ (closed, red) for the $\widehat{\mathfrak{su}}(2)_2$ $(\lambda_0,\lambda_1)$ annulus over $100$ seeds.  At $z=0.5$, $\widetilde G^{\rm exact}=0.4770$ vs ensemble means $0.4775$ (open) and $0.4798$ (closed).}
    \label{fig:wzw_su2_k2_0h}
\end{figure}

\begin{figure}[H]
    \centering
    \begin{subfigure}[b]{0.49\textwidth}
        \centering
        %
        \hspace*{-2.9mm}
    \end{subfigure}
    \caption{NN-predicted reduced annulus correlators $\widetilde G^{(\mathrm o)}(z)$ (open, blue) and $\widetilde G^{(\mathrm c)}(z)$ (closed, red) for the $\widehat{\mathfrak{su}}(2)_2$ $(\lambda_0,\lambda_2)$ annulus over $100$ seeds.  At $z=0.5$, $\widetilde G^{\rm exact}=0.1001$ vs ensemble means $0.0977$ (open) and $0.0969$ (closed).}
    \label{fig:wzw_su2_k2_01}
\end{figure}

\begin{figure}[H]
    \centering
    \begin{subfigure}[b]{0.49\textwidth}
        \centering
        %
        \hspace*{-2.9mm}
    \end{subfigure}
    \caption{NN-predicted reduced annulus correlators $\widetilde G^{(\mathrm o)}(z)$ (open, blue) and $\widetilde G^{(\mathrm c)}(z)$ (closed, red) for the $\widehat{\mathfrak{su}}(3)_1$ $(\lambda_0,\lambda_0)$ annulus over $100$ seeds.  At $z=0.5$, $\widetilde G^{\rm exact}=0.7827$ vs ensemble means $0.7761$ (open) and $0.7747$ (closed).}
    \label{fig:wzw_su3_k1_00}
\end{figure}
\begin{figure}[H]
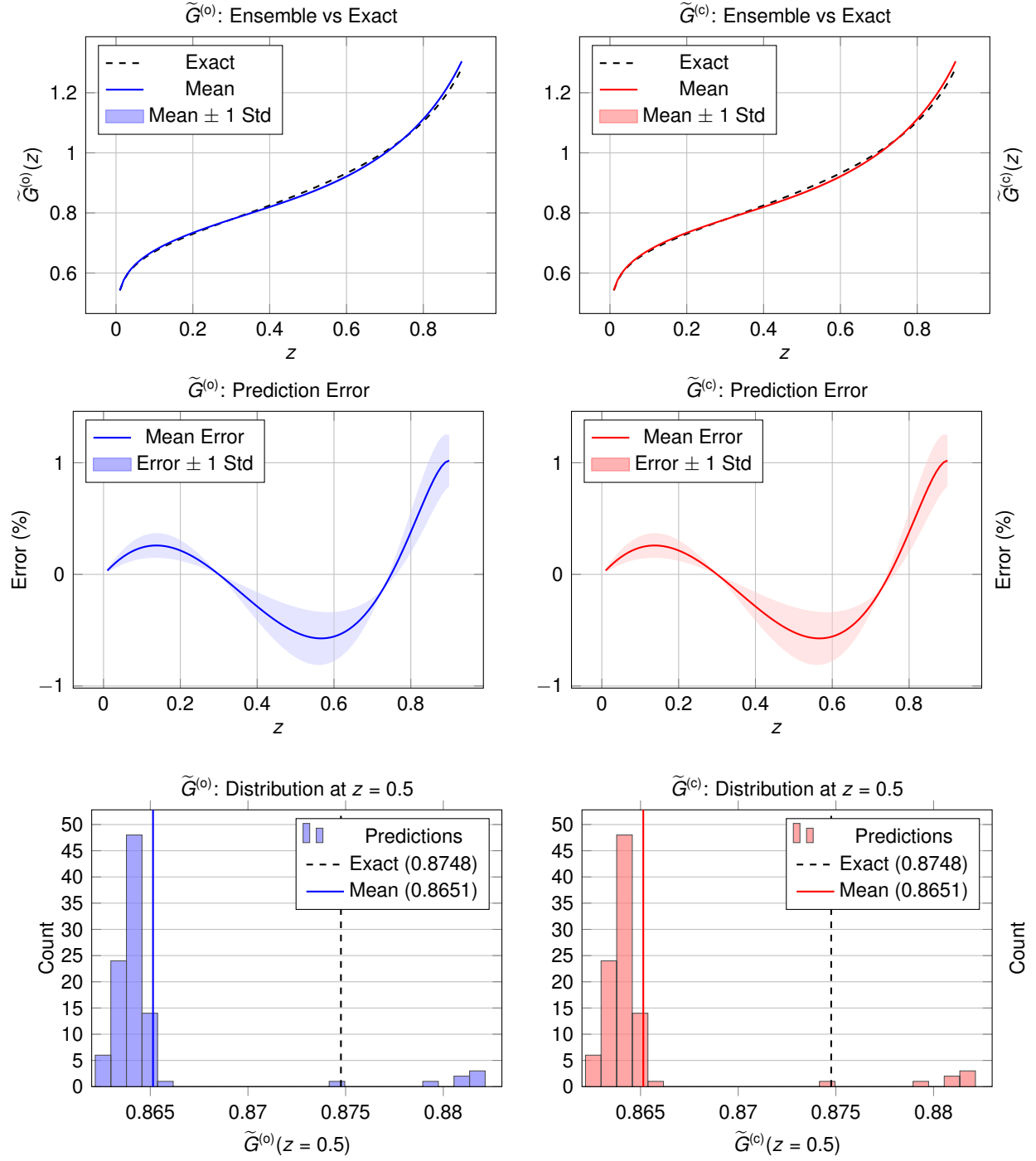

    \centering
    \begin{subfigure}[b]{0.49\textwidth}
        \centering
        %
        \hspace*{-2.9mm}
    \end{subfigure}
    \caption{NN-predicted reduced annulus correlators for the $\widehat{\mathfrak{su}}(2)_2$ $(\lambda_1,\lambda_1)$ pair over $100$ seeds.  The open spectrum $N_{11}{}^j\chi_j=\chi_0+\chi_2$ is $S$-invariant, so $\widetilde G^{(\mathrm o)}=\widetilde G^{(\mathrm c)}$ and both channels coincide.  At $z=0.5$, $\widetilde G^{\rm exact}=0.8748$ vs mean $0.8651$.}
    \label{fig:wzw_su2_k2_hh}
\end{figure}

\bibliography{ancb2}

\end{document}